\documentclass[preprint]{aastex}
\usepackage{float}

\usepackage{graphics}
\usepackage{longtable}

\newcommand{\degree}{\ensuremath{^\circ}}
\newcommand{\chisqnu}{\ensuremath{\chi^2/{\rm dof}}}
\newcommand{\chisq}{\ensuremath{\chi^2}}

\newcommand{\Ho}{$H_{0}$}

\newcommand{\VarCut}{1.5}
\newcommand{\VarClrUp}{$(V-I)>2.5$}
\newcommand{\VarClrDown}{$(V-I)<0$}
\newcommand{\FtestCut}{$3$}
\newcommand{\BlueSigma}{$2\sigma{}$}
\newcommand{\DistCut}{$11.5$}
\newcommand{\PerCut}{$3$ days}

\newcommand{\CenterFirstRA}{$14^{\rm h}03^{\rm m}28.\!\!^{\rm{s}}8$}
\newcommand{\CenterFirstDec}{$+54\degree21'35.\!\!''0$}
\newcommand{\CenterSecondRA}{$14^{\rm h}02^{\rm m}51.\!\!^{\rm{s}}0$}
\newcommand{\CenterSecondDec}{$+54\degree20'22.\!\!{''}0$}

\newcommand{\MacriLMCdist}{$18.41$}
\newcommand{\MacriLMCdistRan}{$0.10$}
\newcommand{\MacriLMCdistSys}{$0.13$}

\newcommand{\Maserdist}{$29.29$}

\newcommand{\LMCreldist}{\Delta \mu_{\textnormal{\small{LMC}}}}
\newcommand{\LMCdist}{\mu_{\textnormal{\small{LMC}}}}
\newcommand{\HII}{H \tiny{II} \normalsize}

\newcommand{\Astrometry}{Astrometry presented is referenced to the HST headers of j9o401010\_drz.fits for Field 1 and j9o413010\_drz.fits for Field 2.  }

\begin{document}

\newcommand{\TRGBedge}{$25.00$}
\newcommand{\TRGBDistMod}{$29.05$}
\newcommand{\TRGBsys}{$0.12$}
\newcommand{\TRGBBootSig}{$0.06$}
\newcommand{\TRGBBootSigUp}{$0.12$}
\newcommand{\TRGBBootSigDown}{$0.00$}
\newcommand{\TRGBSmoothBootSig}{$0.06$}
\newcommand{\TRGBSmoothBootSigUp}{$0.15$}
\newcommand{\TRGBSmoothBootSigDown}{$-0.04$}
\newcommand{\TRGBSmoothBootMean}{$25.11$}
\newcommand{\TRGBSmoothBootMed}{$25.13$}
\newcommand{\TRGBSBootMean}{$25.03$}
\newcommand{\TRGBBootMed}{$25.01$}
\newcommand{\TRGBConInt}{$68$}
\newcommand{\TRGBRadCut}{$4.75$}
\newcommand{\TRGBClrCut}{$1.00$}
\newcommand{\TRGBsourceOrigFone}{$3.75 \times 10^{5}$}
\newcommand{\TRGBsourceRadFone}{$5.64 \times 10^{3}$}
\newcommand{\TRGBsourceClrFone}{$2.65 \times 10^{3}$}
\newcommand{\TRGBsourceOrigFtwo}{$3.90 \times 10^{5}$}
\newcommand{\TRGBsourceRadFtwo}{$3.69 \times 10^{4}$}
\newcommand{\TRGBsourceClrFtwo}{$2.22 \times 10^{4}$}
\newcommand{\TRGBsourceFin}{$2.48 \times 10^{4}$}
\newcommand{\TRGBnumBelow}{$7.14 \times 10^{3}$}
\newcommand{\TRGBboot}{$2.5 \times 10^{4}$}

\newcommand{\WPLConInt}{$68$}
\newcommand{\WPLALONEvFSLOPE}{$99.99$}
\newcommand{\WPLFREEvFSLOPE}{$93.6$}
\newcommand{\WPLALONEvFSLOPEChi}{$15.5$}
\newcommand{\WPLFREEvFSLOPEChi}{$3.2$}
\newcommand{\WPLawMETslope}{3.1}
\newcommand{\WPLawMETslopeUP}{1.8}
\newcommand{\WPLawMETslopeDOWN}{1.7}
\newcommand{\WPLawMETslopeerr}{1.8}
\newcommand{\WPLawMETslopeerrSys}{0.2}
\newcommand{\WPLawMETinter}{-3.1}
\newcommand{\WPLawMETinterUP}{0.1}
\newcommand{\WPLawMETinterDOWN}{0.1}
\newcommand{\WPLawMETintererr}{0.1}
\newcommand{\WPLawMETintererrSys}{0.2}
\newcommand{\WPLbwMETslope}{-0.72}
\newcommand{\WPLbwMETslopeUP}{0.20}
\newcommand{\WPLbwMETslopeDOWN}{0.22}
\newcommand{\WPLbwMETslopeerr}{0.21}
\newcommand{\WPLbwMETslopeerrSys}{0.06}
\newcommand{\WPLbwMETinter}{22.21}
\newcommand{\WPLbwMETinterUP}{0.01}
\newcommand{\WPLbwMETinterDOWN}{0.01}
\newcommand{\WPLbwMETintererr}{0.01}
\newcommand{\WPLbwMETintererrSys}{0.04}
\newcommand{\WPLawMETinterFixSlope}{-3.1}
\newcommand{\WPLawMETinterUPFixSlope}{0.1}
\newcommand{\WPLawMETinterDOWNFixSlope}{0.1}
\newcommand{\WPLawMETintererrFixSlope}{0.1}
\newcommand{\WPLawMETintererrSysFixSlope}{-0.0}
\newcommand{\WPLbwMETslopeFixSlope}{-0.75}
\newcommand{\WPLbwMETslopeUPFixSlope}{0.21}
\newcommand{\WPLbwMETslopeDOWNFixSlope}{0.22}
\newcommand{\WPLbwMETslopeerrFixSlope}{0.22}
\newcommand{\WPLbwMETslopeerrSysFixSlope}{0.06}
\newcommand{\WPLbwMETinterFixSlope}{22.22}
\newcommand{\WPLbwMETinterUPFixSlope}{0.01}
\newcommand{\WPLbwMETinterDOWNFixSlope}{0.01}
\newcommand{\WPLbwMETintererrFixSlope}{0.01}
\newcommand{\WPLbwMETintererrSysFixSlope}{0.04}
\newcommand{\WPLawMETslopeFixInter}{3.8}
\newcommand{\WPLawMETslopeUPFixInter}{1.7}
\newcommand{\WPLawMETslopeDOWNFixInter}{1.7}
\newcommand{\WPLawMETslopeerrFixInter}{1.7}
\newcommand{\WPLawMETslopeerrSysFixInter}{0.3}
\newcommand{\WPLawMETinterFixInter}{-3.1}
\newcommand{\WPLawMETinterUPFixInter}{0.1}
\newcommand{\WPLawMETinterDOWNFixInter}{0.1}
\newcommand{\WPLawMETintererrFixInter}{0.1}
\newcommand{\WPLawMETintererrSysFixInter}{0.2}
\newcommand{\WPLbwMETinterFixInter}{22.20}
\newcommand{\WPLbwMETinterUPFixInter}{0.01}
\newcommand{\WPLbwMETinterDOWNFixInter}{0.01}
\newcommand{\WPLbwMETintererrFixInter}{0.01}
\newcommand{\WPLbwMETintererrSysFixInter}{-0.00}
\newcommand{\WPLawMETinterAlone}{-3.1}
\newcommand{\WPLawMETinterUPAlone}{0.1}
\newcommand{\WPLawMETinterDOWNAlone}{0.1}
\newcommand{\WPLawMETintererrAlone}{0.1}
\newcommand{\WPLawMETintererrSysAlone}{0.0}
\newcommand{\WPLbwMETinterAlone}{22.21}
\newcommand{\WPLbwMETinterUPAlone}{0.01}
\newcommand{\WPLbwMETinterDOWNAlone}{0.01}
\newcommand{\WPLbwMETintererrAlone}{0.01}
\newcommand{\WPLbwMETintererrSysAlone}{0.00}
\newcommand{\WPLawslope}{-2.8}
\newcommand{\WPLawslopeUP}{1.6}
\newcommand{\WPLawslopeDOWN}{1.6}
\newcommand{\WPLawslopeerr}{1.6}
\newcommand{\WPLawinter}{-3.1}
\newcommand{\WPLawinterUP}{0.1}
\newcommand{\WPLawinterDOWN}{0.1}
\newcommand{\WPLawintererr}{0.1}
\newcommand{\WPLbwslope}{0.65}
\newcommand{\WPLbwslopeUP}{0.20}
\newcommand{\WPLbwslopeDOWN}{0.18}
\newcommand{\WPLbwslopeerr}{0.19}
\newcommand{\WPLbwinter}{22.21}
\newcommand{\WPLbwinterUP}{0.01}
\newcommand{\WPLbwinterDOWN}{0.01}
\newcommand{\WPLbwintererr}{0.01}
\newcommand{\WPLFitRedChi}{0.94}
\newcommand{\WPLFitChi}{260.1}
\newcommand{\WPLFitDoF}{277}
\newcommand{\WPLawinterFixSlope}{-3.1}
\newcommand{\WPLawinterUPFixSlope}{0.1}
\newcommand{\WPLawinterDOWNFixSlope}{0.1}
\newcommand{\WPLawintererrFixSlope}{0.1}
\newcommand{\WPLbwslopeFixSlope}{0.68}
\newcommand{\WPLbwslopeUPFixSlope}{0.20}
\newcommand{\WPLbwslopeDOWNFixSlope}{0.18}
\newcommand{\WPLbwslopeerrFixSlope}{0.19}
\newcommand{\WPLbwinterFixSlope}{22.22}
\newcommand{\WPLbwinterUPFixSlope}{0.01}
\newcommand{\WPLbwinterDOWNFixSlope}{0.01}
\newcommand{\WPLbwintererrFixSlope}{0.01}
\newcommand{\WPLFixSlopeRedChi}{0.95}
\newcommand{\WPLFixSlopeChi}{263.4}
\newcommand{\WPLFixSlopeDoF}{278}
\newcommand{\WPLawslopeFixInter}{-3.4}
\newcommand{\WPLawslopeUPFixInter}{1.5}
\newcommand{\WPLawslopeDOWNFixInter}{1.6}
\newcommand{\WPLawslopeerrFixInter}{1.6}
\newcommand{\WPLawinterFixInter}{-3.1}
\newcommand{\WPLawinterUPFixInter}{0.1}
\newcommand{\WPLawinterDOWNFixInter}{0.1}
\newcommand{\WPLawintererrFixInter}{0.1}
\newcommand{\WPLbwinterFixInter}{22.20}
\newcommand{\WPLbwinterUPFixInter}{0.01}
\newcommand{\WPLbwinterDOWNFixInter}{0.01}
\newcommand{\WPLbwintererrFixInter}{0.01}
\newcommand{\WPLFixInterRedChi}{0.98}
\newcommand{\WPLFixInterChi}{273.8}
\newcommand{\WPLFixInterDoF}{278}
\newcommand{\WPLawinterAlone}{-3.1}
\newcommand{\WPLawinterUPAlone}{0.1}
\newcommand{\WPLawinterDOWNAlone}{0.1}
\newcommand{\WPLawintererrAlone}{0.1}
\newcommand{\WPLbwinterAlone}{22.21}
\newcommand{\WPLbwinterUPAlone}{0.01}
\newcommand{\WPLbwinterDOWNAlone}{0.01}
\newcommand{\WPLbwintererrAlone}{0.01}
\newcommand{\WPLAloneRedChi}{1.00}
\newcommand{\WPLAloneChi}{278.8}
\newcommand{\WPLAloneDoF}{279}
\newcommand{\WPLAddErr}{$0.069$}
\newcommand{\WPLNumCeph}{$297$}
\newcommand{\WPLNumboot}{$2.5 \times 10^{4}$}
\newcommand{\WPLScalePho}{$0.2$}
\newcommand{\WPLScaleMet}{$8.57$}
\newcommand{\WPLScaleLogPer}{$1.3$}
\newcommand{\WPLNumScatCalc}{$100$}
\newcommand{\WPLPerLimit}{$100.0$}

\newcommand{\WPLConIntfourty}{$68$}
\newcommand{\WPLALONEvFSLOPEfourty}{$99.98$}
\newcommand{\WPLFREEvFSLOPEfourty}{$89.4$}
\newcommand{\WPLALONEvFSLOPEChifourty}{$13.7$}
\newcommand{\WPLFREEvFSLOPEChifourty}{$2.5$}
\newcommand{\WPLawMETslopefourty}{3.1}
\newcommand{\WPLawMETslopeUPfourty}{1.9}
\newcommand{\WPLawMETslopeDOWNfourty}{1.9}
\newcommand{\WPLawMETslopeerrfourty}{1.9}
\newcommand{\WPLawMETslopeerrSysfourty}{0.2}
\newcommand{\WPLawMETinterfourty}{-3.0}
\newcommand{\WPLawMETinterUPfourty}{0.1}
\newcommand{\WPLawMETinterDOWNfourty}{0.1}
\newcommand{\WPLawMETintererrfourty}{0.1}
\newcommand{\WPLawMETintererrSysfourty}{0.2}
\newcommand{\WPLbwMETslopefourty}{-0.72}
\newcommand{\WPLbwMETslopeUPfourty}{0.19}
\newcommand{\WPLbwMETslopeDOWNfourty}{0.22}
\newcommand{\WPLbwMETslopeerrfourty}{0.21}
\newcommand{\WPLbwMETslopeerrSysfourty}{0.06}
\newcommand{\WPLbwMETinterfourty}{22.21}
\newcommand{\WPLbwMETinterUPfourty}{0.01}
\newcommand{\WPLbwMETinterDOWNfourty}{0.01}
\newcommand{\WPLbwMETintererrfourty}{0.01}
\newcommand{\WPLbwMETintererrSysfourty}{0.04}
\newcommand{\WPLawMETinterFixSlopefourty}{-3.0}
\newcommand{\WPLawMETinterUPFixSlopefourty}{0.1}
\newcommand{\WPLawMETinterDOWNFixSlopefourty}{0.1}
\newcommand{\WPLawMETintererrFixSlopefourty}{0.1}
\newcommand{\WPLawMETintererrSysFixSlopefourty}{-0.0}
\newcommand{\WPLbwMETslopeFixSlopefourty}{-0.76}
\newcommand{\WPLbwMETslopeUPFixSlopefourty}{0.21}
\newcommand{\WPLbwMETslopeDOWNFixSlopefourty}{0.22}
\newcommand{\WPLbwMETslopeerrFixSlopefourty}{0.22}
\newcommand{\WPLbwMETslopeerrSysFixSlopefourty}{0.06}
\newcommand{\WPLbwMETinterFixSlopefourty}{22.21}
\newcommand{\WPLbwMETinterUPFixSlopefourty}{0.01}
\newcommand{\WPLbwMETinterDOWNFixSlopefourty}{0.01}
\newcommand{\WPLbwMETintererrFixSlopefourty}{0.01}
\newcommand{\WPLbwMETintererrSysFixSlopefourty}{0.04}
\newcommand{\WPLawMETslopeFixInterfourty}{3.9}
\newcommand{\WPLawMETslopeUPFixInterfourty}{1.9}
\newcommand{\WPLawMETslopeDOWNFixInterfourty}{1.8}
\newcommand{\WPLawMETslopeerrFixInterfourty}{1.9}
\newcommand{\WPLawMETslopeerrSysFixInterfourty}{0.3}
\newcommand{\WPLawMETinterFixInterfourty}{-3.1}
\newcommand{\WPLawMETinterUPFixInterfourty}{0.1}
\newcommand{\WPLawMETinterDOWNFixInterfourty}{0.1}
\newcommand{\WPLawMETintererrFixInterfourty}{0.1}
\newcommand{\WPLawMETintererrSysFixInterfourty}{0.2}
\newcommand{\WPLbwMETinterFixInterfourty}{22.20}
\newcommand{\WPLbwMETinterUPFixInterfourty}{0.01}
\newcommand{\WPLbwMETinterDOWNFixInterfourty}{0.01}
\newcommand{\WPLbwMETintererrFixInterfourty}{0.01}
\newcommand{\WPLbwMETintererrSysFixInterfourty}{-0.00}
\newcommand{\WPLawMETinterAlonefourty}{-3.0}
\newcommand{\WPLawMETinterUPAlonefourty}{0.1}
\newcommand{\WPLawMETinterDOWNAlonefourty}{0.1}
\newcommand{\WPLawMETintererrAlonefourty}{0.1}
\newcommand{\WPLawMETintererrSysAlonefourty}{0.0}
\newcommand{\WPLbwMETinterAlonefourty}{22.20}
\newcommand{\WPLbwMETinterUPAlonefourty}{0.01}
\newcommand{\WPLbwMETinterDOWNAlonefourty}{0.01}
\newcommand{\WPLbwMETintererrAlonefourty}{0.01}
\newcommand{\WPLbwMETintererrSysAlonefourty}{0.00}
\newcommand{\WPLawslopefourty}{-2.8}
\newcommand{\WPLawslopeUPfourty}{1.7}
\newcommand{\WPLawslopeDOWNfourty}{1.7}
\newcommand{\WPLawslopeerrfourty}{1.7}
\newcommand{\WPLawinterfourty}{-3.0}
\newcommand{\WPLawinterUPfourty}{0.1}
\newcommand{\WPLawinterDOWNfourty}{0.1}
\newcommand{\WPLawintererrfourty}{0.1}
\newcommand{\WPLbwslopefourty}{0.65}
\newcommand{\WPLbwslopeUPfourty}{0.20}
\newcommand{\WPLbwslopeDOWNfourty}{0.17}
\newcommand{\WPLbwslopeerrfourty}{0.19}
\newcommand{\WPLbwinterfourty}{22.21}
\newcommand{\WPLbwinterUPfourty}{0.01}
\newcommand{\WPLbwinterDOWNfourty}{0.01}
\newcommand{\WPLbwintererrfourty}{0.01}
\newcommand{\WPLFitRedChifourty}{0.95}
\newcommand{\WPLFitChifourty}{248.8}
\newcommand{\WPLFitDoFfourty}{263}
\newcommand{\WPLawinterFixSlopefourty}{-3.0}
\newcommand{\WPLawinterUPFixSlopefourty}{0.1}
\newcommand{\WPLawinterDOWNFixSlopefourty}{0.1}
\newcommand{\WPLawintererrFixSlopefourty}{0.1}
\newcommand{\WPLbwslopeFixSlopefourty}{0.68}
\newcommand{\WPLbwslopeUPFixSlopefourty}{0.20}
\newcommand{\WPLbwslopeDOWNFixSlopefourty}{0.19}
\newcommand{\WPLbwslopeerrFixSlopefourty}{0.20}
\newcommand{\WPLbwinterFixSlopefourty}{22.21}
\newcommand{\WPLbwinterUPFixSlopefourty}{0.01}
\newcommand{\WPLbwinterDOWNFixSlopefourty}{0.01}
\newcommand{\WPLbwintererrFixSlopefourty}{0.01}
\newcommand{\WPLFixSlopeRedChifourty}{0.95}
\newcommand{\WPLFixSlopeChifourty}{251.3}
\newcommand{\WPLFixSlopeDoFfourty}{264}
\newcommand{\WPLawslopeFixInterfourty}{-3.5}
\newcommand{\WPLawslopeUPFixInterfourty}{1.6}
\newcommand{\WPLawslopeDOWNFixInterfourty}{1.7}
\newcommand{\WPLawslopeerrFixInterfourty}{1.7}
\newcommand{\WPLawinterFixInterfourty}{-3.1}
\newcommand{\WPLawinterUPFixInterfourty}{0.1}
\newcommand{\WPLawinterDOWNFixInterfourty}{0.1}
\newcommand{\WPLawintererrFixInterfourty}{0.1}
\newcommand{\WPLbwinterFixInterfourty}{22.20}
\newcommand{\WPLbwinterUPFixInterfourty}{0.01}
\newcommand{\WPLbwinterDOWNFixInterfourty}{0.01}
\newcommand{\WPLbwintererrFixInterfourty}{0.01}
\newcommand{\WPLFixInterRedChifourty}{0.99}
\newcommand{\WPLFixInterChifourty}{260.7}
\newcommand{\WPLFixInterDoFfourty}{264}
\newcommand{\WPLawinterAlonefourty}{-3.0}
\newcommand{\WPLawinterUPAlonefourty}{0.1}
\newcommand{\WPLawinterDOWNAlonefourty}{0.1}
\newcommand{\WPLawintererrAlonefourty}{0.1}
\newcommand{\WPLbwinterAlonefourty}{22.20}
\newcommand{\WPLbwinterUPAlonefourty}{0.01}
\newcommand{\WPLbwinterDOWNAlonefourty}{0.01}
\newcommand{\WPLbwintererrAlonefourty}{0.01}
\newcommand{\WPLAloneRedChifourty}{1.00}
\newcommand{\WPLAloneChifourty}{265.0}
\newcommand{\WPLAloneDoFfourty}{265}
\newcommand{\WPLAddErrfourty}{$0.087$}
\newcommand{\WPLNumCephfourty}{$278$}
\newcommand{\WPLNumbootfourty}{$2.5 \times 10^{4}$}
\newcommand{\WPLScalePhofourty}{$0.2$}
\newcommand{\WPLScaleMetfourty}{$8.57$}
\newcommand{\WPLScaleLogPerfourty}{$1.3$}
\newcommand{\WPLNumScatCalcfourty}{$100$}
\newcommand{\WPLPerLimitfourty}{$40.0$}

\newcommand{\MetSlope}{$-0.84$}
\newcommand{\MetSlopeErrRan}{$0.22$}
\newcommand{\MetSlopeErrTot}{$0.23$}
\newcommand{\MetSlopeErrSys}{$0.07$}
\newcommand{\MetSlopeErrUp}{$0.22$}
\newcommand{\MetSlopeErrDown}{$0.22$}
\newcommand{\MetLMCDist}{$10.63$}
\newcommand{\MetDist}{$29.04$}
\newcommand{\MetDistErrRan}{$0.02$}
\newcommand{\MetDistErrTot}{$0.05$}
\newcommand{\MetDistErrSys}{$0.04$}
\newcommand{\MetDistErrUp}{$0.02$}
\newcommand{\MetDistErrDown}{$0.02$}
\newcommand{\Metboot}{$2.5 \times 10^{4}$}
\newcommand{\MetCephs}{$235$}
\newcommand{\MetEVIup}{$0.39$}
\newcommand{\MetEVIdown}{$0.015$}
\newcommand{\MetEBVup}{$0.28$}
\newcommand{\MetEBVdown}{$0.011$}
\newcommand{\MetSigKen}{$2.2$}
\newcommand{\MetStdDev}{$0.17$}
\newcommand{\MetVar}{$0.030$}
\newcommand{\MetStdDevHor}{$0.18$}
\newcommand{\MetChisqHor}{$1142.8$}
\newcommand{\MetChisq}{$1077.1$}
\newcommand{\LMCsysErr}{$0.07$}
\newcommand{\LMCranErr}{$0.04$}
\newcommand{\LMCtotErr}{$0.08$}
\newcommand{\LMCpercentErr}{$4$}
\newcommand{\MsysErr}{$0.18$}
\newcommand{\MranErr}{$0.05$}
\newcommand{\MtotErr}{$0.18$}
\newcommand{\MdistMpc}{$6.4$}
\newcommand{\MdistMpcErrTot}{$0.5$}
\newcommand{\MdistMpcErrRan}{$0.2$}
\newcommand{\MdistMpcErrSys}{$0.5$}
\newcommand{\MpercentErr}{$8$}
\newcommand{\MGOALpercentErr}{$5$}
\newcommand{\MGOALsysErr}{$0.09$}
\newcommand{\MGOALranErr}{$0.06$}
\newcommand{\MGOALtotErr}{$0.11$}

\newcommand{\MetSlopecut}{$-0.66$}
\newcommand{\MetSlopeErrRancut}{$0.25$}
\newcommand{\MetSlopeErrTotcut}{$0.26$}
\newcommand{\MetSlopeErrSyscut}{$0.05$}
\newcommand{\MetSlopeErrUpcut}{$0.25$}
\newcommand{\MetSlopeErrDowncut}{$0.25$}
\newcommand{\MetLMCDistcut}{$10.62$}
\newcommand{\MetDistcut}{$29.03$}
\newcommand{\MetDistErrRancut}{$0.02$}
\newcommand{\MetDistErrTotcut}{$0.04$}
\newcommand{\MetDistErrSyscut}{$0.03$}
\newcommand{\MetDistErrUpcut}{$0.02$}
\newcommand{\MetDistErrDowncut}{$0.02$}
\newcommand{\Metbootcut}{$2.5 \times 10^{4}$}
\newcommand{\MetCephscut}{$235$}
\newcommand{\MetEVIupcut}{$0.39$}
\newcommand{\MetEVIdowncut}{$0.015$}
\newcommand{\MetEBVupcut}{$0.28$}
\newcommand{\MetEBVdowncut}{$0.011$}
\newcommand{\MetSigKencut}{$1.4$}
\newcommand{\MetStdDevcut}{$0.17$}
\newcommand{\MetVarcut}{$0.029$}
\newcommand{\MetStdDevHorcut}{$0.17$}
\newcommand{\MetChisqHorcut}{$991.9$}
\newcommand{\MetChisqcut}{$961.9$}
\newcommand{\LMCsysErrcut}{$0.06$}
\newcommand{\LMCranErrcut}{$0.04$}
\newcommand{\LMCtotErrcut}{$0.07$}
\newcommand{\LMCpercentErrcut}{$3$}
\newcommand{\MsysErrcut}{$0.17$}
\newcommand{\MranErrcut}{$0.05$}
\newcommand{\MtotErrcut}{$0.18$}
\newcommand{\MdistMpccut}{$6.4$}
\newcommand{\MdistMpcErrTotcut}{$0.5$}
\newcommand{\MdistMpcErrRancut}{$0.2$}
\newcommand{\MdistMpcErrSyscut}{$0.5$}
\newcommand{\MpercentErrcut}{$8$}
\newcommand{\MGOALpercentErrcut}{$5$}
\newcommand{\MGOALsysErrcut}{$0.08$}
\newcommand{\MGOALranErrcut}{$0.06$}
\newcommand{\MGOALtotErrcut}{$0.10$}

\newcommand{\CUTPerCut}{$14$}
\newcommand{\CUTnumFinCutTOT}{$290$}
\newcommand{\CUTnumSampTOT}{$827$}
\newcommand{\CUTDistModONE}{$10.58$}
\newcommand{\CUTDistModerrONE}{$0.01$}
\newcommand{\CUTDistModTWO}{$10.57$}
\newcommand{\CUTDistModerrTWO}{$0.02$}
\newcommand{\CUTnumFinCutONE}{$155$}
\newcommand{\CUTnumSampONE}{$411$}
\newcommand{\CUTnumFinCutTWO}{$135$}
\newcommand{\CUTnumSampTWO}{$416$}
\newcommand{\MeanMedDifONE}{$0.03$}
\newcommand{\MeanMedDifErrONE}{$0.01$}
\newcommand{\MeanMedDifSigOffONE}{$1.8$}
\newcommand{\MeanMedDifTWO}{$0.03$}
\newcommand{\MeanMedDifErrTWO}{$0.02$}
\newcommand{\MeanMedDifSigOffTWO}{$2.0$}

\title{A New Cepheid Distance to the Giant Spiral M101 Based On Image Subtraction of HST/ACS Observations}
\shorttitle{Cepheid Distance to M101}
\shortauthors{Shappee \& Stanek}

\author{
{Benjamin J. Shappee}\altaffilmark{1,2}, 
and
{K. Z. Stanek}\altaffilmark{1}
}

\email{shappee@astronomy.ohio-state.edu, kstanek@astronomy.ohio-state.edu}

\altaffiltext{1}{Department of Astronomy, Ohio State University, Columbus, OH 43210, USA}
\altaffiltext{2}{NSF Graduate Fellow}

\date{\today}

\begin{abstract}

We accurately determine a new Cepheid distance to M101 (NGC 5457) using archival HST/ACS \textit{V} and \textit{I} time series photometry of two fields within the galaxy. We make a slight modification to the ISIS image subtraction package to obtain optimal differential light curves from HST data.  We  discovered \CUTnumSampTOT{} Cepheids with periods between 3 and 80 days, the largest extragalactic sample of Cepheids observed with HST by a factor of 2.  With this large Cepheid sample we find that the relative distance of M101 from the LMC is $\LMCreldist{} = $ \MetLMCDist{} $\pm $  \LMCranErr{} (random)  $  \pm  $ \LMCsysErr{} (systematic) mag.  If we use the geometrically determined maser distance to NGC 4258 as our distance anchor, the distance modulus of M101 is $\mu_{0} = $ \MetDist{} $ \pm  \ $\MranErr{} (random) $  \pm $ \MsysErr{} (systematic) mag or D $ = $ \MdistMpc{} $ \pm  \ $\MdistMpcErrRan{} (random) $  \pm $ \MdistMpcErrSys{} (systematic) Mpc.  The uncertainty is dominated by the maser distance estimate ($\pm{} 0.15$ mag), which should improve over the next few years. We determine a steep metallicity dependence, $\gamma$, for our Cepheid sample through two methods, yielding $\gamma \ = $ \MetSlope{}  $\pm{} $  \MetSlopeErrRan{} (random) $\ \pm{}  $ \MetSlopeErrSys{} (systematic) mag dex$^{-1}$ and      $\gamma \ =\WPLbwMETslope ^{+ \WPLbwMETslopeUP }_{- \WPLbwMETslopeDOWN{}  }$ (random) $\pm{} \ \WPLbwMETslopeerrSys{} $ (systematic) mag dex$^{-1}$. We see marginal evidence for variations in the Wesenheit P-L relation slope as a function of deprojected galactocentric radius.  We also use the TRGB method to independently determine the distance modulus to M101 of $\mu_{0} = $ \TRGBDistMod{} $ \pm  $ \TRGBBootSig{} (random) $  \pm $ \TRGBsys{} (systematic) mag.

\end{abstract}


\section{Introduction}
\label{sec:introduc}

\textit{Hubble Space Telescope} (HST) observations of Cepheid variables have proven to be invaluable for extragalactic distance estimates outside the local group, determining distances out to  $\sim$ 35 Mpc (e.g. \citealp{freedman01, riess09b, riess09}).  Type Ia supernovae and other secondary distance indicators used at greater distances rely on calibrations using extragalactic Cepheid distances.  Through these calibrations, measurements of the Hubble constant (\Ho) depend directly on the uncertainties in the Cepheid distance anchor. Utilizing the Wide Field and Planetary Camera 2 (WFPC2) aboard HST, the HST Key Project \citep{freedman94, freedman01} and the Supernova Ia HST Calibration Program \citep{saha01, sandage06} were both able to determine \Ho{} with $<10\%$ uncertainties.  The two projects found estimates of \Ho{} = $72\pm8$ km s$^{-1}$ Mpc$^{-1}$ and \Ho{} = $62.3\pm1.3$ (random) $\pm5.0$ (systematic)  km s$^{-1}$ Mpc$^{-1}$, respectively, and the measurements were only consistent at the 2-sigma level.  
  
A dominant source of systematic error for both these measurements is their dependence on the distance to the Large Magellanic Cloud (LMC) for the absolute calibration of the Cepheid distances.  The distance to the LMC is only known to a precision of $\sim10\%$, with some estimates differing by $\pm0.25$ mag (e.g. \citealp{benedict02}).  With such a large uncertainty in the distance, it is important to replace the LMC as our extragalactic distance anchor.  There have been some attempts to provide a replacement anchor, for example the eclipsing binary distance to M33 of \citet{bonanos06}.  NGC 4258, with a geometrically determined maser distance \citep{herrnstein99}, is one of the most promising alternatives to the LMC.  \citet{macri06} used HST and the Advanced Camera for Surveys/Wide Field Camera (ACS/WFC; \citealp{ford03}) observations of Cepheids in NGC 4258 to make the maser host galaxy a viable distance anchor for extragalactic Cepheid distance measurements.

Extragalactic Cepheid distance measurements assume the Cepheids are at a common distance and apply the period-luminosity (P-L) relation of Cepheids variables. These measurements are complicated by the compositional dependence of the Cepheid P-L relation.  The magnitude of this metallicity dependence has been under debate for over two decades with conflicting claims from theory \citep{stothers88, stift90, fiorentino02, valle09} as well as Galactic and extragalactic observations \citep{caldwell86, caldwell87, freedman90, gould94, stift95, sasselov97, kochanek97, kennicutt98, groenewegen04, sakai04, macri06}.  Furthermore, \citet{tammann03} and \citet{sandage08} (and references therein) argued that the slope and zero-point of the P-L relation must be dependent on metallicity, although \citet{madore09} demonstrated that the slope of the ``reddening free'' Wesenheit P-L relation is insensitive to changes in the slope of the underlying \textit{V} and \textit{I} P-L relations. \citet{riess09} do not see any dependence in the slope of fit Wesenheit P-L relations in a sample of 7 galaxies observed with WFPC2 and ACS/WFC with average metallicities spanning 0.2 dex.

The purpose of this paper is to measure the distance to M101 and test the universality of the Wesenheit P-L relation.  We employ ISIS image subtraction \citep{alard98, alard00} on archival F555W and F814W time series ACS/WFC data of M101 to identify a large sample of Cepheids.  Using NGC 4258 as our distance anchor we are then able to measure the distance to M101.  With our large sample of Cepheid variables we explore possible variations in the slope and zero-point of the Wesenheit P-L relation as a function of deprojected galactocentric radius.  We investigate the possibility of a constant slope and zero-point, a linearly varying zero-point with a constant slope, a linearly varying slope with a constant zero-point, and a linearly varying slope and zero-point.

The paper is organized as follows: Section \ref{sec:observations} describes the observations; Section \ref{sec:reduction} details the data reduction and image subtraction; Section \ref{sec:CephVarSearch} explains our Cepheid search algorithms; Section \ref{sec:CephDistance} presents our Cepheid distance to M101; Section \ref{sec:WPL} investigates the universality of the Wesenheit P-L relation with deprojected galactocentric radius; Section \ref{sec:TRGB} presents our Tip of the Red Giant Branch (TRGB) distance to M101; Section \ref{sec:VarObjs} shows interesting non-Cepheid variable objects; Section \ref{sec:Disgusted} discusses our results; and Section \ref{sec:Conclusion} presents our six main conclusions.

\section{Observations}
\label{sec:observations}

\begin{figure}[H]
\plotone{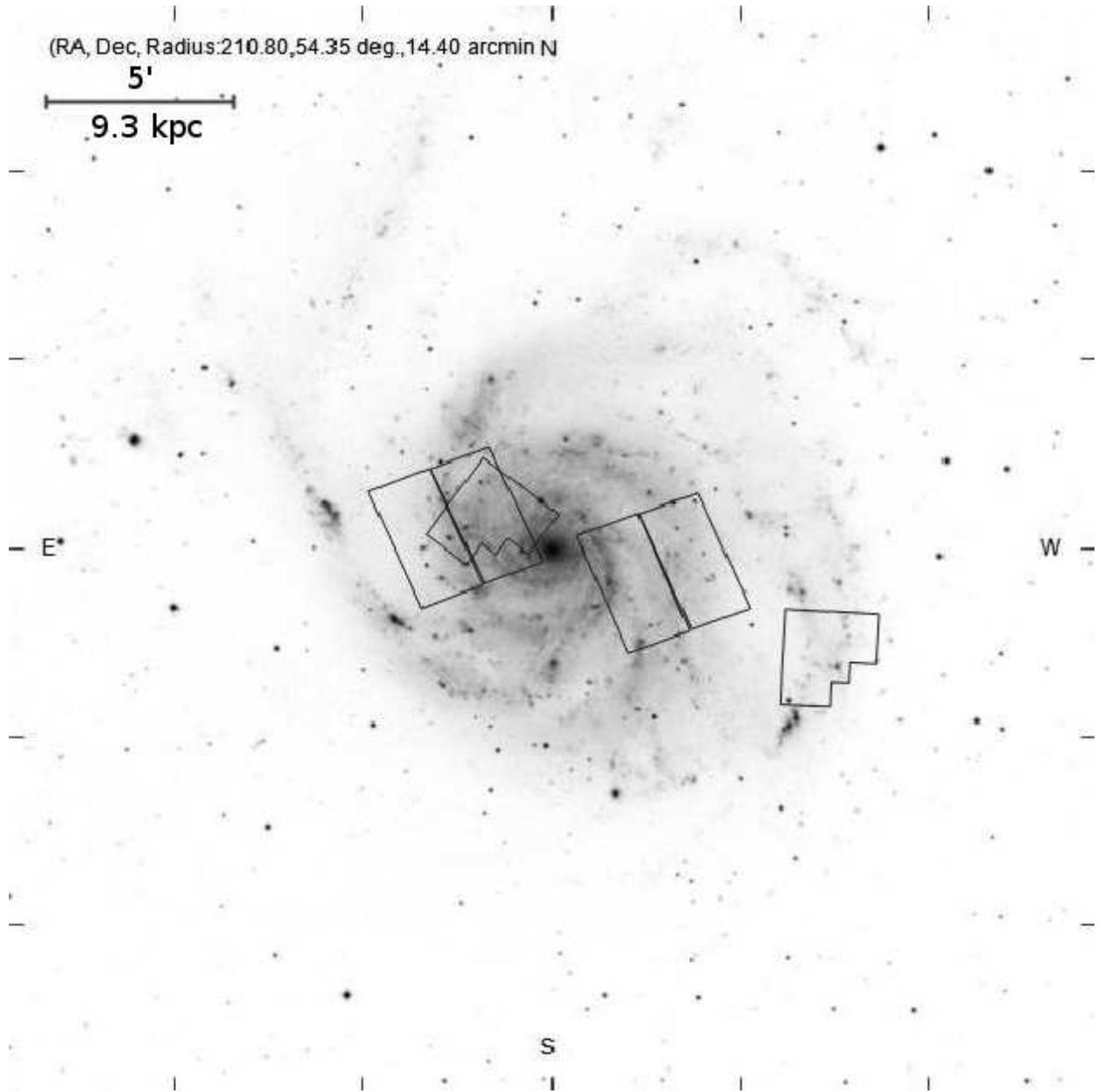}
\caption{
SDSS image of M101 showing the two fields observed with ACS/WFC (rectangular shape) which are used in this study and the two fields observed with WFPC2 (`stealth fighter' shape) for the Hubble Key Project. For both projects, Field 1 is in the East and Field 2 is in the West.  Figure discussed in \S\ref{sec:observations}.
}
\label{fig:Field}
\end{figure}

For this work we used archival ACS/WFC data (GO program 10918; PI Wendy Freedman) of M101 (NGC 5457). M101 has been classified as a SAB(rs)cd III-IV spiral galaxy \citep{devaucouleurs91} and a S(s)c I spiral galaxy \citep{sandage81}. The images were first processed by the STScI ACS calibration pipeline (see \citealp{pavlovsky05}).  We obtained the calibrated and flat-fielded images (\_flt.fits) and the drizzled images (\_drz.fits) from the Space Telescope Science Institute (STScI) HST Archive. The program observed two fields in M101 with Field 1 centered at $(\alpha,\delta) = $ (\CenterFirstRA, \CenterFirstDec) and Field 2 centered at (\CenterSecondRA, \CenterSecondDec).  Fig.~\ref{fig:Field} shows the position of these two fields as well as the WFPC2 fields used in the Hubble Key Project superimposed on the Sloan Digital Sky Survey (SDSS) image of M101 obtained using the Hubble Legacy Archive.\footnote{http://hla.stsci.edu/} 

The fields were observed in both F555W(\textit{V}-band) and F814W(\textit{I}-band) for 12 epochs following a power law sampling in time to minimize possible period aliasing \citep{freedman94, madore05}.  The observations have a baseline of $\approx{}30$ days and were completed between 2006 Dec 23 and 2007 Jan 21 (Field 1) and between 2006 Dec 25 and 2007 Jan 23 (Field 2).  Table~\ref{tab:observlog} provides a log of the observations.  The observations were taken in a two point dither pattern with total exposure times at each epoch of 1330s for F555W and 724s for F814W.  

\section{Data Reduction}
\label{sec:reduction}

To make the most of these excellent data, we reduced them using a procedure that took advantage of both the increased sensitivity to variability that image subtraction provides and the high photometric precision inherent in instrument-dedicated photometry packages, such as the ACS module of the DOLPHOT package.  Surprisingly, image subtraction has been underutilized on HST data even though it is well established in variable object searches at a variety of wavelengths (e.g. \citealp{henderson10, khan10, kozlowski10, nataf10, peeples07b, wozniak00}). Additionally, \citet{bonanos03} demonstrated the advantage of using image subtraction over more conventional photometric procedures in Cepheid variable observations, finding a nine-fold increase in the number of Cepheids found in Very Large Telescope observations of M83. To our knowledge, this work is the first paper applying image subtraction to HST data although the idea is not novel \citep{bersier01}.  The method we describe below uses image subtraction to produce light curves for all objects in our field in the instrumental magnitude system.  We then use the ACS module of the DOLPHOT package to measure and correct the magnitudes of these sources to the \citet{landolt92} \textit{V} and \textit{I} magnitude system.

\subsection{Image Subtraction}
\label{sec:image_sub}

We used the image-subtraction package ISIS\footnote{The ISIS package and a tutorial are available at http://www2.iap.fr/users/alard/package.html.} \citep{alard98, alard00}. ISIS performs image subtraction utilizing a spatially variable kernel, allowing for accurate differential photometry even in crowded fields.  

Before we ran ISIS, we used standard IRAF\footnote{IRAF is distributed by the National Optical Astronomy Observatories, which are operated by the Association of Universities for Research in Astronomy, Inc., under cooperative agreement with the NSF.} routines to multiply the drizzled images downloaded from the HST archive by their exposure time to change their units from flux to counts so that we can accurately determine the uncertainties in the photometry.  We then interpolated individual frames to a single drizzled image in each field, so all exposures had the same astrometric positions.  This is usually done with ISIS, but ACS data is especially difficult as it has large numbers of resolved sources.  ISIS was originally developed for microlensing and assumes that all sources are point sources. Instead we used Sexterp (Siverd 2011, in preparation)\footnote{http://www.astronomy.ohio-state.edu/$\sim$siverd/soft/is3/index.html}, a hybrid code which utilizes the interpolation method of ISIS with source identification and centroiding done by Sextractor \citep{bertin96}.  Using Sexterp we interpolated all the images to match the first F814W image in each field (j9o401010\_drz.fits for Field 1 and j9o413010\_drz.fits for Field 2).  As a result, all of our quoted astrometric coordinates will be referenced to the header astrometry of these two images.

\begin{figure}[H]
\plotone{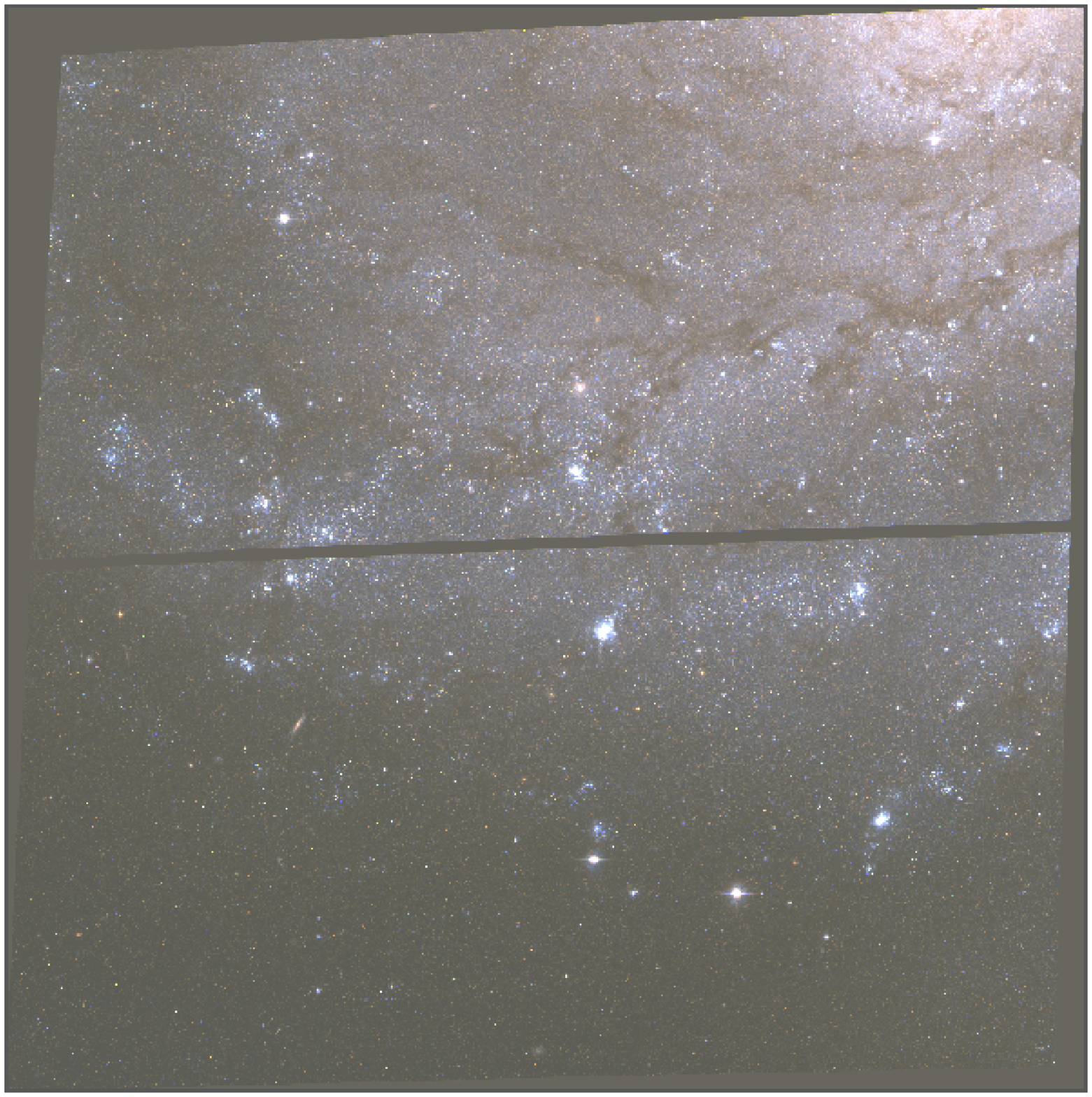}
\caption{
HST ACS/WFC F555W and F814W 2-band color composite image of the ISIS reference images for Field 1 in M101. The absolute orientation of the field is shown in Fig.~\ref{fig:Field}.  The ACS/WFC field of view is 202$''$ on each side, corresponding to a physical size of 6.3 kpc. Figure discussed in \S\ref{sec:image_sub}.
}
\label{fig:ColorF1}
\end{figure}

\begin{figure}[H]
\plotone{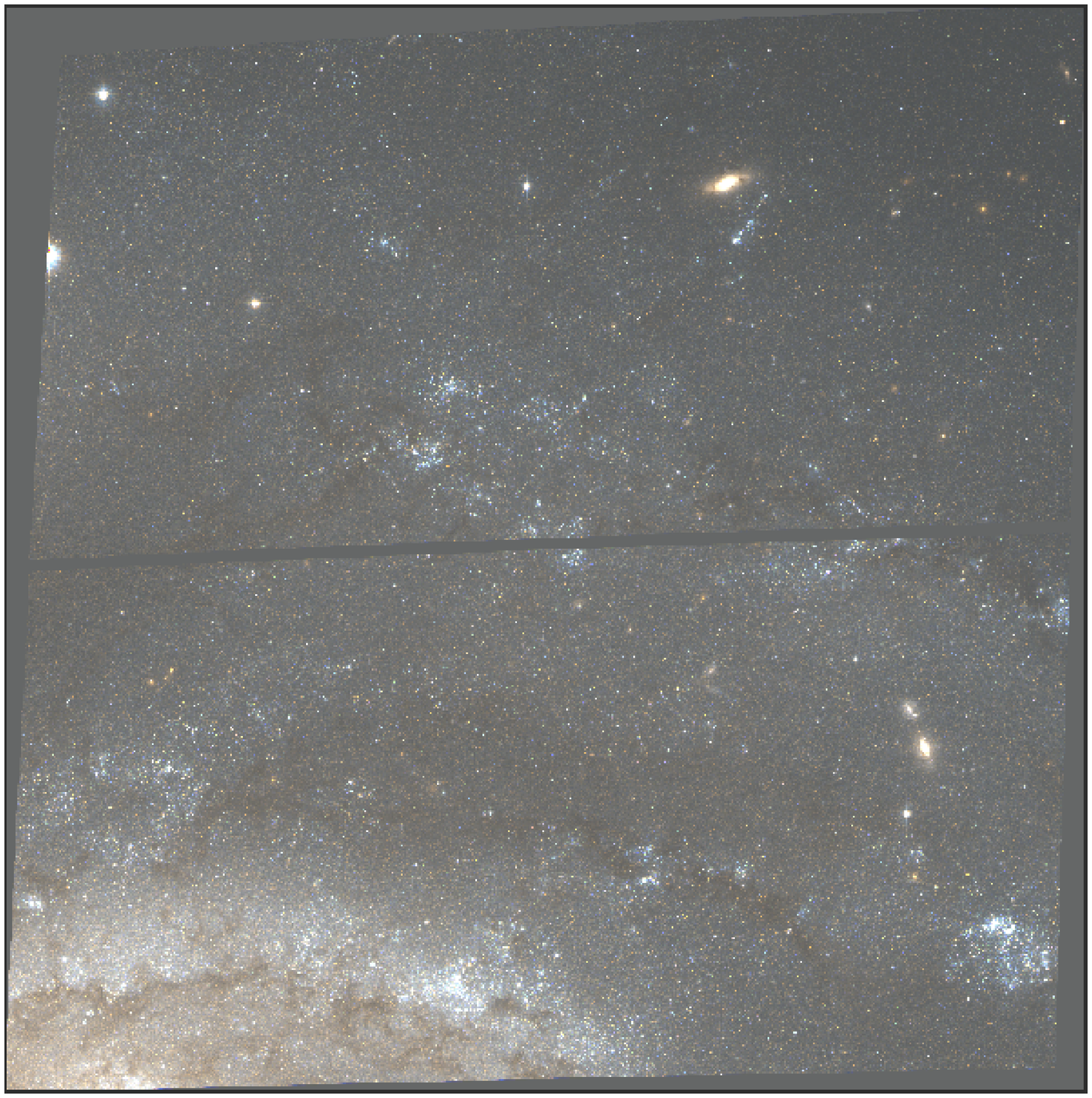}
\caption{
HST ACS/WFC F555W and F814W 2-band color composite image of the ISIS reference images for Field 2 in M101. The absolute orientation of the field is shown in Fig.~\ref{fig:Field}.  The ACS/WFC field of view is 202$''$ on each side, corresponding to a physical size of 6.3 kpc. Figure discussed in \S\ref{sec:image_sub}.
}
\label{fig:ColorF2}
\end{figure}

We used ISIS to create a reference image for each band in each field.  First, ISIS transforms all the images to the same point-spread function (PSF) and background level by convolving the images with a space-varying convolution kernel.  ISIS then stacks the resulting images using a 3-sigma rejection limit from the median.  Fig.~\ref{fig:ColorF1} and Fig.~\ref{fig:ColorF2} show color composite images of the F555W and F814W reference images.  

We then used DAOPHOT and ALLSTAR \citep{stetson87, stetson92, stetson94} for photometry on the four ISIS reference images, to find more then $250,000$ objects on each image. We then matched objects in each field between the two filters and found that $142,018$ and $160,551$ matched in Field 1 and Field 2 respectively.  

We next used ISIS to subtract each individual epoch from the corresponding reference image by first convolving the reference image with a space-varying kernel to match it to the image before subtracting it. In these subtracted images, all constant sources will disappear, leaving only the variable objects. ISIS makes a variability image by taking the mean absolute deviation of the subtracted images.  A small region of the reference image and the variability image are shown in Fig.~\ref{fig:F2Sub}.

We then used ISIS to extract the light-curves of all the sources identified in the reference images by DAOPHOT. However, the resolved sources inherent in HST observations cause ISIS to make a poor PSF itself.  Instead we fed ISIS the PSF created in DAOPHOT using carefully selected isolated stars.  This procedure requires a slightly altered version of ISIS that we describe in Appendix \ref{sec:isis_psf}.  Using this model PSF, we performed photometry on all the objects matched between the filters on all the subtracted images.  ISIS outputs the light curves in differential flux units which we convert to an instrumental magnitude system using VARTOOLS \citep{hartman08}. 

\subsection{DOLPHOT magnitude calibration}
\label{sec:dolphot}

The ACS module for DOLPHOT 1.1\footnote{http://purcell.as.arizona.edu/dolphot/} was used on the calibrated and flat-fielded images of M101 following to the recommended settings in the DOLPHOT User's Guide\footnote{http://purcell.as.arizona.edu/dolphot/dolphot.ps.gz} and DOLPHOT ACS User's Guide\footnote{http://purcell.as.arizona.edu/dolphot/dolphotACS.ps.gz}.  Photometry for each epoch and each field was performed separately with the same astrometric reference images as were used for ISIS. The recommended cuts were applied to remove non-stellar detections and to eliminate objects with uncertain photometry. For each epoch, we only accepted objects for which DOLPHOT reported a magnitude in both filters.  After these cuts, DOLPHOT reported $\approx{}320,000$ objects in Field 1 and $\approx{}380,000$ objects in Field 2 for each epoch.  

The output of DOLPHOT contains an instrumental and transformed magnitude for each filter as well as measurements for each of the individual, dithered frames.  We used the combined transformed magnitudes for each filter.  The final magnitudes are corrected for charge transfer efficiency (CTE) losses.  DOLPHOT uses transformations given in  \citet{sirianni05} to transform from the HST filters F555W and F814W to the \citet{landolt92} \textit{V} and \textit{I} magnitudes. However, \citet{sirianni05} warn that these transformations have a large uncertainty and that theoretical and empirical transformations differ.  Unfortunately, these transformations are required to make use of any published Cepheid P-L relations.  Many systematic biases can be avoided when using the maser-host galaxy NGC 4258 as a distance anchor, since the \citet{macri06} Cepheid sample was also observed with HST ACS/WFC and applied the same transformations.  In such comparisons these photometric issues cancel.

To calibrate the light curves from image subtraction, we matched the DAOPHOT objects to DOLPHOT objects in each epoch, making sure to note the offset in positions due to the differing coordinate conventions between the two packages.  We accepted DAOPHOT objects that matched an object in at least 1 DOLPHOT epoch, leaving $130,522$ objects in Field 1 and $141,817$ objects in Field 2.  Differences between the DOLPHOT transformed magnitudes and ISIS instrumental magnitudes were then calculated individually for each object at each epoch where DOLPHOT had a matching object.  We then used a weighted mean with iterative sigma clipping to find the magnitude offset for each object.  Finally, each light curve was calibrated with the addition of this offset magnitude.  Uncertainties in this offset were then added in quadrature to the Cepheid template fit mean magnitudes presented in \S\ref{sec:FTest}.

Table~\ref{tab:stdstars} contains the positions and the calibrated \textit{V} and \textit{I} magnitudes of bright non-variable secondary standards.  The magnitudes were measured on the first epoch for each field.

\section{Cepheid Variable Search}
\label{sec:CephVarSearch}

In the following subsections we describe our variability search and Cepheid selection criteria, which is similar to that of \citet{macri06} and \citet{riess09}. A variability index is first used to drastically cut the number of sources.  Light curves are then fit with a Cepheid template and poor fits are determined by an F-test and rejected.  The \citet{udalski99} P-L relations were adopted and color cuts applied to refine the Cepheid sample.  Lastly, a minimum period cut and an iterative distance modulus fit removed Population II Cepheids and other outliers.

\subsection{Variability Search}
\label{sec:VarSearch}

The \citet{welch93} variability index $I_{V}$ was calculated on each of the light curves.  The variability index is given by,
\begin{equation}
I_V=\sqrt{\frac{1}{n(n-1)}}\sum_{i=1}^n \delta V_i \delta I_i
\end{equation}
where $\delta V_i$ and $\delta I_i$ are the residuals from the mean $V$ and $I$ magnitudes normalized by measurement uncertainties.  Correlated deviations from the mean magnitude will contribute positively to the variability index whereas anti-correlated deviations contribute negatively.  Thus, a constant star with random noise will have a variability index of $\approx{}0$ and a truly variable object will have a positive variability index.  Fig.~\ref{fig:VarF1} shows the variability index as a function of V magnitude for Field 1.  We then selected objects with $I_{V}>\VarCut{}$ as variable objects. Lastly, we removed objects with mean \VarClrDown{} mag and mean \VarClrUp{} mag. These color cuts are well outside the expected range of the instability strip and lessen the number of variable objects that had to be fit with a Cepheid template.

\subsection{Cepheid Template Fit and F-Test}
\label{sec:FTest}

We then fit the Cepheid templates of \citet{yoachim09} to the variability-selected light curves from \S\ref{sec:VarSearch}. The \citet{yoachim09} templates are a principle component analysis of a Fourier decomposition of Galatic Cepheids \citep{ochsenbein00, berdnikov97, berdnikov01, gieren81, moffett84, coulson85, berdnikov95, henden96, barnes97}, LMC Cepheids and SMC Cepheids \citep{udalski99, udalski99b, sebo02, moffett98}. \citet{yoachim09} provides\footnote{http://www.astro.washington.edu/users/yoachim/Ceph\_code.tar.gz} an Interactive Data Language (IDL) procedure to fit these Cepheid templates to light curves.  As the amplitudes of different Cepheids with the same period are known to vary \citep{stetson96}, we modified the IDL Cepheid code to allow the amplitude of the Cepheid templates to vary by a multiplicative factor. Naively, one would allow the first principle component of the Cepheid template to be left as a free parameter in the fit, as the majority of this component is amplitude variations \citep{yoachim09}.  However, we found that even when bounded to values within the scatter seen in Figure 8 of \citet{yoachim09}, this resulted in unrealistic Cepheid light curve fits. 	

The Cepheid fitting code implemented the IDL Levenberg-Marquardt least-squares minimization procedure mpfit.pro\footnote{Available at http://www.physics.wisc.edu/~craigm/idl/fitting.html} \citep{markwardt09} to fit the light curves giving an initial period and an initial phase. We started the fitting procedure at the two periods identified by the Minimum String Length Method \citep{burke70,dworetsky83} for the \textit{V} and \textit{I} light curves.  However, no periodicity searching method will work for time baselines longer then the observational baseline.  Fortunately, the accuracy of HST photometry and the use of Cepheid templates allows us to determine the properties of Cepheid variables with periods significantly longer than this baseline.  This requires a brute force template fitting search for periods longer than the baseline.  In addition to the Minimum String Length periods, we also tried 250 initial periods from 3\textendash100 days logarithmically spaced for each Cepheid. The lowest initial period bound of 3 days we set because Cepheids with shorter periods would be too close to our detection limit to be useful. We set an upper bound of 100 days in the initial periods to avoid the possible change in the P-L relation as shown in \citet{bird09} for ultra-long period Cepheids. For each initial period we tried 10 different initial phases of the Cepheid template. We accepted the overall best-fit template for each variable based on the fit's \chisq value. Our method is computationally expensive but thorough. Representative light curves for Cepheids in our final sample (as described in \S\ref{sec:sampleSel}) are shown in Fig.~\ref{fig:F1LightCurves} and Fig.~\ref{fig:F2LightCurves} for Field 1 and Field 2 respectively.

We then performed a version of the \textit{F}-test to reject non-Cepheid variable objects. We compared the \chisqnu{} of the Cepheid template fit ($\chisqnu{}_{\rm{template}}$) with that of a constant magnitude least-squares fit ($\chisqnu{}_{\rm{constant}}$) and a linear least-squares fit ($\chisqnu{}_{\rm{linear}}$) for both \textit{V} and \textit{I} light curves.  We then define,
\begin{equation}
F = \frac{\rm{min}(\chisqnu{}_{\rm{contant}}, \chisqnu{}_{\rm{linear}})}{\chisqnu{}_{\rm{template}}}.
\end{equation}
A stringent $F > \ \FtestCut{}$ cut was then applied.  Fig.~\ref{fig:FtestF1} shows the distribution of $F$ as a function of fit period for Field 1.  The objects that survive this cut are the Cepheid candidates presented in Tables \ref{tab:cephbasic} and \ref{tab:cephreject} along with their positions, variability index, fit period, fit mean magnitude and fit amplitudes.

\subsection{Fiducial Period-Luminosity Relations}	
\label{sec:AdoptePL}

We adopted the updated OGLE-II P-L relations \citep{udalski99} (as updated on the OGLE website\footnote{ftp://sirius.astrouw.edu.pl/ogle/ogle2/var\_stars/lmc/cep/catalog/README.PL.}) which have been the standard choice in most extragalactic studies (e.g. \citealp{freedman01}; \citealp{macri06}).  Because we used the distance modulus to the LMC determined from the maser host NGC 4258 presented in \citet{macri06}, we adopted the same P-L relation for our final distance measurement.  This choice prevents possible systematic errors that could be caused by the adoption of a different P-L relation.  The \citet{udalski99} P-L relations were calculated on a sample of $\approx{}650$ Cephieds in the LMC, and are given by:
\begin{eqnarray}
V_{0} & = & 14.287(0.021) - 2.779(0.031)\ [\log P - 1] \label{eqn:plv}
\label{eq:PLEqV} \\
I_{0} & = & 13.615(0.014) - 2.979(0.021)\ [\log P - 1] \label{eqn:pli}
\label{eq:PLEqI}
\end{eqnarray}
where P is the Cepheid period in days. We can then define the \textit{V} and \textit{I} LMC relative distance modulus:
\begin{eqnarray}
\Delta\mu_{V} = V - V_{0},  \ \ \textrm{and} \ \ \
\Delta\mu_{I} = I - I_{0}
\end{eqnarray}
where $V$ and $I$ are the measured Cepheid magnitudes.  We then create the fiducial Cepheid color:
\begin{eqnarray}
(V-I)_{0} = V_{0} - I_{0}
\end{eqnarray}
using the adopted P-L relations.  We can now define the color excess,
\begin{eqnarray}
E(V-I) = (V-I) - (V-I)_{0}.
\end{eqnarray}
The LMC relative Wesenheit distance modulus \citep{madore82} is,
\begin{eqnarray}
\LMCreldist{} = \Delta\mu_{I} - (R-1)E(V-I).
\end{eqnarray}
where $R \equiv A_{V}/(A_{V} - A_{I})$. Finally, assuming a selective to total extinction coefficient $R_{V}=3.1$, and using the $A_{\lambda}$ from Table 6 of \citet{schlegel98} we calculate $R=2.45$.

\subsection{Selection Criteria and Final Sample of Cepheid Variables}
\label{sec:sampleSel}

We made the following cuts to our Cepheid candidate sample to identified true Cepheid variables and to minimize the effects of blending and extinction. We chose these cuts following the criteria of \citet{macri06}.

\noindent{  (1) \emph{Amplitude ratios}} - Objects with a ratio of \textit{I} amplitude to \textit{V} amplitude greater then 0.75 or less than 0.25 were removed from the sample. 

\noindent{  (2) \emph{  Blue blends}} - Objects with $E(V-I)<$ \BlueSigma{} below the Galactic foreground extinction value were removed from the sample.  Cepheids that appear significantly bluer then the Galactic extinction are likely to have blue blends. We adopted $E(B-V)=0.009$ mag or equivalently $E(V-I)=0.011$ mag \citep{schlegel98} as our foreground extinction.  

\noindent{  (3) \emph{  Red blends and highly extinguished objects}} - Objects with $E(B-V)>0.5$ mag or equivalently $E(V-I)>0.68$ mag were removed from the sample. Cepheids that appear extremely red are either highly extincted or have red blends. Although the Wesenheit magnitude and Wesenheit distance modulus are resistant to reddening, highly extinguished sources may systematically differ if the true $R_{V}$ value differs from our assumed $R_{V}$ value.  Additionally, if the color excess is due to blends these objects would clearly bias our distance.   

\noindent{  (4) \emph{  Population II Cepheids}} - Objects with a distance modulus relative to the LMC of $\LMCreldist > $ \DistCut{} were removed from the sample to exclude W Virginis and RV Tauri variables.  These Population II Cepheids have similar light curve shapes and colors as those of fundamental mode Cepheids, and some would have passed our initial sample selection. 

\noindent{  (5) \emph{  Period Cut}} - Object with a period $<$ \PerCut{} were removed from the sample because their light curves are too noisy for accurate identification of Cepheid candidates.

\noindent{  (6) \emph{  Iterative Sigma Clipping}} - We then computed a weighted median LMC-relative distance modulus for each field using a least-absolute-deviation technique as also applied by \citet{macri06}.  We perform iterative sigma clipping to remove outliers.  This weighted median was used in place of the typical mean modulus because of its outlier resistance.  Additionally, any Cepheid distance moduli distribution is skewed due to blending and the median will be a more robust statistic against these systematic problems.  The difference between these two fitting methods is investigated in \S\ref{sec:MinPer}.

The effects of the sample cuts are shown in Table~\ref{tab:samplecuts}.  A period histogram for the final Cepheid sample is shown in Fig.~\ref{fig:PerHist}. Cepheids and rejected Cepheids candidates are listed in Tables \ref{tab:cephbasic} and \ref{tab:cephreject}, respectively.  The quantities derived using our fiducial PL for the final Cepheid sample are listed in Table~\ref{tab:cephprop}.  The locations of the Cepheids are shown in Fig.~\ref{fig:F1Ceph} and Fig.~\ref{fig:F2Ceph} for Fields 1 and 2, respectively.  The color-magnitude diagrams (CMDs) of the fields with the Cepheids marked are presented in Fig.~\ref{fig:CMD}.  The \textit{V}, \textit{I} and \textit{W} luminosities for the final Cepheid sample are shown as functions of period with the predicted Wesenheit P-L relation overlain are shown in  Fig.~\ref{fig:PLAll}. Finally, the \textit{V}, \textit{I} and \textit{W} LMC-relative distance moduli are shown as a functions of period in  Fig.~\ref{fig:DistModAll}.

\section{The Cepheid Distance to M101}
\label{sec:CephDistance}

In the following subsections we describe how we determined the distance modulus of M101 from our Cepheid sample.  We first refined our Cepheid sample through cuts in period and color.  We then assume the \citet{bresolin07} oxygen abundance gradient for M101 and compute the metallicities of our Cepheid sample.  The distance modulus relative to the LMC was taken to be the distance modulus at $\textrm{[O/H]} = 8.50 \textrm{ dex}$ of a linear least-squares fit to the individual Cepheid distance moduli as a function of metallicity.  Finally, we adopted the \citet{macri06} LMC distance modulus, which was determined from the \citet{herrnstein99} NGC 4258 maser distance, to compute our final distance modulus to M101.

\subsection{Refined Cepheid Sample for Distance Determination}
\label{sec:MinPer}

Because our goal is to determine the most accurate distance to M101, we require a refined sample of fundamental mode Cepheids.  First, we exclude Cepheids with periods below 6 days because we are unable to distinguish between overtone and fundamental mode pulsators with such sparsely sampled light curves.  Overtone pulsators with periods $<6$ days have similar light curve shapes but are significantly brighter than their fundamental mode counterparts (e.g. \citealp{udalski99}), and thus would bias our distance measurement.

We computed the median distance modulus relative to the LMC under minimum period cuts from 6 to 20 days following the procedure described in \S\ref{sec:sampleSel}.  This analysis is similar to the period cuts performed in \S4.2.1 of \citet{macri06} and we share the motivation therein.  Fig.~\ref{fig:DistPcut} shows the median distance moduli relative to the LMC as a function of the minimum period for Fields 1 and 2.  We find that lower period cutoffs lead to smaller distance moduli. To avoid this bias, we adopt a minimum period of \CUTPerCut{} days for both fields.  Fig.~\ref{fig:PDistF1} and Fig.~\ref{fig:PDistF2} show the distance modulus relative to the LMC as a function of period for Fields 1 and 2, respectively.  Fig.~\ref{fig:PEV_IF1} and Fig.~\ref{fig:PEV_IF2} show the E(V-I) color excess as a function of period for Field 1 and Field 2 respectively.  These four plots also show the rejected Cepheid candidates from Table~\ref{tab:cephreject} for reference.  These period cuts lead to a median distance modulus relative to the LMC of $\LMCreldist{} = $ \CUTDistModONE$\pm$\CUTDistModerrONE{} mag using \CUTnumFinCutONE{} Cepheids in Field 1 and $\LMCreldist{} = $ \CUTDistModTWO$\pm$\CUTDistModerrTWO{} mag based on \CUTnumFinCutTWO{} Cepheids in Field 2. For each of the minimum period cuts, we also show the difference between the mean and median fit distance moduli relative to the LMC in Fig.~\ref{fig:PMean_Med}.  At a period cut of \CUTPerCut{} days, there is a small difference in the distance moduli  \MeanMedDifONE$\pm$\MeanMedDifErrONE{} mag and \MeanMedDifTWO$\pm$\MeanMedDifErrTWO{} mag for Fields 1 and 2, respectively. 

For our final distance measurement Cepheid sample we make more stringent cuts than were performed in \S\ref{sec:sampleSel} as were also applied by \citet{macri06} to remove blue blends, remove red blends and ensure that our sample cover the same range in extinction.  We discard Cepheids with color excesses below the Galactic foreground reddening $E(B-V) < $ \MetEBVdown{} mag, or, equivalently, $E(V-I) < $ \MetEVIdown{} mag.  We also remove Cepheids that show large reddening,  $E(B-V) > $ \MetEBVup{} mag, or, equivalently, $E(V-I) > $ \MetEVIup{} mag, leaving \MetCephs{} Cepheids in our refined Cepheid sample.

\subsection{Deprojection and Metallicity Dependence}
\label{sec:DepandMet}

Extragalactic Cepheid studies make the implicit assumption that the galactic oxygen abundance gradient tracks the metallicity of Cepheids at the same galactocentric radius. With this assumption we obtain a differential measurement of the metallicity dependence of the Wesenheit P-L relation.  M101 displays one of the strongest radial abundance gradients found among nearby spiral \citep{kennicutt96}.  We adopted the following oxygen abundance gradient for M101 from Eq. 5 in \citet{bresolin07}:
\begin{eqnarray}
\textrm{[O/H]}=(8.75(0.05) - 0.90(0.07) \rho) \ \textrm{dex}
\label{eq:MetEq}
\end{eqnarray}
where $\rho$ is the fractional isophotal.  \citet{bresolin07} extended the 20 \HII{} oxygen abundance measurements of \citet{kennicutt03} to within $1.\!\!'5$ of the galactic center with two additional measurements in the inner regions of M101.  Figure 2 of \citet{bresolin07} shows the oxygen abundance gradient fit for M101 where the \HII{} regions span more then a dex in metallicity.  We then computed the deprojected galactocentric distances for the Cepheids were calculated using the following corrected\footnote{Eq. 11 and Eq. 12 in \citet{macri06} are missing the geometric factor of $\cos{\delta_0}$.} equations presented in \citet{macri06}:
\begin{eqnarray}
x & = & (\alpha-\alpha_0) \cos{\delta_0} \cos{\phi} + (\delta - \delta_0) \sin{\phi} \\
y & = & \frac{(\delta - \delta_0) \cos{\phi} -(\alpha-\alpha_0) \cos{\delta_0} \sin{\phi}}{\cos{i}} \\
\rho & = & \frac{(x^2+y^2)^{\frac{1}{2}}}{R_0}
\end{eqnarray}
where $R_0$ is the isophotal radius.  We adopted the nuclear position of M101 from \citet{corwin94} of (J2000) $(\alpha,\delta)=(14^{h}03^{m}12.\!\!^{s}53, +54\degree20'55.\!\!''6)$.  We also adopted the major axis position angle $\phi=35.\!\!\degree0 \pm{} 1.\!\!\degree0$ and inclination of $i=17.\!\!\degree0 \pm{} 1.\!\!\degree0$ from \citet{zaritsky90}.  Additionally, we assumed isophotal radius $R_0=14.\!\!'4$ from \citet{devaucouleurs91} as was also assumed in \citet{bresolin07}.  Finally, we computed the metallicities of the Cepheids in our final sample as reported in Table~\ref{tab:cephprop}.  

Fig.~\ref{fig:MetalAll} shows the LMC relative distance modulus as a function of the deprojected galactocentric distance and metallicity for the refined Cepheid sample.  A linear least-squares fit and a constant least-squares fit to the data are shown in Fig.~\ref{fig:MetalAll}. The standard deviation about the linear fit and the constant fit are \MetStdDev{} mag and \MetStdDevHor{} mag, respectively.  The \chisq{} values of the linear fit and constant fit are \chisq=\MetChisq{} and \chisq=\MetChisqHor{}, respectively.  We see that the linear least-squares fit is preferred by the data.  Fig.~\ref{fig:MetalEBV} displays the residuals from the linear least-squares fit against $E(V-I)$, where no dramatic trends are seen. 	 

A linear least-squares fit to the data yields a slope of $\gamma \ = $ \MetSlope{} $\pm{} $  \MetSlopeErrRan $_{r}  \pm{}  $ \MetSlopeErrSys $_{s} \ $ mag dex$^{-1}$  and a zero-point determination at $\textrm{[O/H]} = 8.50 \textrm{ dex}$ of $\LMCreldist{} = $ \MetLMCDist{}  $\pm{} $  \MetDistErrRan $_{r}  \pm{}  $ \MetDistErrSys $_{s} \ $ mag\footnote{In this paper we denote random and systematic uncertainties with subscript \textit{r} and \textit{s} respectively, i.e., $3.00 \pm 0.10_{r} \pm 0.10_{s}$ mag.}. We determined the random uncertainties of the linear best-fit parameters by bootstrap re-sampling the Cepheids with \Metboot{} simulations. In each simulation we randomly re-sampled the refined Cepheid sample with replacement, to create a new sample with the same size. For each re-sample we calculated the linear least-squares best-fit slope and zero-point.  We then took the standard deviation of the distribution of \Metboot{} best-fit slopes/zero-points as the random uncertainty in the original determined slope/zero-point, respectively. As both fields in the data are at approximately the same deprojected galactocentric distance, we have a relatively small range in metallicity with which to determine the correlation between metallicity and luminosity, leading to the large uncertainty we see in $\gamma$. The systematic uncertainties are calculated from the uncertainties in the \citet{bresolin07} oxygen abundance gradient. The random and systematic uncertainties in the zero-point are included in the distance error budget shown in Table~\ref{tab:err}, represented by R5 and S4, respectively.

To test if the Cepheids at the smallest galactocentric radii dominated the linear least-squares fit, we removed Cepheids with metallicities less than $8.65$ dex from our sample and refit. The new best-fit slope $\gamma \ = $ \MetSlopecut{} $\pm{} $  \MetSlopeErrRancut $_{r}  \pm{}  $ \MetSlopeErrSyscut $_{s} \ $ mag dex$^{-1}$ is shallower but consistent with the slope determined before the metallicity cut.  Even with a large uncertainty in $\gamma$ there is tension (\MetSigKen{} $\sigma$) between our measured value and the Hubble Key Project determination of $\gamma \ = -0.24 \ \pm 0.16$ mag dex$^{-1}$ for M101 using WFPC2 data \citep{kennicutt98}.  The implications of our measured Cepheid metallicity dependence are discussed in \S\ref{sec:Disgusted}.

\subsection{Final Cepheid Distance to M101}
\label{sec:FinDistance}

We then adopted the \citet{macri06} LMC distance modulus $\LMCdist{} = $ \MacriLMCdist{} $\pm{} $  \MacriLMCdistRan  $_{r} \ \pm{}  $ \MacriLMCdistSys $_{s} \ $mag determined from the \citet{herrnstein99} NGC 4258 maser distance to compute our final distance modulus to M101.  Our error budget is shown in Table~\ref{tab:err}.  We report a distance modulus relative to the LMC of $\LMCreldist{} = $ \MetLMCDist{} $\pm{} $  \LMCranErr  $_{r} \ \pm{}  $ \LMCsysErr $_{s} \ $ mag and a M101 distance modulus of $\mu_{0} = $ \MetDist{} $ \pm{}  $ \MranErr$_{r} \ \pm{} $\MsysErr$_{s}$ mag.  We compare these distances to other determinations of distance modulus relative to the LMC and estimates of the distance to M101 in Table~\ref{tab:CompDist}.

\section{Slope and Zero-Point of the Wesenheit P-L Relation}
\label{sec:WPL}

We can also use the large number of extragalactic Cepheids in our sample to test the universality of the Wesenheit P-L relation.  Let us define the classic ``Wesenheit reddening-free''  mean magnitude by:
\begin{equation}
W=V-R(V-I),
\label{eq:W}
\end{equation}
where \textit{R} is defined in \S\ref{sec:AdoptePL}.  Based on the mean $V$ and $I$ magnitudes of our final Cepheid sample given in Table~\ref{tab:cephprop} we can define a expected mean Wesenheit magnitude ($W_{0}$) for each Cepheid.  We then want to fit for a Wesenheit P-L relation:
\begin{equation}
W_{0}=a_{w}\log{P}+b_{w}
\label{eq:Wo}
\end{equation}
where the slope $a_{w}$ and zero-point $b_{w}$ may depend on deprojected radius as a proxy for metallicity. Based on Eq. \ref{eq:Wo}, the expected mean Wesenheit magnitude as a function of period and $\rho$ for each these models is given by:
\begin{equation}
W_{0}(P)=[a_{w}^{I}+a_{w}^{S}(\rho-\rho')](\log{P}-\log{P'})+[b_{w}^{I}+b_{w}^{S}(\rho-\rho')]
\label{eq:ExpW}
\end{equation}
where $a_{w}^{I}$ and $b_{w}^{I}$ are the values of the slope and zero-point, respectively, of the Wesenheit P-L relation at some characteristic fraction of the isophotal radius ($\rho'$).  $a_{w}^{S}$ and $b_{w}^{S}$ are the rates at which the slope and zero-point of the Wesenheit P-L relation change with $\rho$.  The zero-point of the Wesenheit P-L relation is defined at a characteristic pulsation period ($P'$). In order to minimize the covariance between $a_{w}^{I}$,  $a_{w}^{S}$, $b_{w}^{I}$ and $b_{w}^{S}$ we use the mean log period ($\log{P'} = $ \WPLScaleLogPer{} log(days) $ \approx{} \langle \log{P} \rangle$) and mean factional isophotal radius ($\rho' = $ \WPLScalePho{} $ \approx{} \langle \rho \rangle$) for the characteristic period and radius, respectively.

We use the Cepheid sample in Table \ref{tab:cephbasic} and imposed the same minimum period cut of $P >$ \CUTPerCut{} days used in \S\ref{sec:MinPer} and either impose no maximum period cut or impose a maximum period cut of $P <$ \WPLPerLimitfourty{} days to define a full and restricted sample, respectively.  There are two reasons a maximum period may avoid systematic deviations.  First, the $\approx30$ day baseline of our data makes the period and magnitude determination suspect for very long period Cepheids. Many of them are missed in any case because they fail our F-test in \S\ref{sec:FTest}.  Eliminating the long period Cepheids is of little concern when determining the distance modulus to M101 because the mean magnitude and period errors will also be larger on these objects.  However, our ability to estimate the radius-dependent slopes of the Wesenheit P-L relation could be strongly affected by the stochastic nature that the inclusion of a very few long period Cepheids would induce.  However, adopting this maximum cut limits our sensitivity to changes in the P-L slope. As there is no obvious solution to address both concerns, we performed and present both fits. After trimming, there are \WPLNumCeph{} Cepheids in our full sample and \WPLNumCephfourty{} Cepheids in our restricted sample.  

We then fit Eq. \ref{eq:ExpW} to both the full and restricted samples using the following procedure.  First, we fit a model with no radial dependence, and iteratively remove outliers.  We took the 100 Cepheids closest in radius to each Cepheid, computed their scatter about the best-fit Wesenheit P-L, and rejected the Cepheid if it was more then 3 times the scatter from the best-fit P-L relation.  This procedure takes into account the increased scatter of Cepheids at smaller $\rho$.  We then renormalized the errors to have \chisqnu{} $= 1$ for the model with no radial dependence to account for the intrinsic scatter of the Cepheids. We then used the IDL Levenberg-Marquardt \chisq{} minimization procedure mpfit.pro \citep{markwardt09} to fit four model Wesenheit P-L relations, Model 1 had a constant slope and zero-point, Model 2 had a linearly varying zero-point with a constant slope, Model 3 had a linearly varying slope with a constant zero-point and Model 4 had a linearly varying slope and zero-point. For each model initial $a_{w}^{I}$ and $b_{w}^{I}$ values were taken from the best-fit model with no radial dependence and the initial slopes $a_{w}^{S}$ and $b_{w}^{S}$ were taken to be 0.  $a_{w}^{S}$ and $b_{w}^{S}$ were fixed to 0 for the constant slope models and the constant zero-point models, respectively.  To determine the \WPLConInt$\%$ confidence intervals for each model we performed a bootstrap re-sampling with \WPLNumboot{} simulations.  Fig.~\ref{fig:WPLfits} and Fig.~\ref{fig:WPLfitsfourty} present the Model 1, Model 2 and Model 4 fits on top of the Wesenheit P-L relations for six sub-samples with equal number of Cepheids stepping out in deprojected radius for the full and trimmed sample, respectively.  The best-fit slope and best-fit zero-point, both as a function of $\rho{}$, for the full sample are shown in Figures \ref{fig:WPLslope} and \ref{fig:WPLinter}, respectively. The best fit slope and intercepts of the six sub-samples in Fig.~\ref{fig:WPLfits} are also shown for comparison in Figures \ref{fig:WPLslope} and \ref{fig:WPLinter}, respectively. Best-fit parameters and \chisq{} values for the four model are presented in Table~\ref{tab:WPLmodels}.  Using the \citet{bresolin07} abundance gradient we transform these fit parameter into functions of metallicity as shown in Table~\ref{tab:WPLmodelsMET}.  The characteristic radius is then equivalent to a characteristic metallicity of \WPLScaleMet{} dex. 

We then compared the models and tested the validity of the additional terms between models fit to both samples.  Based upon the \chisq values presented in Table~\ref{tab:WPLmodels} the data seems to prefer models 2 and 4 over model 1 for both samples.  We then computed the F-test confidence to quantify our justification for additional parameters between models 1 and 2 and between models 2 and 4.  The calculated confidence levels are shown in Table~\ref{tab:WPLmodelsComp}.  Both Cepheid samples are better fit by Model 2 than Model 1.  This is expected because a changing Wesenheit P-L zero-point ($b_{w}^{S}$) is equivalent to the change in the distance modulus as a function of metallicity ($\gamma$) presented in Fig.~\ref{fig:MetalAll}.  When comparing Model 2 and Model 4 we note a marginal detection of a change in the Wesenheit P-L relationship slope with $\rho$.  This detection is at a \WPLFREEvFSLOPE \% confidence level for the full sample, but the significance decreases to \WPLFREEvFSLOPEfourty \% for the restricted sample.  We postpone further discussion until \S~\ref{sec:Disgusted}.

\section{TRGB Distance to M101}
\label{sec:TRGB}

We also determined the distance to M101 using the TRGB method \citep{lee93}.  The underlying physics of this method is well understood and described in \citet{salaris02}.  The effects of metallicity are also well understood \citep{bellazzini01, bellazzini04}, and many methods exist to correct for these effects \citep{mager08, madore09}.  The TRGB method must be applied at the largest possible galactocentric radii to limit contamination from the intermediate mass asymptotic giant branch (AGB) stars \citep{sakai96}.  AGB stars have a continuous distribution at similar colors but extend to higher luminosities than the TRGB, diluting the abrupt edge of the old, low mass stars at the TRGB.

The TRGB method works best given deep and accurate photometry.  To accomplish this we reran DOLPHOT for each field, but include all epochs in a single run.  DOLPHOT then combines each epoch's photometry and computes one transformed magnitude per object per band. This second run of DOLPHOT yielded \TRGBsourceOrigFone{} and \TRGBsourceOrigFtwo{} sources in Fields 1 and 2, respectively.  We then trimmed the sample for optimally application of our TRGB detection method.  We first eliminated sources at galactocentric radii smaller then \TRGBRadCut{}$'$ to limit the contamination from AGB stars, yielding \TRGBsourceRadFone{} and \TRGBsourceRadFtwo{} sources from Fields 1 and 2, respectively. We removed sources with $V-I < $ \TRGBClrCut{} mag as they are too blue to be TRGB stars. Our final TRGB sample size contained \TRGBsourceClrFone{} and \TRGBsourceClrFtwo{} sources from Fields 1 and 2, respectively.  

We then applied reddening corrections to account for the small amount of foreground extinction.  Based on the \citet{schlegel98} dust maps, the foreground absolute extinction correction of \textit{V} and \textit{I} are $A_{V} = 0.028$ mag and  $A_{I} = 0.017$ mag. 

We then used these corrected magnitudes to calculate the \citet{madore09} \textit{T} magnitude for all our TRGB sources:
\begin{eqnarray}
T\equiv{}I_{\circ}-\beta{}[(V-I)_{\circ}-\gamma_{\textrm{\tiny{TRGB}}}],
\end{eqnarray}
where $\beta$ is the \textit{I} versus $V-I$ slope of the TRGB, and $\gamma_{\textrm{\tiny{TRGB}}}$ is the fiducial color where the $T$ magnitudes are normalized to the $I$ magnitude of a TRGB star of that color.  The \textit{T} magnitude is constructed to correct for the slope of the tip \textit{I} magnitudes as a function of $V-I$, forcing the tip \textit{T} magnitudes to be constant over a large range of colors or equivalently metallicities\footnote{See Figure 1 in \citet{madore09} for a schematic representation.}.  With the TRGB slant removed, edge detection methods more efficiently determine the TRGB position with fewer sources. \citet{madore09} adopt $\beta=0.20\pm0.05$ for the intrinsic color range $1.5 < (V-I)_{\circ} < 3.0$, which is the color range observed for our TRGB.  This slope is consistent with both observational values of $\beta=0.22\pm0.02$ \citep{rizzi07} and theoretical predictions of $\beta=0.15$ \citep{mager08}.  With a fiducial color of $\gamma_{\textrm{\tiny{TRGB}}}  =(V-I)_{\circ}=1.5$ mag, \citet{madore09} define the distance modulus based on the TRGB by:
\begin{eqnarray}
\mu_{\textrm{\tiny{TRGB}}} \equiv T - M_{\textrm{\tiny{TRGB}}} = T + 4.05(0.12) \ \textrm{mag},
\label{eq:TRGBdist}
\end{eqnarray}
where $M_{\textrm{\tiny{TRGB}}}$ is the TRGB absolute $I$ magnitude at the fiducial color $\gamma$. The uncertainty in $M_{\textrm{\tiny{TRGB}}}$ was taken from \citet{mager08}, a companion paper to \citet{madore09}.
The absolute TRGB adopted in both \citet{madore09} and \citet{mager08} are 0.01 mag different at the same fiducial color, so we adopt the uncertainty presented in \citet{mager08}.

The TRGB magnitude is determined using a edge detection procedure similar to that of \citet{sakai04}, but using the newly constructed \textit{T} magnitude in place of the typical \textit{I} magnitude.  A continuous luminosity function $\phi(m)$ was first computed using equation (A1) in the appendix of \citet{sakai96}. We then used equation (4) of \citet{mendez02} to compute a logarithmic edge-detection response function $E(m)$.  This function is a continuous version of a Sobel filter.  Similar to what is done in discrete edge detectors, we weight the edge-detector response by the Poisson noise of our luminosity function, $E(m)[\phi (m)]^{\frac{1}{2}}$ \citep{mendez02} and took the maximum as the TRGB.  We then estimated the random uncertainty of tip determination by bootstrap re-sampling of the TRGB sample with \TRGBboot{} simulations.  In each simulation we randomly re-sampled the TRGB sample with replacement to create new sample of the same size. For each re-sample we determine the TRGB $T$ magnitude with our edge detection method.  We then took the standard deviation of the distribution of \TRGBboot{} TRGB $T$ magnitudes as the random uncertainty in the original tip detection.  Fig.~\ref{fig:TRGBD} presents the \textit{T} magnitude CMD in the left panel with the tip detection response function in the right panel.  The tip detection at \textit{T} $ = $ \TRGBedge $\pm$ \TRGBBootSig{} mag is marked on both panels for reference.  With this detection, we use Eq. \ref{eq:TRGBdist} to determine the TRGB distance modulus of $\mu_{0} = $\TRGBDistMod$ \pm  $\TRGBBootSig$_{r}  \pm $\TRGBsys$_{s}$ mag.  The systematic uncertainty is dominated by the uncertainty of $M_{\textrm{\tiny{TRGB}}}$.

\section{Non-Cepheid Variable Objects}
\label{sec:VarObjs}

There have been many studies searching for luminous variable stars, including a recent study of the LMC which found 117 variable objects with absolute magnitudes of $\textrm{M}_{V} \gtrsim  -4.4$ mag \citep{szczygiel10}.  We performed a non-exhaustive (by eye) search of the brightest, most variable objects in both fields which are not in our final Cepheid sample shown in Table~\ref{tab:cephbasic}.  The majority of these objects appeared to be rejected Cepheid candidates, which usually displayed an obvious reason for their rejection, namely blends from colors or luminosity.  Additionally, many possible long period Cepheids were rejected since they had poor light curve coverage due to the restricted time span of the data.  

Some of the most luminous and interesting variable objects are shown in Fig.~\ref{fig:VarObjone} with their properties given in Table~\ref{tab:VarObj}.  Possible periodic objects are shown in Fig.~\ref{fig:PerObjone} with their properties listed in Table~\ref{tab:PerObj}.  Fig.~\ref{fig:VarCMD} presents the CMD of both fields with the luminous variables marked.  

\section{Discussion}
\label{sec:Disgusted}

In \S\ref{sec:CephDistance} and \S\ref{sec:TRGB} we determined the distance to M101 using Cepheids and the TRGB method, respectively, and found that they are in good agreement with $\mu_{\textrm{\tiny{Ceph}}} = $ \MetDist{} $ \pm{}  $ \MranErr$_{r} \ \pm{} $\MsysErr$_{s}$ mag and $\mu_{\textrm{\tiny{TRGB}}} = $\TRGBDistMod$ \pm  $\TRGBBootSig$_{r}  \pm $\TRGBsys$_{s}$ mag.  We compared our M101 distance moduli against other recent studies in Table~\ref{tab:CompDist}. We see that our LMC relative distance modulus is in excellent agreement with the recent measurements of \citet{saha06} and \citet{freedman01} and our absolute M101 distance moduli differ within the uncertainties.  This difference is due to the applied luminosity calibration for the adopted Cepheid P-L relations. \citet{saha06} used a hybrid method to determine the metallicity dependence of the P-L relation determined using Galactic Cepheids from \citet{tammann03} and LMC Cepheids from \citet{sandage04}.  However, the Wesenheit P-L slopes measured for the seven galaxy sample of \citet{riess09} are significantly shallower  than the Wesenheit P-L slope of \citet{tammann03}.  \citet{riess09} points out the many difficulties with the method of the \citet{tammann03} Galactic Cepheid P-L determination including the use of non-parallax based distances and high mean extinctions.

Despite our large sample of Cepheids, our M101 distance modulus measurement suffers from large systematic uncertainties.  As seen in Table~\ref{tab:err}, the majority of this uncertainty originates from the \citet{herrnstein99} maser distance to NGC 4258. Efforts to improve the uncertainty of the maser distance to NGC 4258 to $~3\%$ are underway using Very Long Baseline Array observations \citep{humphreys07, humphreys08}.  However, to our knowledge, \citet{herrnstein99} is the latest actual measurement in the literature. Table~\ref{tab:err} contains a goal error budget calculated under the assumption of a $3\%$ maser distance to NGC 4258. Additionally, as this is a differential measurement, systematic errors in the photometric calibrations could be negated with a reanalysis of the \citet{macri06} Cepheid observations of NGC 4258 with the same photometric procedure. 

We note that the \Ho{} measurement with a total uncertainty of $4.8\%$ from \citet{riess09b} uses the published maser distance of $\mu_{\textrm{\tiny{N4258}}} = $ \Maserdist{} mag from \citet{herrnstein99} but assumes the error bars ($3\%$) of the yet unpublished result of Humphreys et al. (2009) rather than the published uncertainties ($7\%$) of \citet{herrnstein99}.

The universality of the Wesenheit P-L relation has been a point of concern and contention in the literature since it has important implications for deriving extragalactic distances.  With our large sample of M101 Cepheids we investigate the Wesenheit P-L as a function of deprojected galactocentric radius, and by assumption, as a function of metallicity. We determine the metallicity dependence of the Wesenheit P-L zero-point in \S\ref{sec:DepandMet} and \S\ref{sec:WPL} to be $\gamma \ = $ \MetSlope{} $\pm{} $  \MetSlopeErrRan $_{r}  \pm{}  $ \MetSlopeErrSys $_{s} \ $ mag dex$^{-1}$ and $\gamma =b_{w}^{S} \ =\WPLbwMETslope ^{+ \WPLbwMETslopeUP }_{- \WPLbwMETslopeDOWN } \textrm{(random)} \pm \WPLbwMETslopeerrSys _{s} \ $ mag dex$^{-1}$, respectively.  These metallicity dependences are in contention with current values in the literature.  Using the TRGB method as a metallicity independent distance indicator \citet{sakai04} report $\gamma = -0.25 \ \pm{} \ 0.09$  mag.  However, the \citet{sakai04} results have recently been challenged in \citet{rizzi07}.  The HST Key Project reported $\gamma = -0.24 \pm 0.16$  mag  dex$^{-1}$ \citep{kennicutt98} using an WFPC2 data for an inner field and an outer field in M101. There is even contention with the more recent determination from \citet{macri06} who found $\gamma = -0.29 \pm{} 0.09_{r} \pm 0.05_{s}$  mag  dex$^{-1}$ using HST ACS data of NGC 4258. However, \citet{bono08} shows that by removing a single outlier \HII{} region used in the determination of the metallicity gradients of NGC 4258 leads to a far shallower metallicity gradient:
\begin{eqnarray}
\textrm{[O/H]}=(8.89 - 0.16(\rho - 0.4) \ \textrm{dex},
\label{eq:MetEqBono}
\end{eqnarray}
then assumed by \citet{macri06}:
\begin{eqnarray}
\textrm{[O/H]}=(8.97 \pm 0.06) - (0.49 \pm{}0.08)(\rho - 0.4) \ \textrm{dex}.
\label{eq:MetEqMacri}
\end{eqnarray}  
With the shallower metallicity gradient of \citet{bono08}, the $\gamma$ measurement of \citet{macri06} for NGC 4258 steepens to $\gamma = -0.89$ mag dex$^{-1}$, and is consistent with our estimate for M101.

There have been arguments that the change in the zero-point of the Wesenheit P-L as a function of deprojected galactocentric radius may be partially due to changes in  blending as well as metallicity.  For example, \citet{bono08} argues that blending could explain the negative values of $\gamma$ that are seen because Cepheids in the inner region will be more blended on average, decreasing their apparent distance modulus. The idea that blending could undermine the community's efforts to use Cepheids as a distance indicator has been discussed in many studies including \citet{stanek99}, \citet{mochejska00} and \citet{macri01}.

In addition to the varying zero-point of the Wesenheit P-L, the slope is also expected to show variations at some level.  \citet{sandage08} argue that slope of the \textit{V}-band P-L relation decreases monotonically with metallicity.  However, \citet{madore09} show that the Wesenheit P-L slope is largely insensitive to changes in the slopes of the underlying monochromatic P-L relations.  Through an algebraic argument and reasonable assumptions about the observed properties of Cepheids, \citet{madore09} show that the change in slope of the Wesenheit P-L could be $\approx{} -0.3$ times the change in the slope of the \textit{V} P-L.  However, \citet{madore09} notes that the observational properties used are not currently known to a 10\% precision, so their conclusions must be taken with caution. 

Our Cepheid sample is marginally better fit by a Wesenheit P-L relation model with both slope and zero-point linearly varying as a function of deprojected isophotal radius than by a model with a linearly varying zero-point alone. The evidence for a linearly varying slope as a function of $\rho{}$ degrades when using a sample with an applied maximum period cut of \WPLPerLimitfourty{} days.  However, this degradation was expected as the shortest and longest period Cepheids are those that best constrain the slope of the Wesenheit P-L.  We see three possible explanations for our observations:

1.  If Cepheid blending is a larger effect at smaller radii $\rho{}$, then one would naively expect a change in slope with our observed sign. To illustrate this point, imagine adding a constant flux to all Cepheids as a function of $\rho{}$; one would then expect more than just a change in the zero-point of the Wesenheit P-L relation.  At smaller $\rho{}$ the less luminous, shorter period Cepheids would increase in magnitude more then their more luminous, longer period counterparts at the same $\rho$.  Thus the slope of the Wesenheit P-L relation would become shallower in the inner regions of a galaxy.  This type of blending is not trivial or obvious, as the smooth increase in the stellar density would simply be subtracted in our photometric procedures.  An increase in stellar density associated with Cepheids as a function of $\rho{}$ would have to be invoked.  A detailed and quantitative analysis of the magnitude of this effect is beyond the scope of this study.

2.  From the recent work of \citet{sandage08} and \citet{madore09}, it is expected that the metallicity dependence of the Wesenheit P-L slope should be small.  First, assume that the change in the slope of the Wesenheit P-L relation is a negative fraction of the change in the slope of the \textit{V} P-L relation \citep{madore09}.  Also assume that the \textit{V} P-L slope decreases monotonically with metallicity \citep{sandage08}.  From this we would expect the observed Wesenheit P-L relation to become shallower towards higher metallicities in the inner regions of a galaxy.  

3.  The observed change in slope of the Wesenheit P-L relation is tenuous, and as such could simply be a statistical fluctuation in the data.   

M101 is a large face-on spiral galaxy, has bright \HII{} regions at all radii and displays one of the strongest radial abundance gradients found among nearby spiral \citep{kennicutt96}, which makes it an extremely useful object to study the systematic effects on the Wesenheit P-L relation for Cepheids.  The two fields observed in M101 used in this study are at approximately the same deprojected galactocentric radius, so the range in $\rho{}$ is relatively small.  This makes any study of the effects that $\rho{}$ or metallicity have on the observed Wesenheit P-L relation difficult.  It would be worth while to obtain another equivalent set of Cepheid observation of M101 at a greater deprojected galactocentric radius to further test the effects of metallicity on the Wesenheit P-L relation.  Additionally, it would be useful and of relatively little cost to reobserve one of the fields in M101 used in this work for 2-3 additional epochs with HST/ACS.  These observations would help to identify and constrain the periods of Cepheids with periods greater then 30 days.  With a large sample of Cepheids covering a large range in period it would be possible to test the change in the P-L relation slope for Ultra Long Period Cepheids at periods $>$ 100 days \citep{bird09}.

\section{Summary}
\label{sec:Conclusion}

We accurately determine a Cepheid distance to M101 (NGC 5457) using archival HST ACS \textit{V} and \textit{I} time series photometry of two fields within the galaxy. We used ISIS image subtraction to obtain differential light curves and developed a new method where ISIS uses an externally determined PSF model. We used DAOPHOT to create this PSF model and DOLPHOT to calibrate the photometry. Our six primary results are:

1.  We discovered  \CUTnumSampTOT{} Cepheids using 12 epochs of HST ACS/WFC F555W and F814W data for two fields within M101.  

2.  We determine a Cepheid based LMC-relative distance modulus to M101 of $\LMCreldist{} = $ \MetLMCDist $\pm{} $  \LMCranErr  $_{r}  \pm{}  $ \LMCsysErr $_{s}  $ mag if we assume the OGLE II LMC $V$-band and $I$-band PL relations. 

3.  Based on the \citet{herrnstein99} geometrically determined maser distance to NGC 4258 and the \citet{macri06} Cepheid based LMC-relative distance modulus to NGC 4258, we determine an absolute M101 distance modulus of $\mu_{0} = $ \MetDist{} $ \pm{}  $ \MranErr$_{r}  \pm $ \MsysErr$_{s}  $ mag.

4.  We also determine a M101 distance modulus based on the TRGB method of $\mu_{0} = $ \TRGBDistMod{} $ \pm  $ \TRGBBootSig$_{r}  \pm{} $ \TRGBsys$_{s}$ mag.

5.  We find a steep Cepheid metallicity dependence through 2 methods, yielding $\gamma \ = $ \MetSlope{} $\pm{} $  \MetSlopeErrRan $_{r}  \pm{}  $ \MetSlopeErrSys $_{s}  $ mag dex$^{-1}$ and      $\gamma =b_{w}^{S} \ =\WPLbwMETslope ^{+ \WPLbwMETslopeUP }_{- \WPLbwMETslopeDOWN } $ (random) $ \pm{} \WPLbwMETslopeerrSys _{s}  $ mag dex$^{-1}$.

6.  Our Cepheid sample is marginally better fit by a P-L relation where both the slope and the zero-point linearly vary with galactocentric radius, over models with only a linearly varying zero-point.

\acknowledgments

	We would like to thank Christopher Kochanek, Lucas Macri, and Jennifer van Saders for their invaluable suggestions and discussions.  We would also like to thank Joe Antognini, Thomas Beatty, Andrew Dolphin and Jill Gerke for other useful suggestions.  This research gratefully made use of the NASA/IPAC Extragalactic Database (NED).  Benjamin J. Shappee was supported by a Graduate Research Fellowship from the National Science Foundation.

\appendix
\section{Creating a PSF for ISIS from Daophot}
\label{sec:isis_psf}

Here we describe how to make ISIS use an externally determined PSF model.

\begin{figure}[H]
\plotone{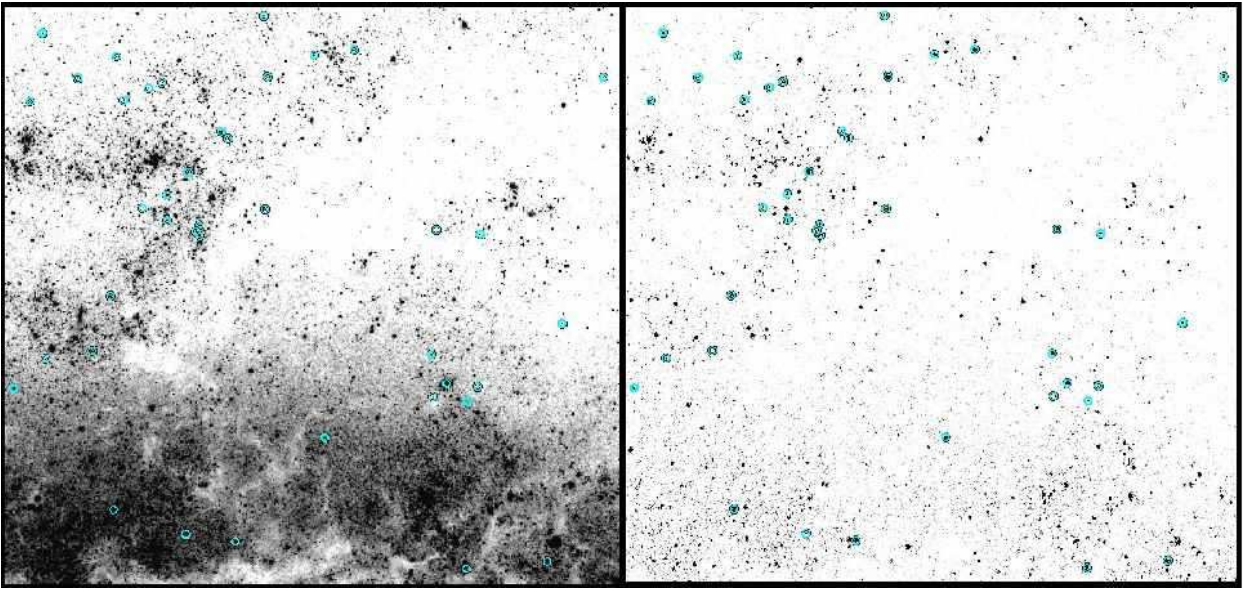}
\caption{
Inner region of Field 2 in M101 from our F555W ISIS reference image (left) and our F555W ISIS variability map (right).  Cepheids from the final sample found in \S\ref{sec:CephVarSearch} are marked by teal annuli.  The region is $\approx62''$ on each side, corresponding to a physical size of $\approx1.9$ kpc.  Figure discussed in \S\ref{sec:image_sub}.
}
\label{fig:F2Sub}
\end{figure}

\begin{figure}[H]
\includegraphics[scale=0.75, angle=90]{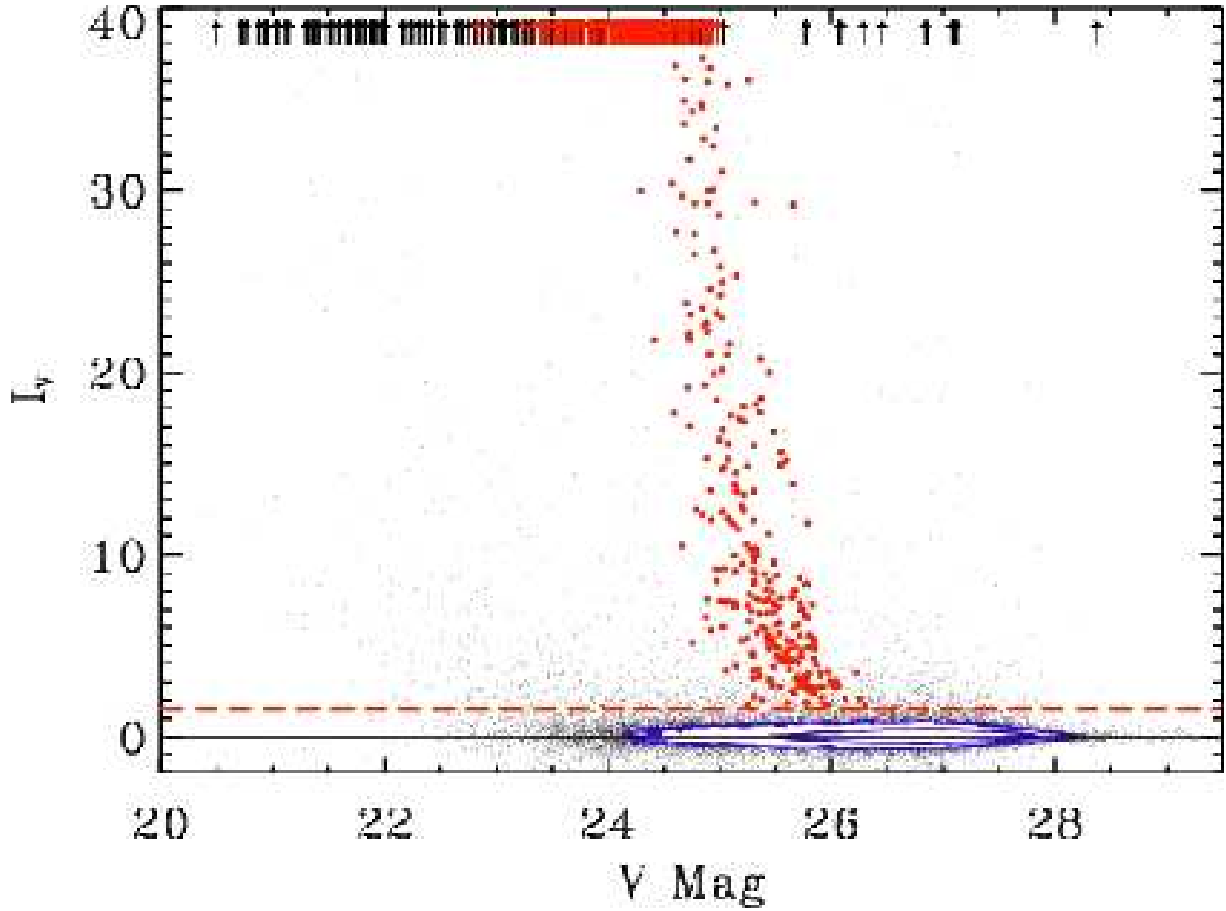}
\caption{
Welch-Stetson variability index $I_{V}$ \citep{welch93} versus \textit{V}-band magnitude for the objects in Field 1.  The black solid line marks a variability index of 0. The red dashed line marks the $I_{V} \geq $ \VarCut{} criteria used to select variables.  Objects with a $I_{V}$ $>$ 40 are marked with arrows. The blue lines represent the 68\% and 90\% contours.  Red circles and red arrows mark the locations of the final Cepheid sample found in \S\ref{sec:CephVarSearch}.  Figure discussed in \S\ref{sec:VarSearch}.
}
\label{fig:VarF1}
\end{figure}

\begin{figure}[H]
\includegraphics[scale=0.75, angle=90]{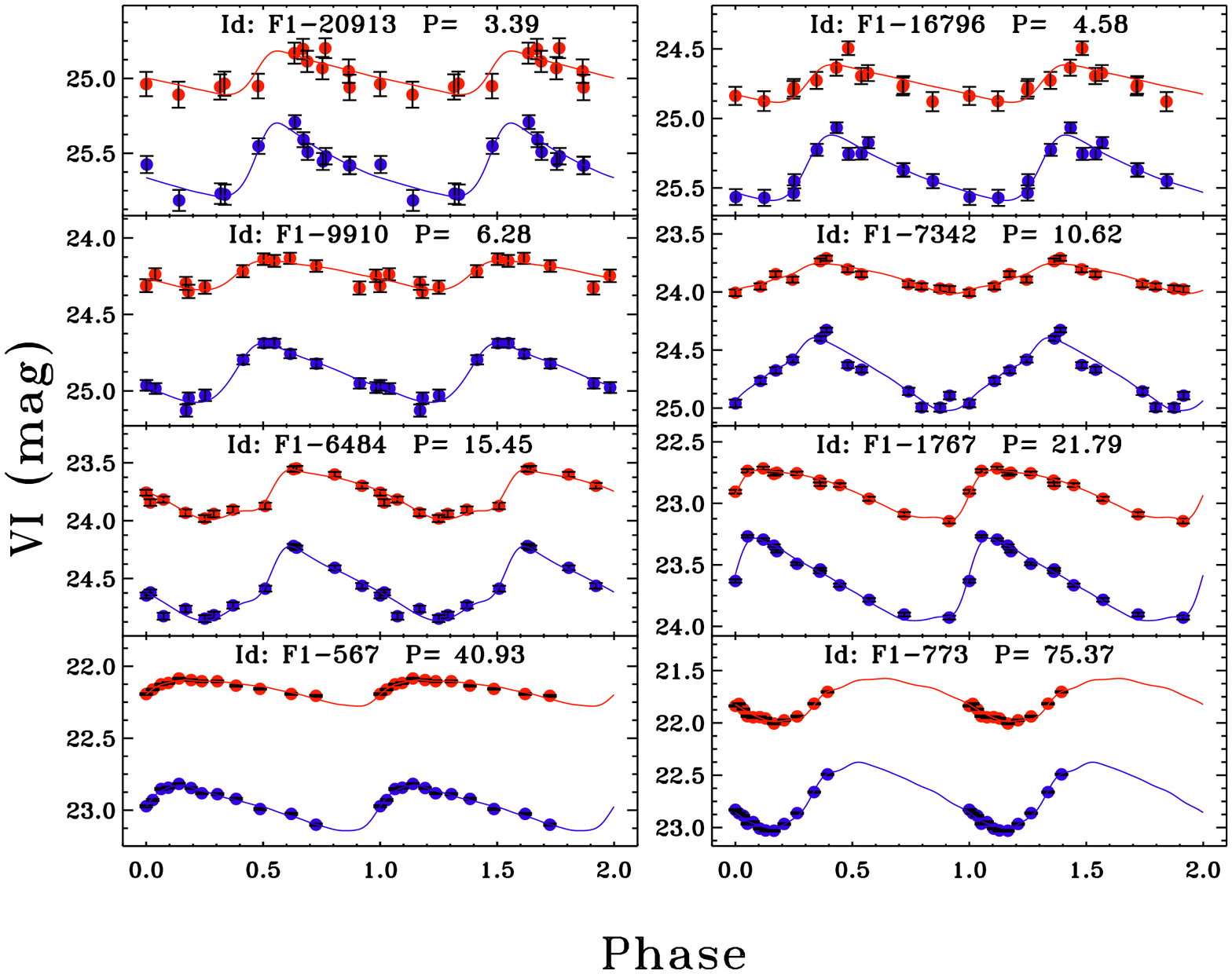}
\caption{
Cepheid light curves of various periods from Field 1.  Blue (bottom): \textit{V} filter light curves.  Red (top): \textit{I} filter light curves.  Solid line shows the best-fit light curve template from \citet{yoachim09} with additional fit amplitude parameters.   Figure discussed in \S\ref{sec:FTest}.
}
\label{fig:F1LightCurves}
\end{figure}

\begin{figure}[H]
\includegraphics[scale=0.75, angle=90]{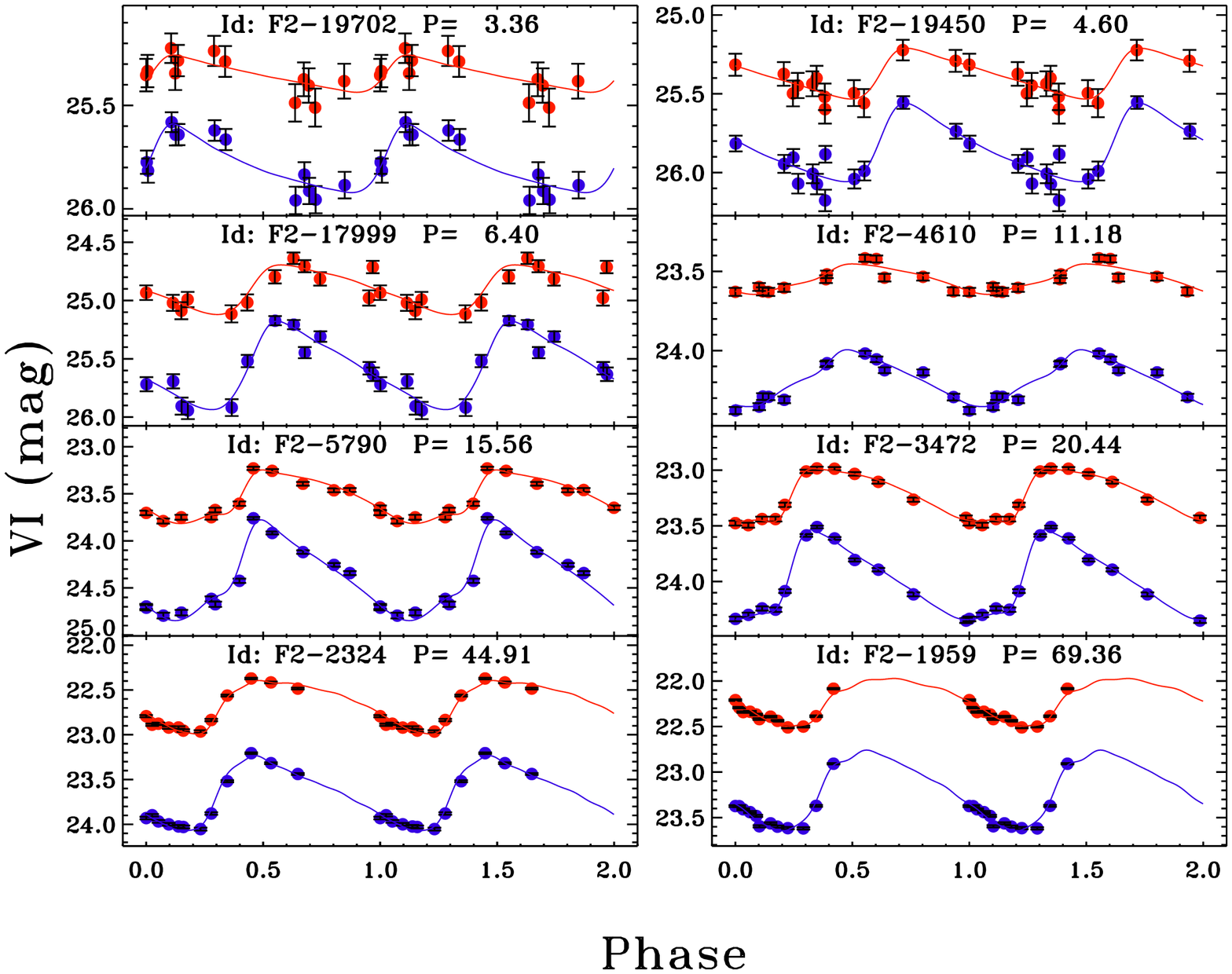}
\caption{
Cepheid light curves of various periods from Field 2.  Blue (bottom): \textit{V} filter light curves.  Red (top): \textit{I} filter light curves.  Solid line shows the best-fit light curve template from \citet{yoachim09} with additional fit amplitude parameters.   Figure discussed in \S\ref{sec:FTest}.
}
\label{fig:F2LightCurves}
\end{figure}

\begin{figure}[H]
\includegraphics[scale=0.75, angle=90]{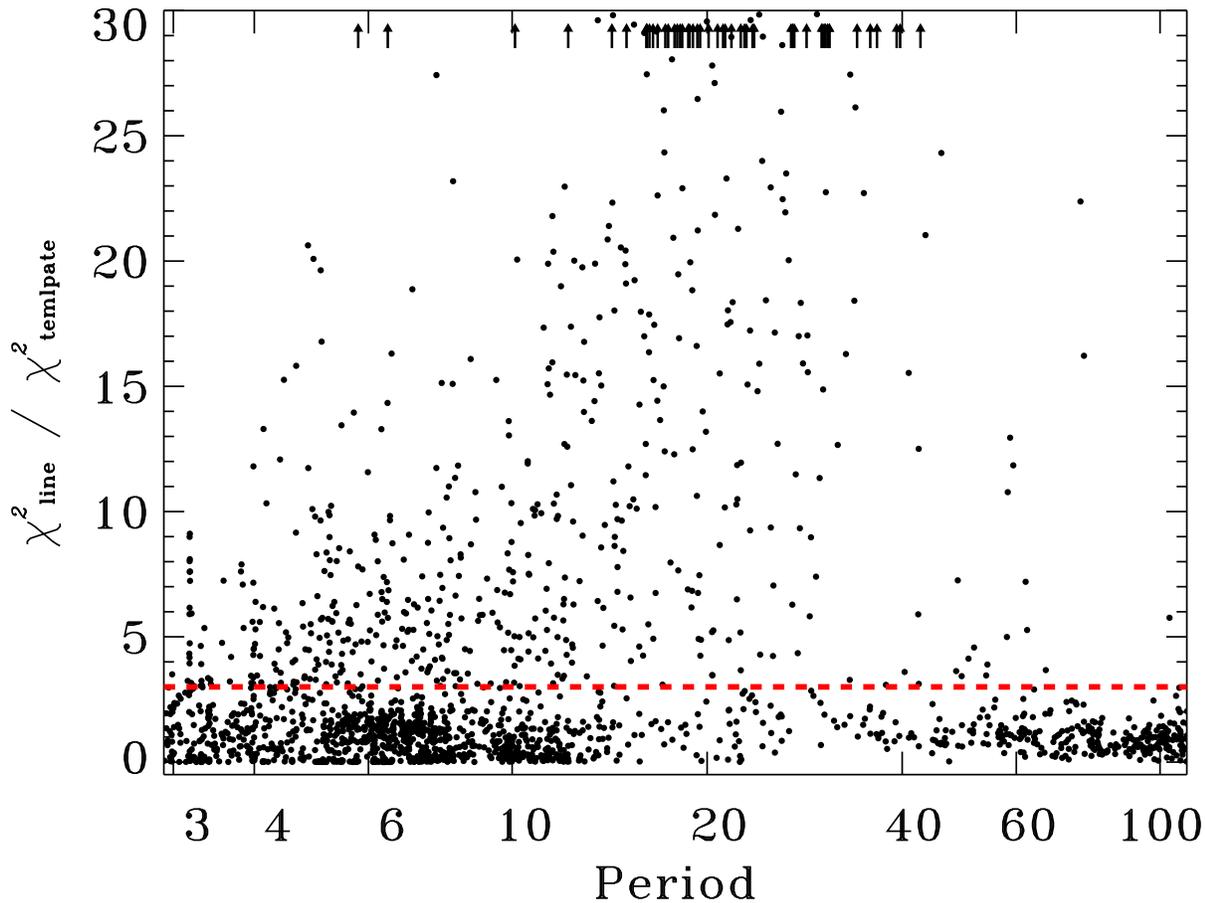}
\caption{
Distribution of $\Delta{}\chisqnu{}$ as a function of the template model period for objects in Field 1.  Objects with $\Delta{}\chisqnu{}> 30$ are represented by arrows.  Figure discussed in \S\ref{sec:FTest}.
}
\label{fig:FtestF1}
\end{figure}

\begin{figure}[H]
\plotone{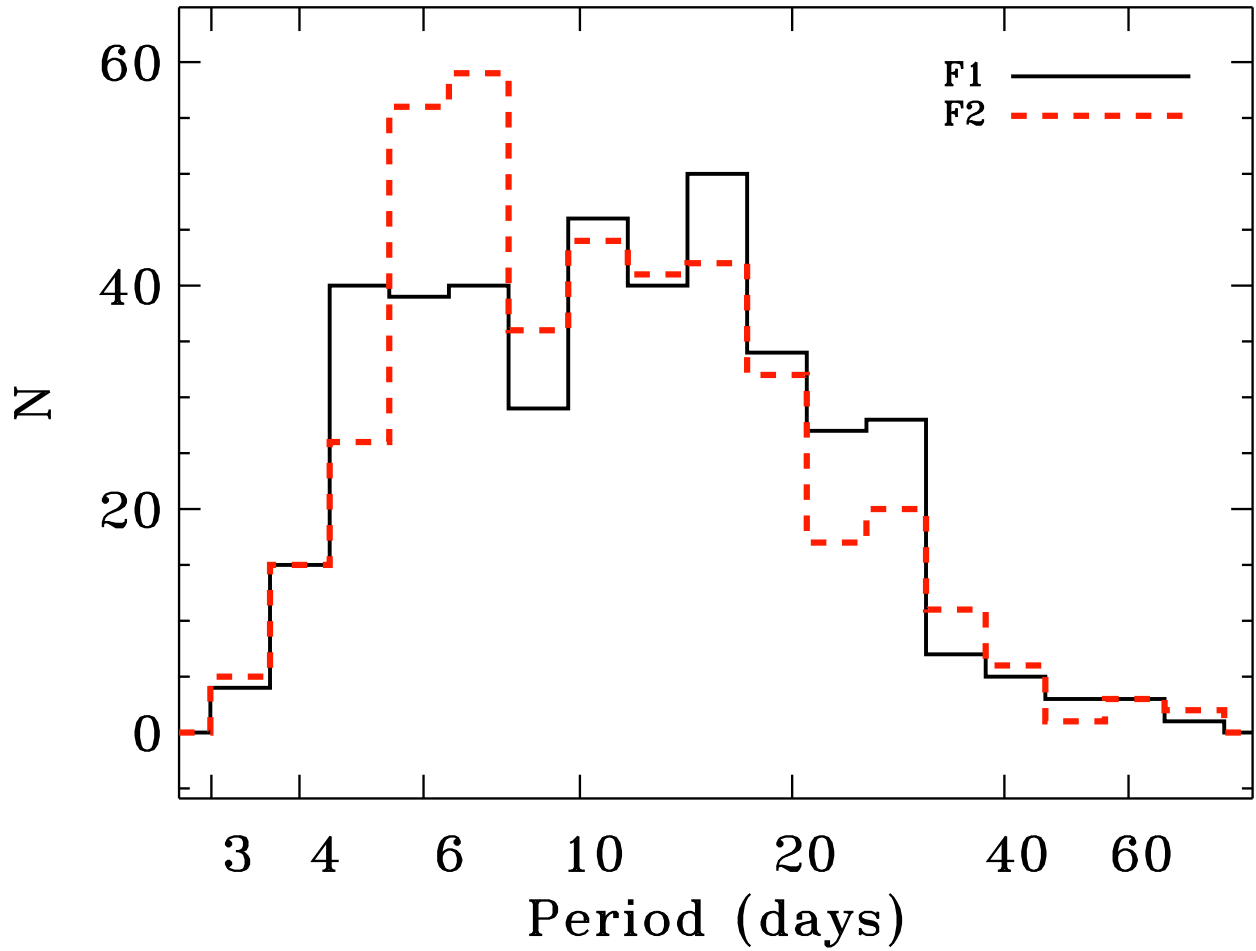}
\caption{
Distribution of periods in Field 1 (solid) and Field 2 (dashed) for the final Cepheid sample shown in Table~\ref{tab:cephbasic}.  Figure discussed in \S\ref{sec:sampleSel}.
}
\label{fig:PerHist}
\end{figure}

\begin{figure}[H]
\plotone{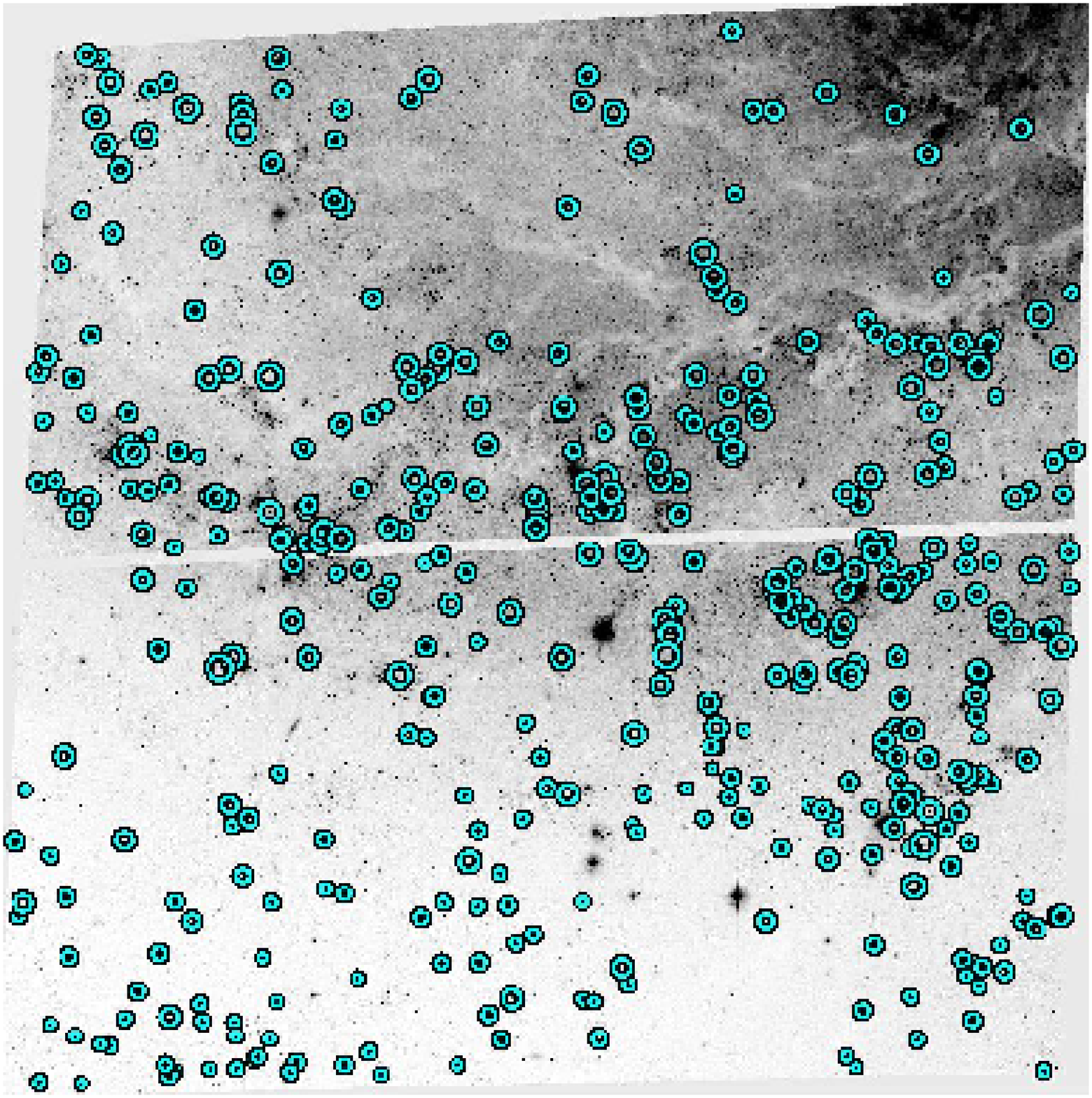}
\caption{
ISIS reference image of the HST ACS/WFC F555W data for Field 1 in M101.  Cepheids from the final sample shown in Table~\ref{tab:cephbasic} are marked by teal open circles. The area of each circle is proportional to the logarithm of the Cepheid's period. The shortest and longest period Cepheids have periods of 3.2 days and 75.4 days, respectively.  Figure discussed in \S\ref{sec:sampleSel}.
}
\label{fig:F1Ceph}
\end{figure}

\begin{figure}[H]
\plotone{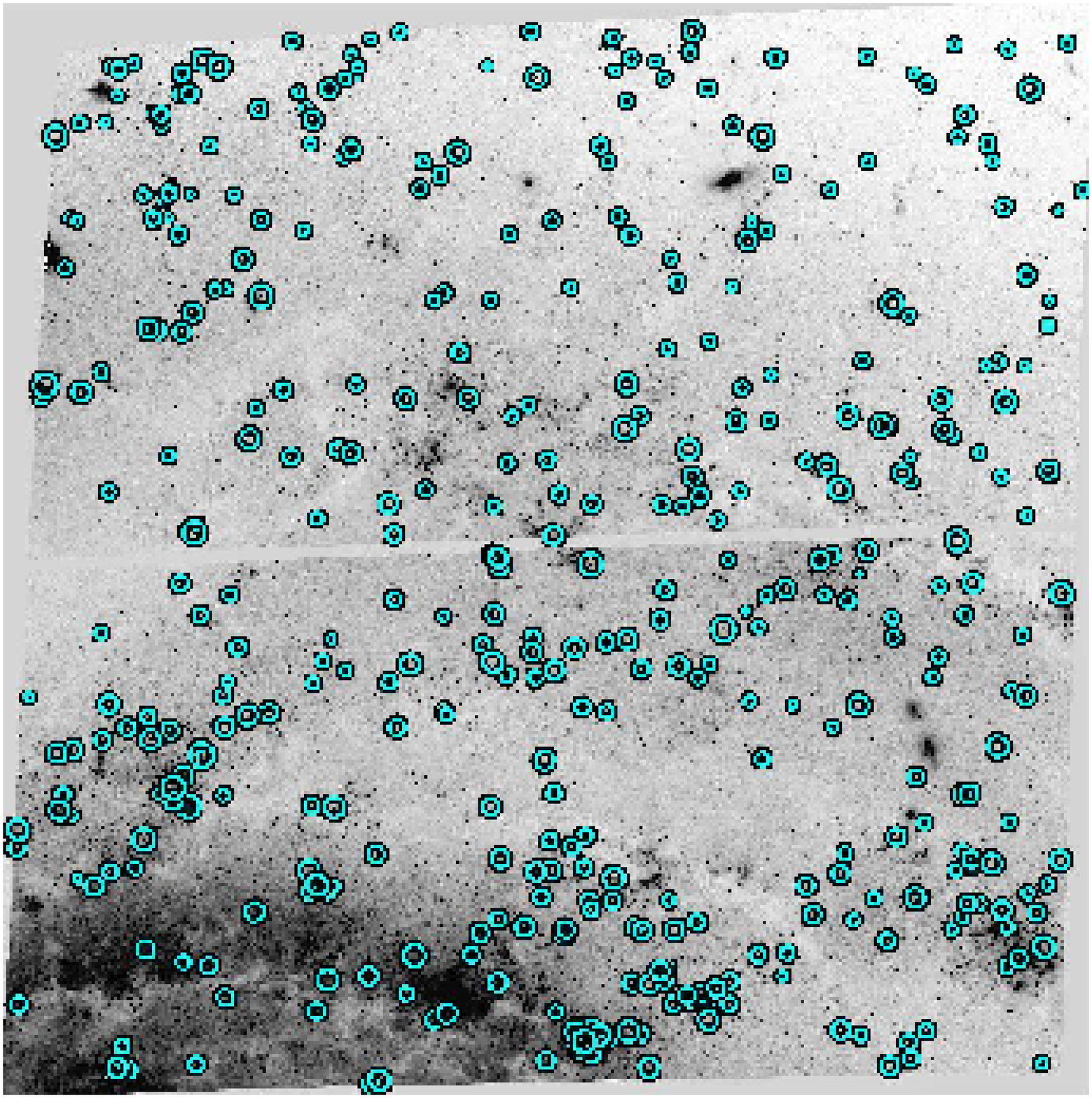}
\caption{
ISIS reference image of the HST ACS/WFC F555W data for Field 2 in M101.  Cepheids from the final sample shown in Table~\ref{tab:cephbasic} are marked by teal open circles. The area of each circle is proportional to the logarithm of the Cepheid's period. The shortest and longest period Cepheids have periods of 3.0 days and 69.4 days, respectively.   Figure discussed in \S\ref{sec:sampleSel}.
}
\label{fig:F2Ceph}
\end{figure}

\begin{figure}[H]
\includegraphics[scale=0.75, angle=90]{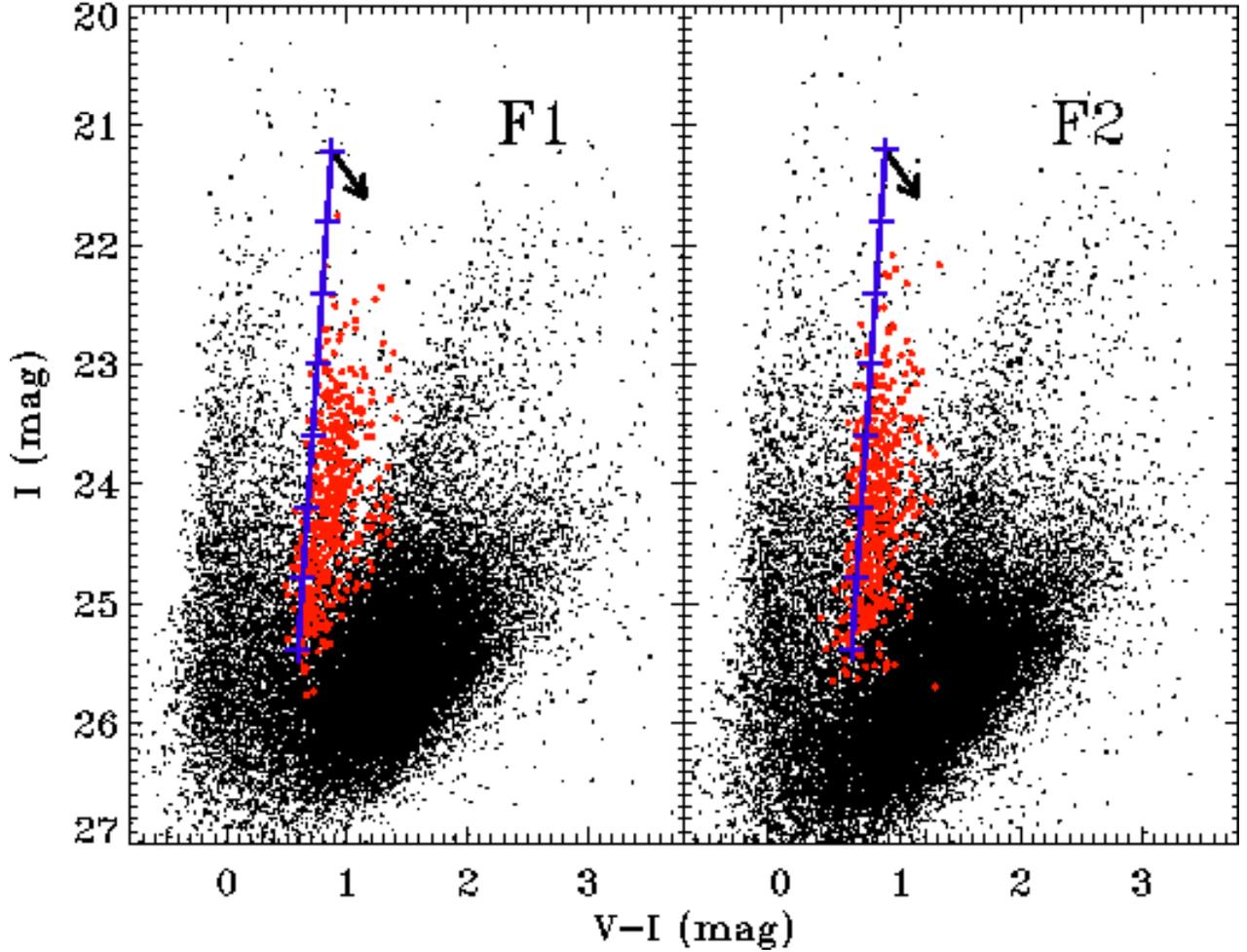}
\caption{
\textit{I} versus $(V-I)$ color-magnitude diagram for Field 1 (left) and Field 2 (right), displaying every fifth source. The solid blue lines are the predicted position of the unreddened LMC P-L relations of \citet{udalski99} for Cepheids of periods 4-100 days, shifted to the distance modulus fit in Fig.~\ref{fig:PDistF1} and Fig.~\ref{fig:PDistF2}.  The tick marks are every 0.2 log (days) along the P-L relation.  The arrow indicates a reddening of $E(B-V)=0.2$ mag.  Red circles represent the final Cepheid sample presented in Table~\ref{tab:cephprop}.  Figure discussed in \S\ref{sec:sampleSel}.
}
\label{fig:CMD}
\end{figure}

\begin{figure}[H]
\plotone{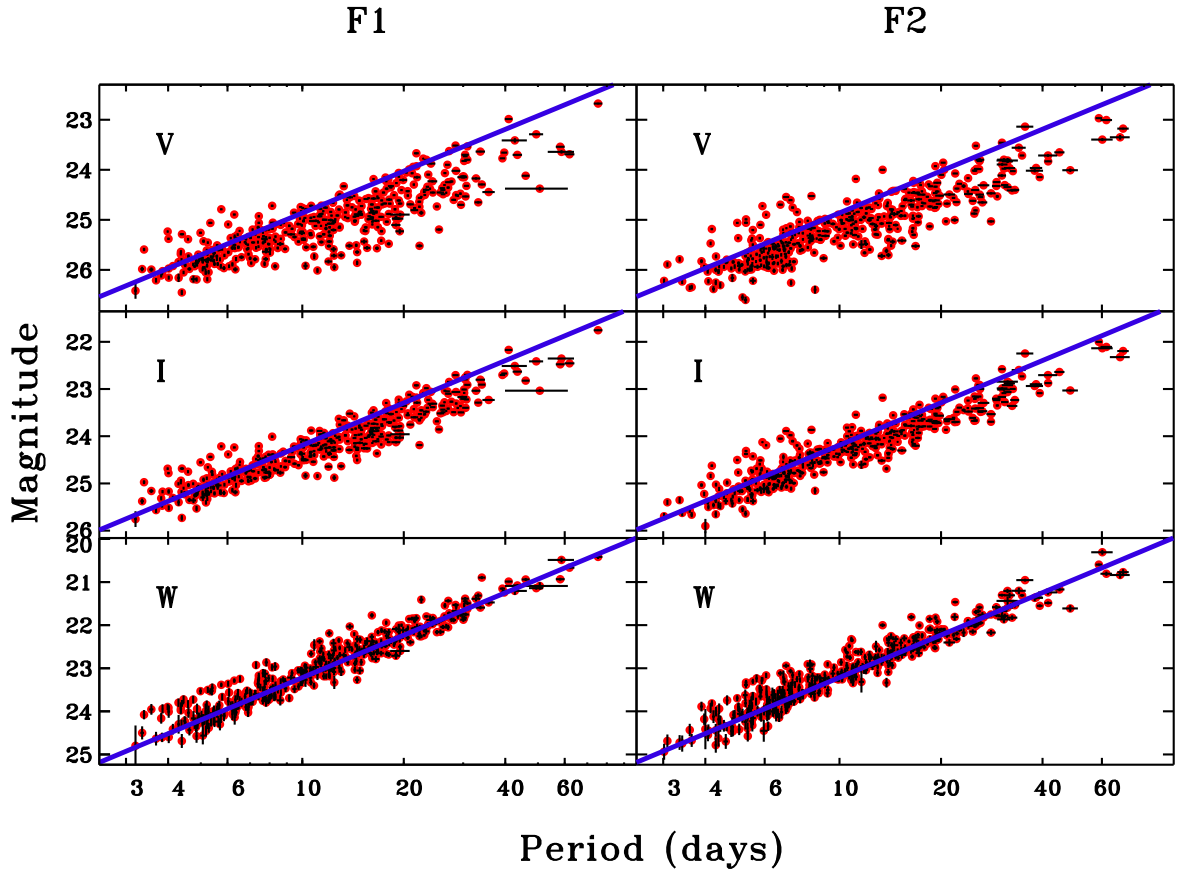}
\caption{
\textit{V}, \textit{I} and Wesenheit P-L relations for Field 1 (left) and Field 2 (right). The blue solid lines are the LMC P-L relations of \citet{udalski99} shifted to the distance modulus fit in Fig.~\ref{fig:PDistF1} and Fig.~\ref{fig:PDistF2}. Random errors for both period and magnitude are shown. Reddening can be seen in the \textit{V} and \textit{I} P-L relations.   Figure discussed in \S\ref{sec:sampleSel}.
}
\label{fig:PLAll}
\end{figure}

\begin{figure}[H]
\plotone{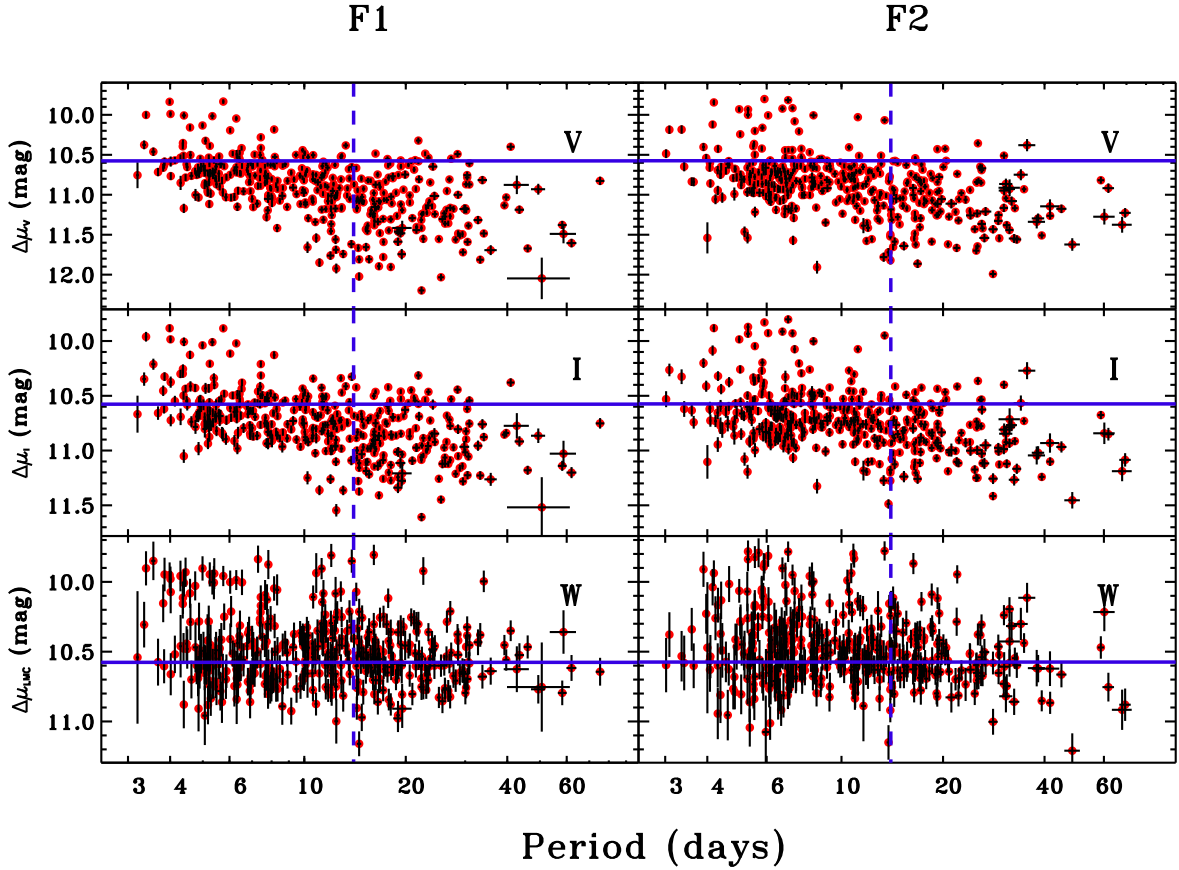}
\caption{
\textit{V}, \textit{I} and Wesenheit LMC relative distance moduli of the Cepheids as a function of period for Field 1 (left) and Field 2 (right). The vertical dashed blue line marks the minimum period cut used in each field to refine the final Cepheid sample used to determine the distance modulus. \CUTnumFinCutONE{} Cepheids in Field 1 and \CUTnumFinCutTWO{} Cepheids in Field 2 have periods larger then the period cut.  The horizontal solid blue line represents the distance modulus fit in Fig.~\ref{fig:PDistF1} and Fig.~\ref{fig:PDistF2}. Random errors for both period and relative distance modulus are shown. Reddening can be seen in the \textit{V} and \textit{I} P-L relations for Cepheids with periods above the period cut.  Figure discussed in \S\ref{sec:sampleSel}.
}
\label{fig:DistModAll}
\end{figure}

\begin{figure}[H]
\includegraphics[scale=0.75, angle=90]{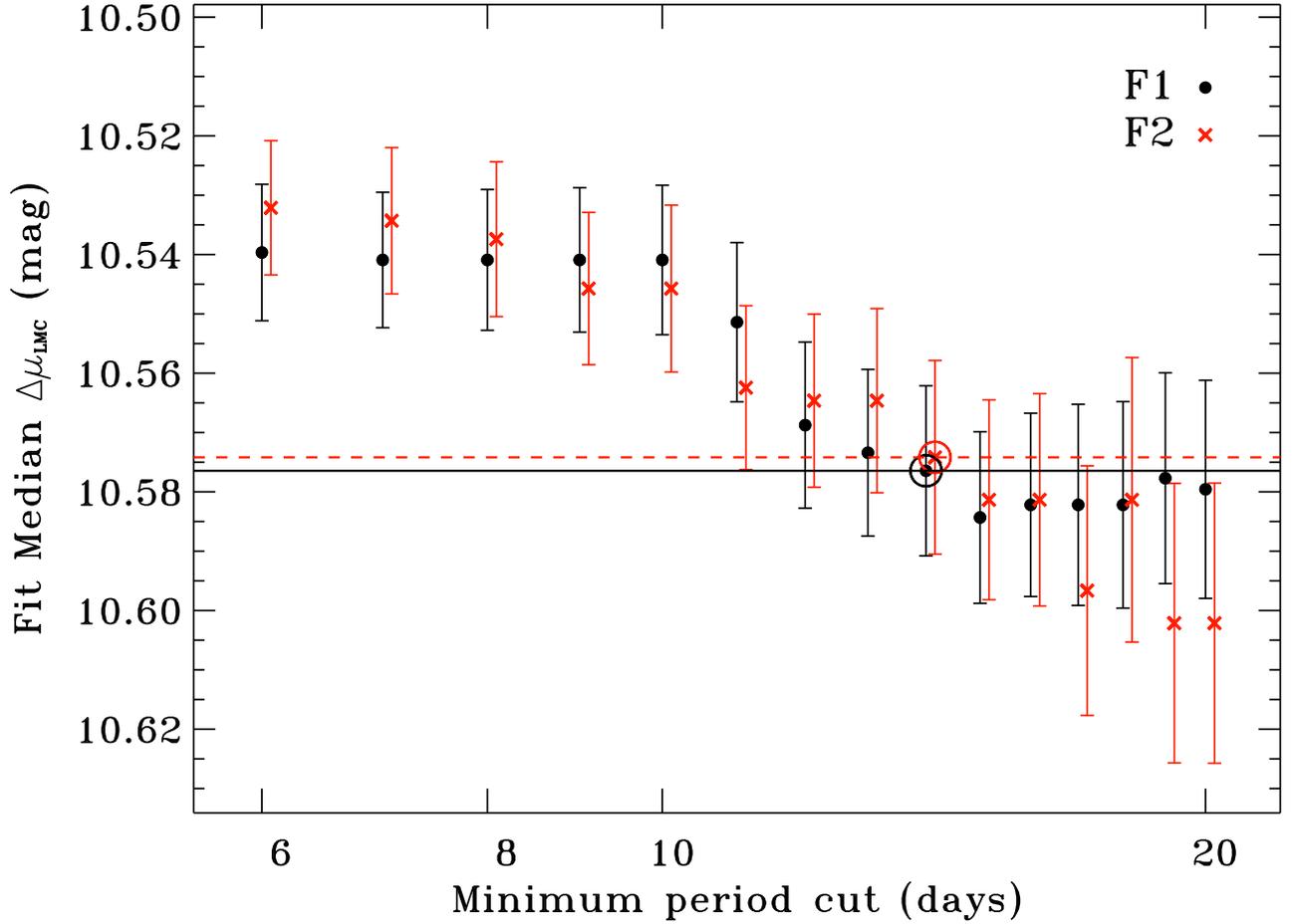}
\caption{
Median relative Wesenheit distance moduli as a function of the minimum period cut applied. Filled circles represent Field 1. Red crosses represent Field 2 and are shifted to higher periods by $0.01$ log(days). The solid line and the dashed red line represents the median distance modulus for Field 1. Final choices for period cut are shown by open circles.  \CUTnumFinCutONE{} Cepheids in Field 1 and \CUTnumFinCutTWO{} Cepheids in Field 2 have periods larger then the period cut.  The dashed line represents the median distance modulus at the period cut for Fields 1 and 2, respectively. Figure discussed in \S\ref{sec:MinPer}.
}
\label{fig:DistPcut}
\end{figure}

\begin{figure}[H]
\includegraphics[scale=0.75, angle=90]{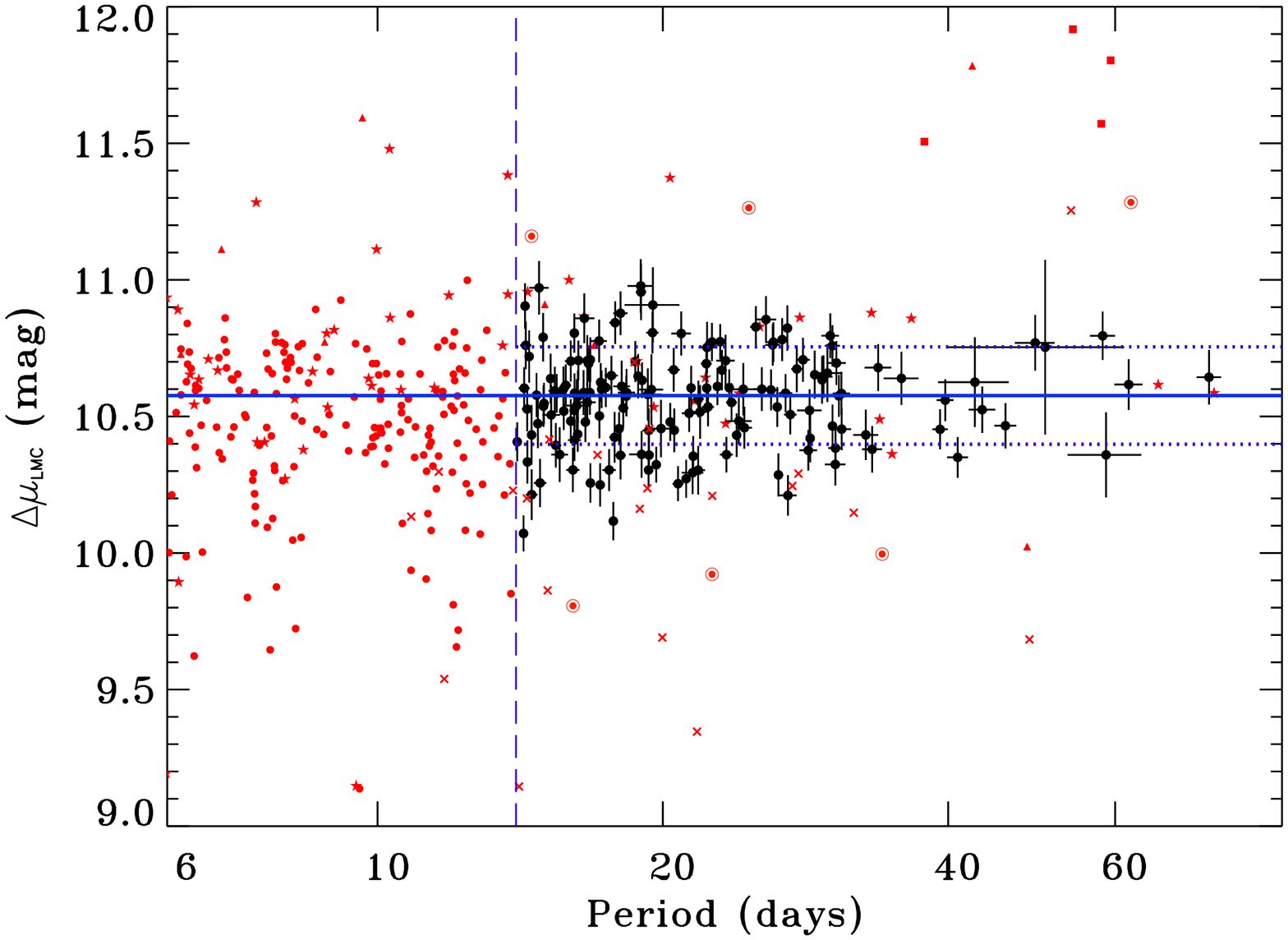}
\caption{
Relative Wesenheit distance moduli as a function of period for Field 1. Random errors are shown for Cepheids with periods larger then the period cut.   Figure discussed in \S\ref{sec:MinPer}.  \textbf{Filled Circles:}  \CUTnumFinCutONE{} Cepheids with periods larger then the period cut. \textbf{Solid line:} Median relative distance modulus. \textbf{Dotted line:} 1 $\sigma$ dispersion about the median in final Cepheid sample. \textbf{Dashed line:} Represents period cut. \textbf{Red Symbols:} Cepheids removed from various cuts. \textbf{Stars:} Amplitude ratio cut. \textbf{Triangles:} Blue blends cut.  \textbf{Crosses:} Red blend or large extinction cut.  \textbf{Squares:} Population II Cepheid or egregious outlier cut. \textbf{Red Circle:} Period cut. \textbf{Circled Red Point:} Iterative sigma clipping.  
}
\label{fig:PDistF1}
\end{figure}

\begin{figure}[H]
\includegraphics[scale=0.75, angle=90]{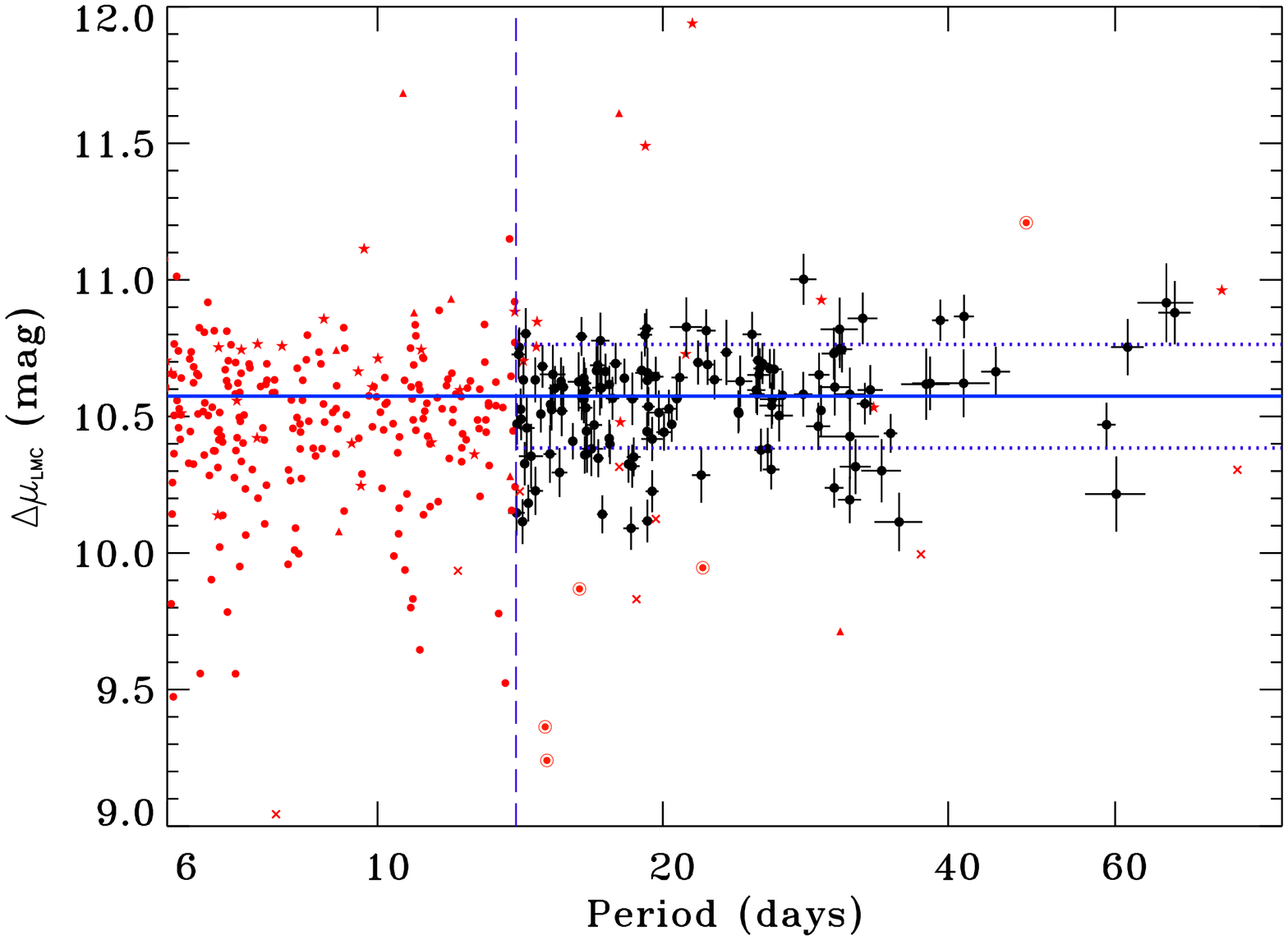}
\caption{
Relative Wesenheit distance moduli as a function of period for Field 2. Random errors are shown for Cepheids with periods larger then the period cut.   Figure discussed in \S\ref{sec:MinPer}.  \textbf{Filled Circles:} \CUTnumFinCutTWO{}  Cepheids with periods larger then the period cut. \textbf{Solid line:} Median relative distance modulus. \textbf{Dotted line:} 1 $\sigma$ dispersion about the median in final Cepheid sample. \textbf{Dashed line:} Represents period cut. \textbf{Red Symbols:} Cepheids removed from various cuts. \textbf{Stars:} Amplitude ratio cut. \textbf{Triangles:} Blue blends cut.  \textbf{Crosses:} Red blend or large extinction cut.  \textbf{Squares:} Population II Cepheid or egregious outlier cut. \textbf{Red Circle:} Period cut. \textbf{Circled Red Point:} Iterative sigma clipping.  
}
\label{fig:PDistF2}
\end{figure}

\begin{figure}[H]
\includegraphics[scale=0.75, angle=90]{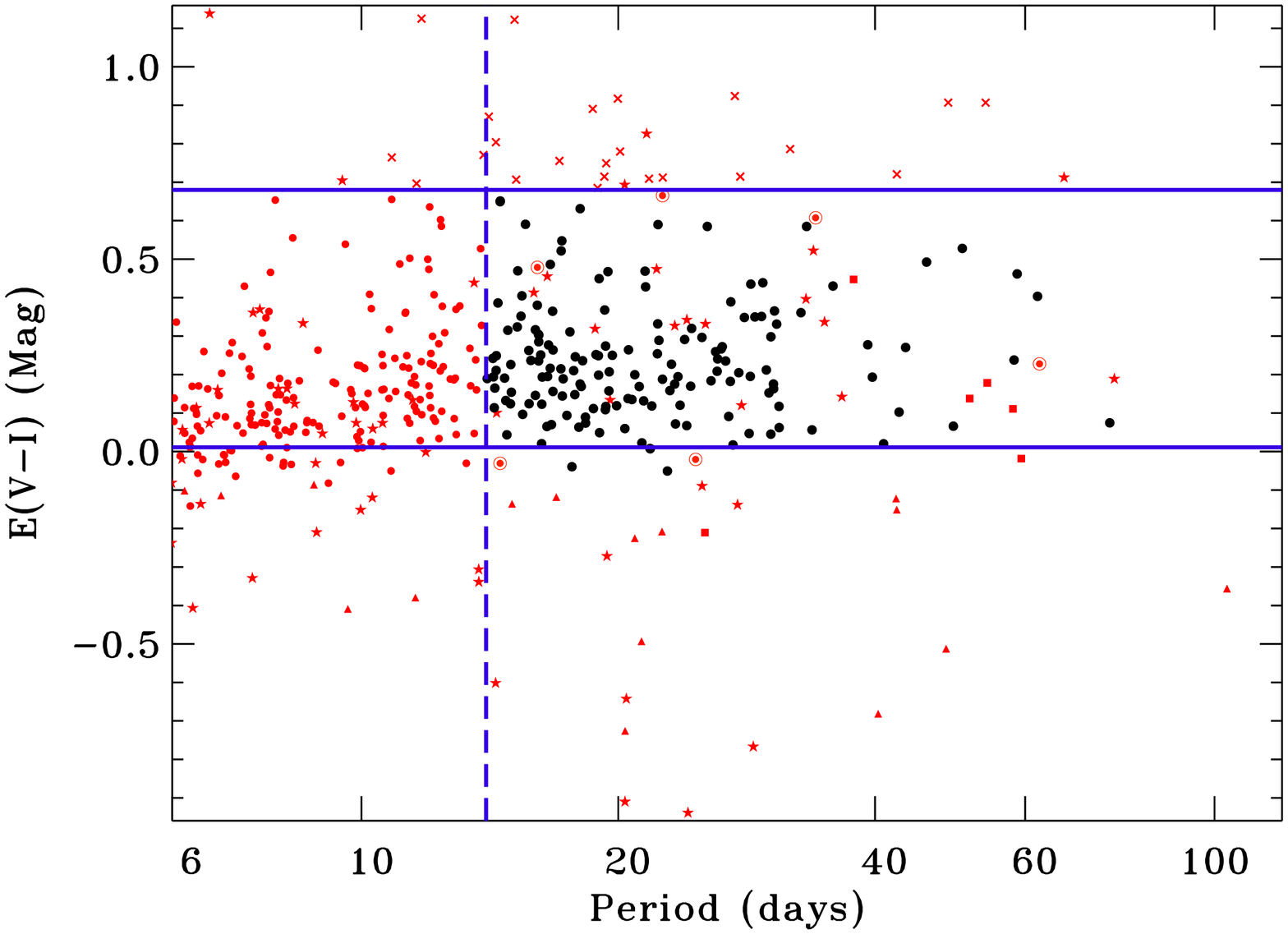}
\caption{
$E(V-I)$ as a function of period for Field 1. \textbf{Filled Circles:} \CUTnumFinCutONE{} Cepheids with periods larger then the period cut. \textbf{Solid line:} $E(V-I)$ values where the large extinction and blue blend cuts were applied.  \textbf{Vertical Dashed line:} Represents period cut off.  \textbf{Red Symbols:} Cepheids removed from various cuts. \textbf{Stars:} Amplitude ratio cut. \textbf{Triangles:} Blue blends cut.  \textbf{Crosses:} Red blend or large extinction cut.  \textbf{Squares:} Population II Cepheid or egregious outlier cut. \textbf{Red Circle:} Period cut. \textbf{Circled Red Point:} Iterative sigma clipping.  Figure discussed in \S\ref{sec:MinPer}.
}
\label{fig:PEV_IF1}
\end{figure}

\begin{figure}[H]
\includegraphics[scale=0.75, angle=90]{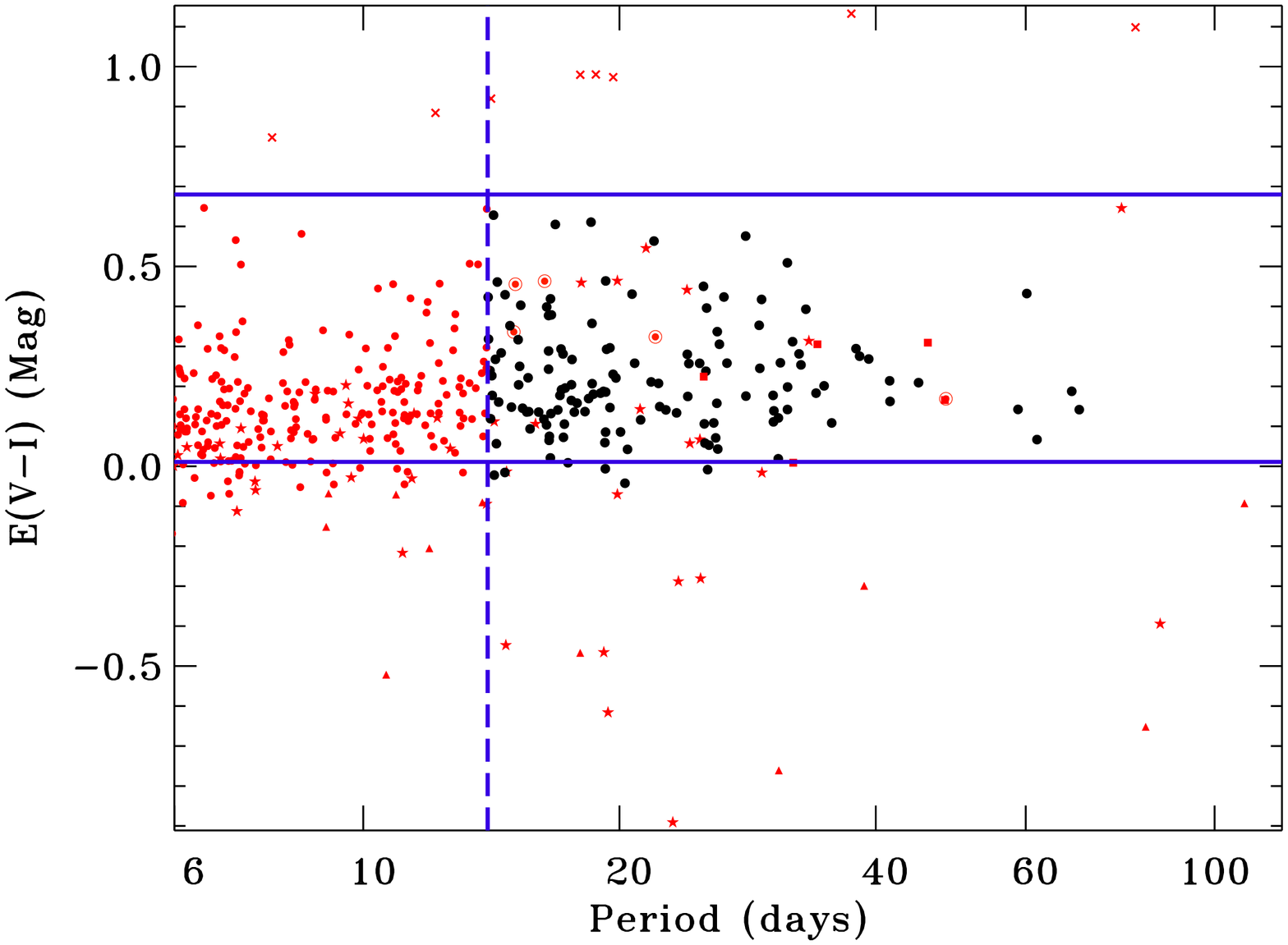}
\caption{
$E(V-I)$ as a function period for Field 2. \textbf{Filled Circles:} \CUTnumFinCutTWO{} Cepheids with periods larger then the period cut. \textbf{Solid line:} $E(V-I)$ values where the large extinction and blue blend cuts were applied.  \textbf{Vertical Dashed line:} Represents period cut off.  \textbf{Red Symbols:} Cepheids removed from various cuts. \textbf{Stars:} Amplitude ratio cut. \textbf{Triangles:} Blue blends cut.  \textbf{Crosses:} Red blend or large extinction cut.  \textbf{Squares:} Population II Cepheid or egregious outlier cut. \textbf{Red Circle:} Period cut. \textbf{Circled Red Point:} Iterative sigma clipping.    Figure discussed in \S\ref{sec:MinPer}.
}
\label{fig:PEV_IF2}
\end{figure}

\begin{figure}[H]
\includegraphics[scale=0.75, angle=90]{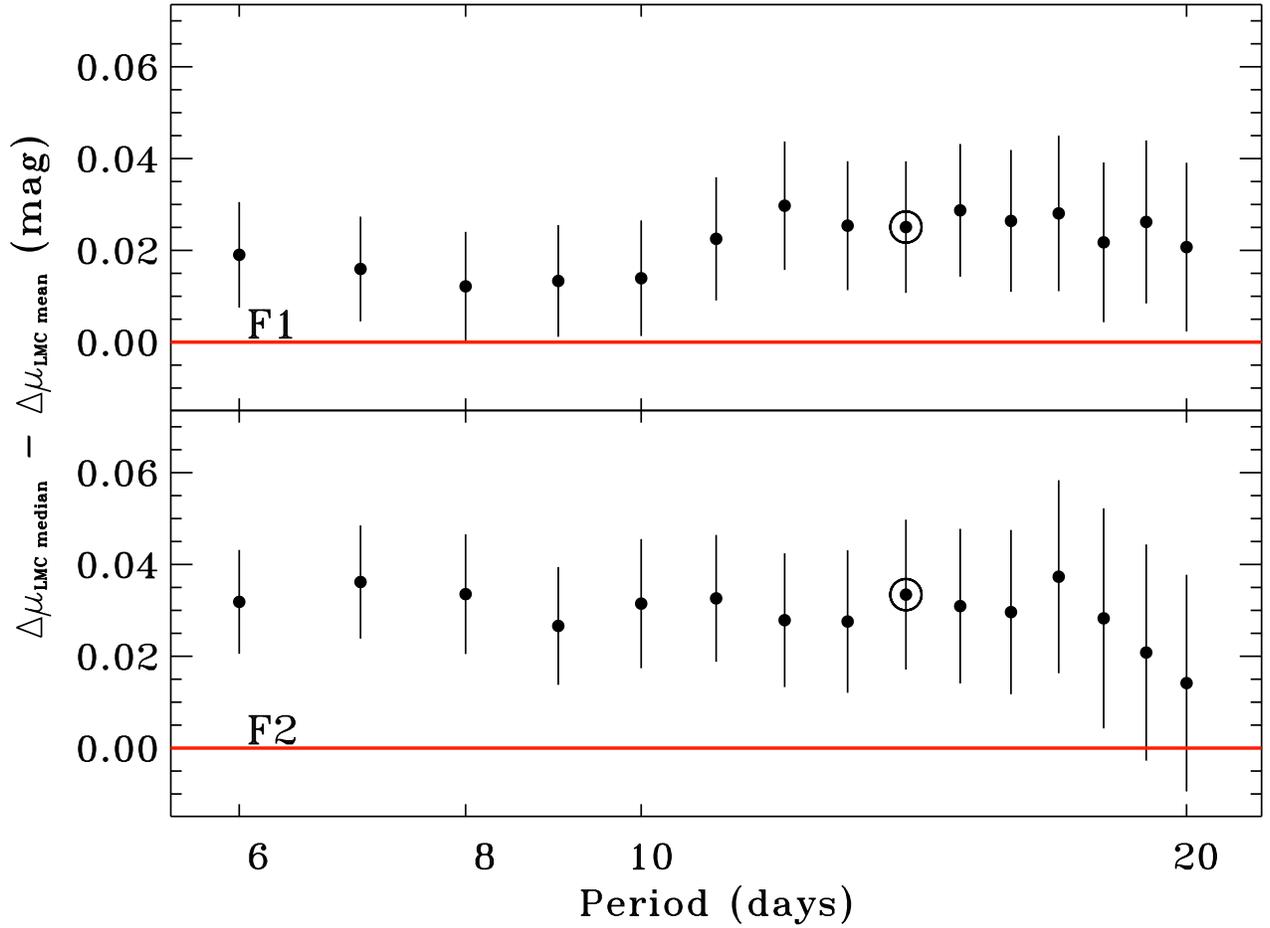}
\caption{
Difference between median and mean relative Wesenheit distance moduli as a function minimum period cut applied.  Top: Field 1.  Bottom: Field 2. Solid red line is at 0 to guide the eye.  Errors estimates shown are the quadrature addition of the scatter about the median and the scatter about the mean.  Figure discussed in \S\ref{sec:MinPer}.
}
\label{fig:PMean_Med}
\end{figure}

\begin{figure}[H]
\includegraphics[scale=0.75, angle=90]{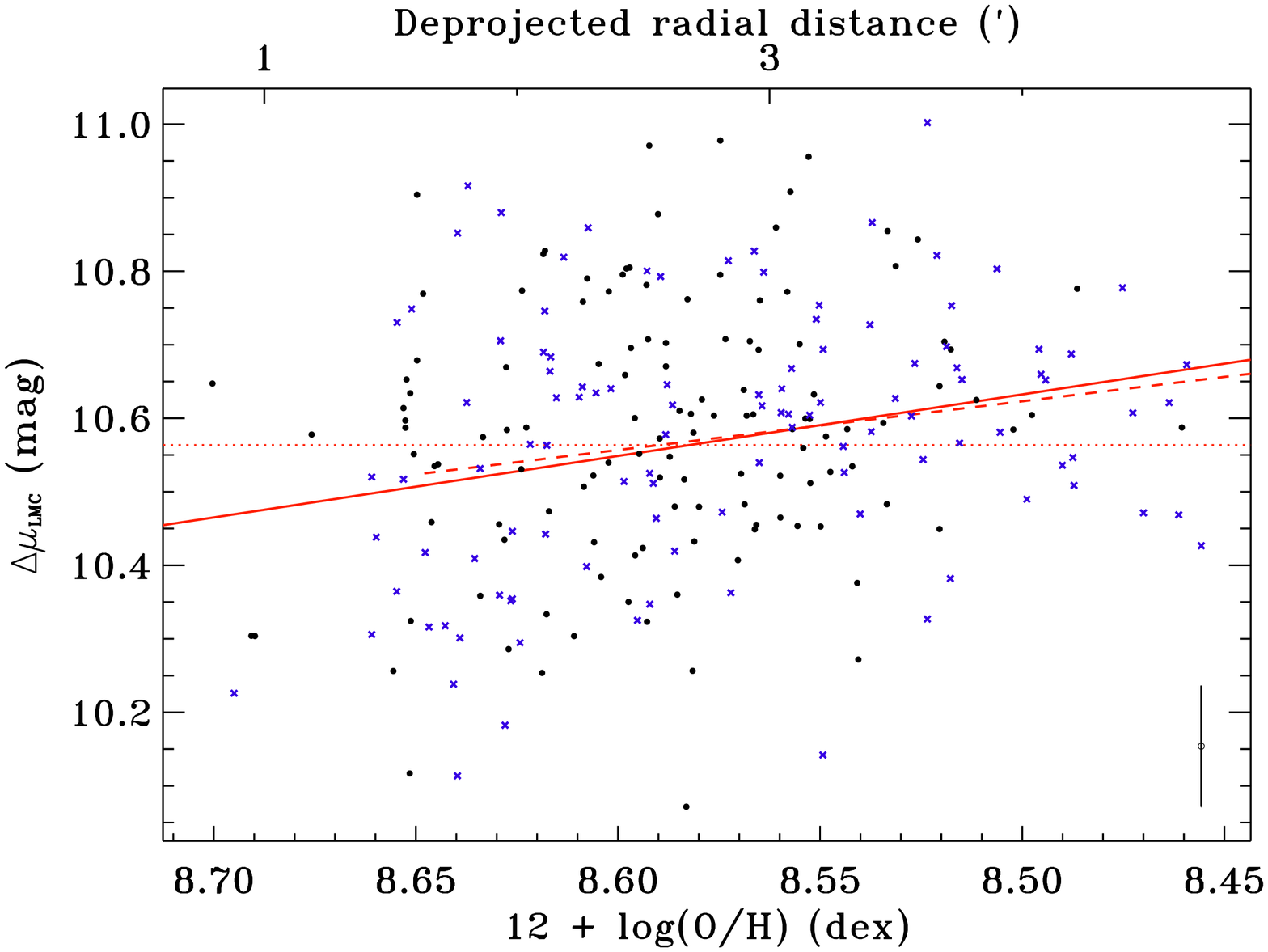}
\caption{
Metalicity dependence of the relative Wesenheit distance modulus.  Filled circles and blue crosses represent Cepheids in the refined sample in Fields 1 and 2, respectively. 
The solid red best-fit line is show with a slope of \MetSlope{} $\pm{}$ \MetSlopeErrRan $_r \ \pm{}$ \MetSlopeErrSys $_s$ mag  dex$^{-1}$.  The dotted red best-fit constant is shown for reference.  The dashed red best-fit line to a restricted sample of Cepheid with 12 + log(O/H) $< 8.65$ dex is shown with a slope of \MetSlopecut{} $\pm{}$ \MetSlopeErrRancut $_r \ \pm{}$ \MetSlopeErrSyscut $_s$ mag  dex$^{-1}$.  At 12 + log(O/H) = 8.5 dex, the LMC relative distance modulus of M101 is $\LMCreldist{} \ = \ $\MetDist{}$ \ \pm{} \ $\MetDistErrRan$_{r} \ \pm{} $\MetDistErrSys$_{s}$.  The error bar on the open circle in lower right displays the average individual random uncertainties.  Figure discussed in \S\ref{sec:DepandMet}.
}
\label{fig:MetalAll}
\end{figure}

\begin{figure}[H]
\includegraphics[scale=0.75, angle=90]{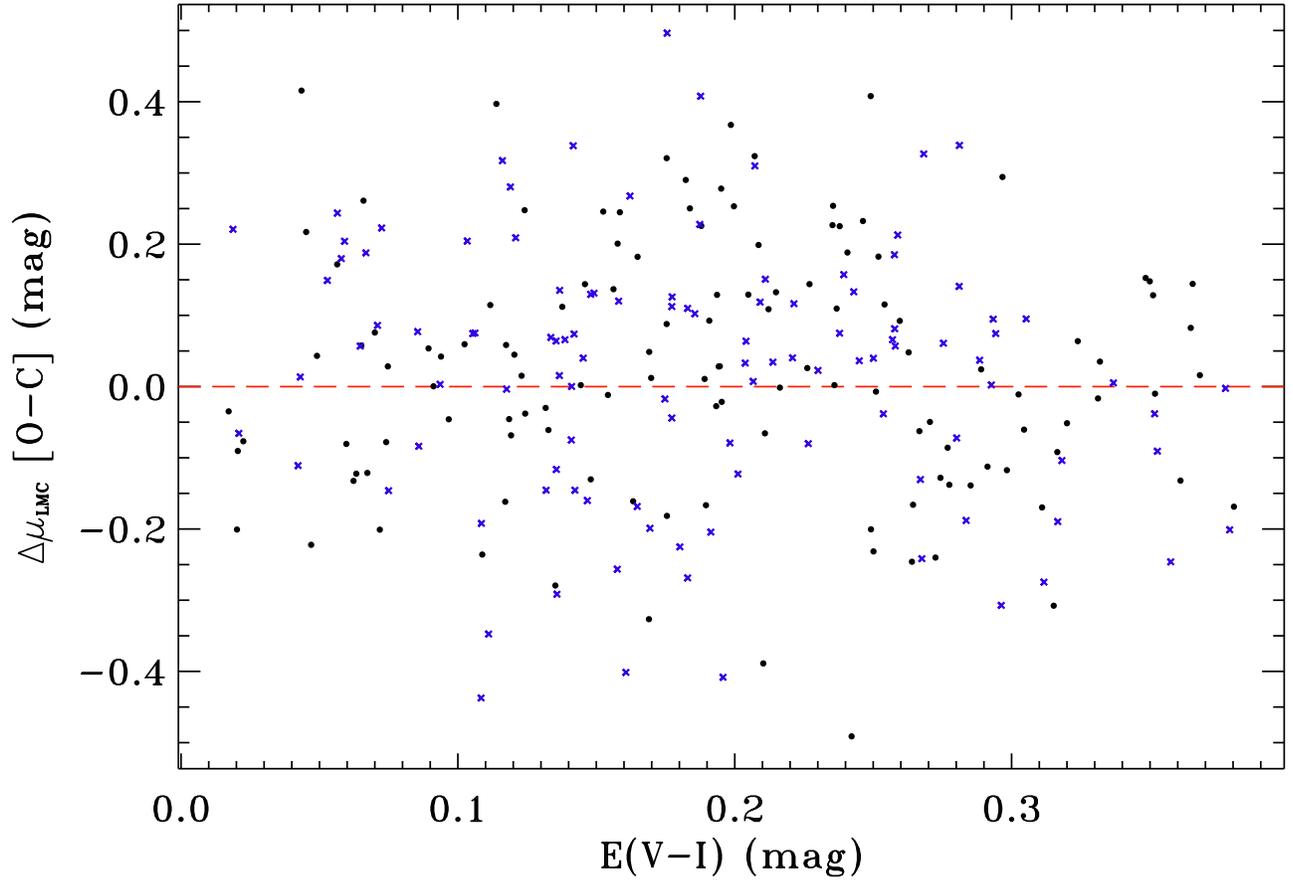}
\caption{
Residuals about the best-fit metallicity relation shown in Fig.~\ref{fig:MetalAll}. Filled circles and blue crosses represent Cepheids in the refined sample in Fields 1 and 2, respectively. Dashed red line shown at $\LMCreldist{}[\textrm{O}-\textrm{C}] \ = \ 0$ to aid the eye.    Figure discussed in \S\ref{sec:DepandMet}.
}
\label{fig:MetalEBV}
\end{figure}

\begin{figure}[H]
\includegraphics[scale=0.75, angle=90]{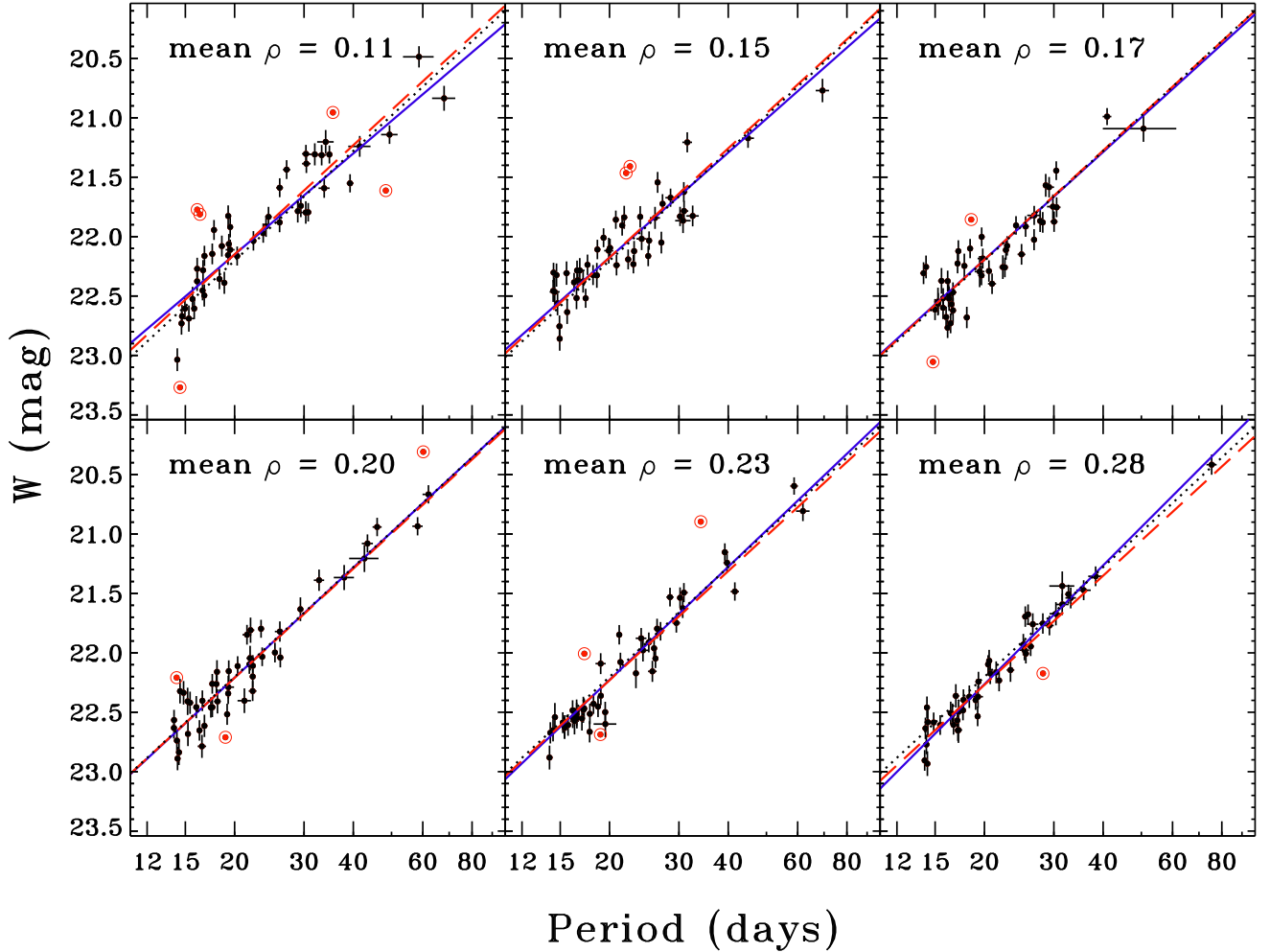}
\caption{
Wesenheit P-L relations with equal numbers of Cepheids in bins of deprojected radius.  Filled circles represent Cepheids selected in \S\ref{sec:sampleSel} with an additional minimum period cut of \CUTPerCut{} days.  The dotted black line indicates fit Wesenheit P-L relation with no variation as function of deprojected radius (Model 1).  The dashed red line represents the best-fit Wesenheit P-L with constant a slope and a linearly varying zero-point as a function of deprojected radius (Model 2).  The solid blue line is the best-fit Wesenheit P-L allowing linear variations in the slope and zero-point as a function of deprojected radius (Model 4).  Cepheids were removed by iterative sigma clipping from Model 1 and are represented by red points encircled by red rings. The mean Cepheid deprojected isophotal radius fraction is shown for each frame.  Fit parameters shown in Table~\ref{tab:WPLmodels}.  Figure discussed in \S\ref{sec:WPL} and \S\ref{sec:Disgusted}.
}
\label{fig:WPLfits}
\end{figure}

\begin{figure}[H]
\includegraphics[scale=0.50, angle=90]{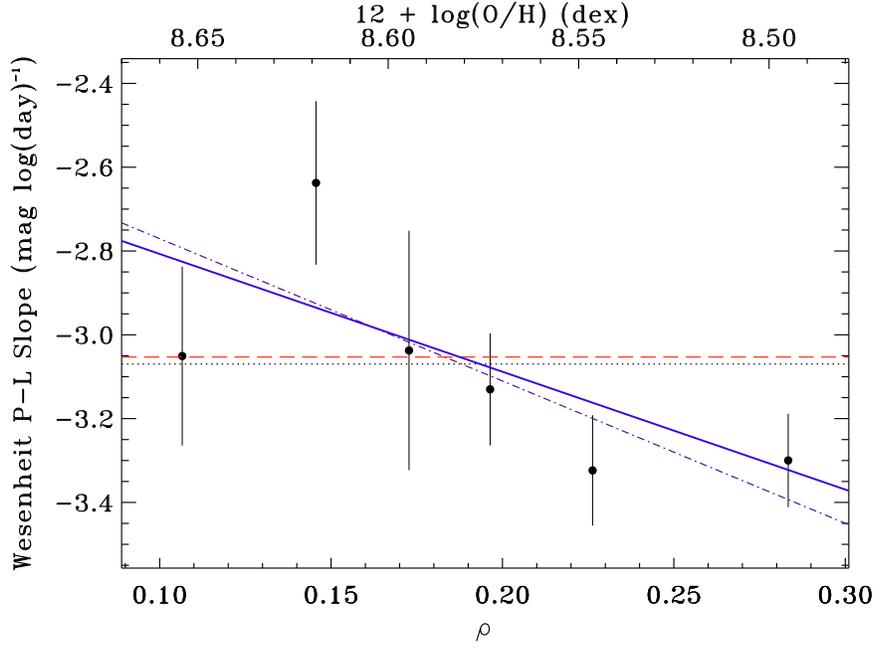}
\caption{
The best-fit slope of the Wesenheit P-L relation as a function of fractional isophotal radius. The dotted black line indicates fit Wesenheit P-L relation with constant slope and zero-point (Model 1).  The dashed red line represents fit Wesenheit P-L with constant slope and linearly varying zero-point as a function of deprojected radius (Model 2). The dot dashed purple line indicates fit Wesenheit P-L with constant zero-point and linear varying slope as a function of deprojected radius (Model 3). The dashed red line represents fit Wesenheit P-L with linear varying slope and zero-point as a function of deprojected radius (Model 4).  The filled black dots represent the slopes of the individually fit Wesenheit P-L relations for smaller samples shown in Fig.~\ref{fig:WPLfits} with error bars determined using bootstrap re-sampling. It is important to note that these individually fit sub-samples are plotted for comparison purposes only.  Figure discussed in \S\ref{sec:WPL}.
}
\label{fig:WPLslope}
\end{figure}

\begin{figure}[H]
\includegraphics[scale=0.50, angle=90]{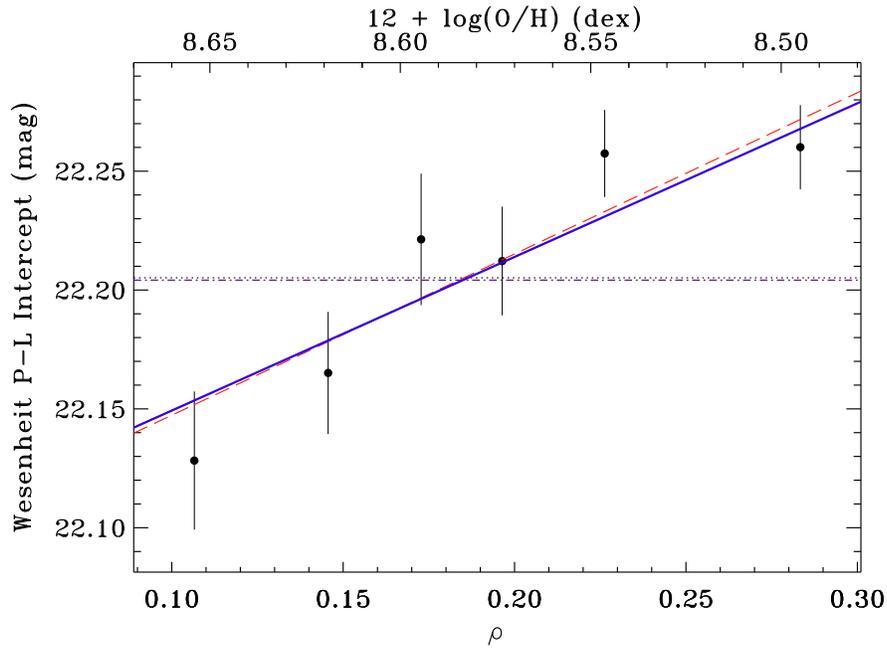}
\caption{
Same as Fig.~\ref{fig:WPLslope} but showing the Wesenheit P-L zero-point as a function of fractional isophotal radius.   Figure discussed in \S\ref{sec:WPL}.
}
\label{fig:WPLinter}
\end{figure}

\begin{figure}[H]
\includegraphics[scale=0.75, angle=90]{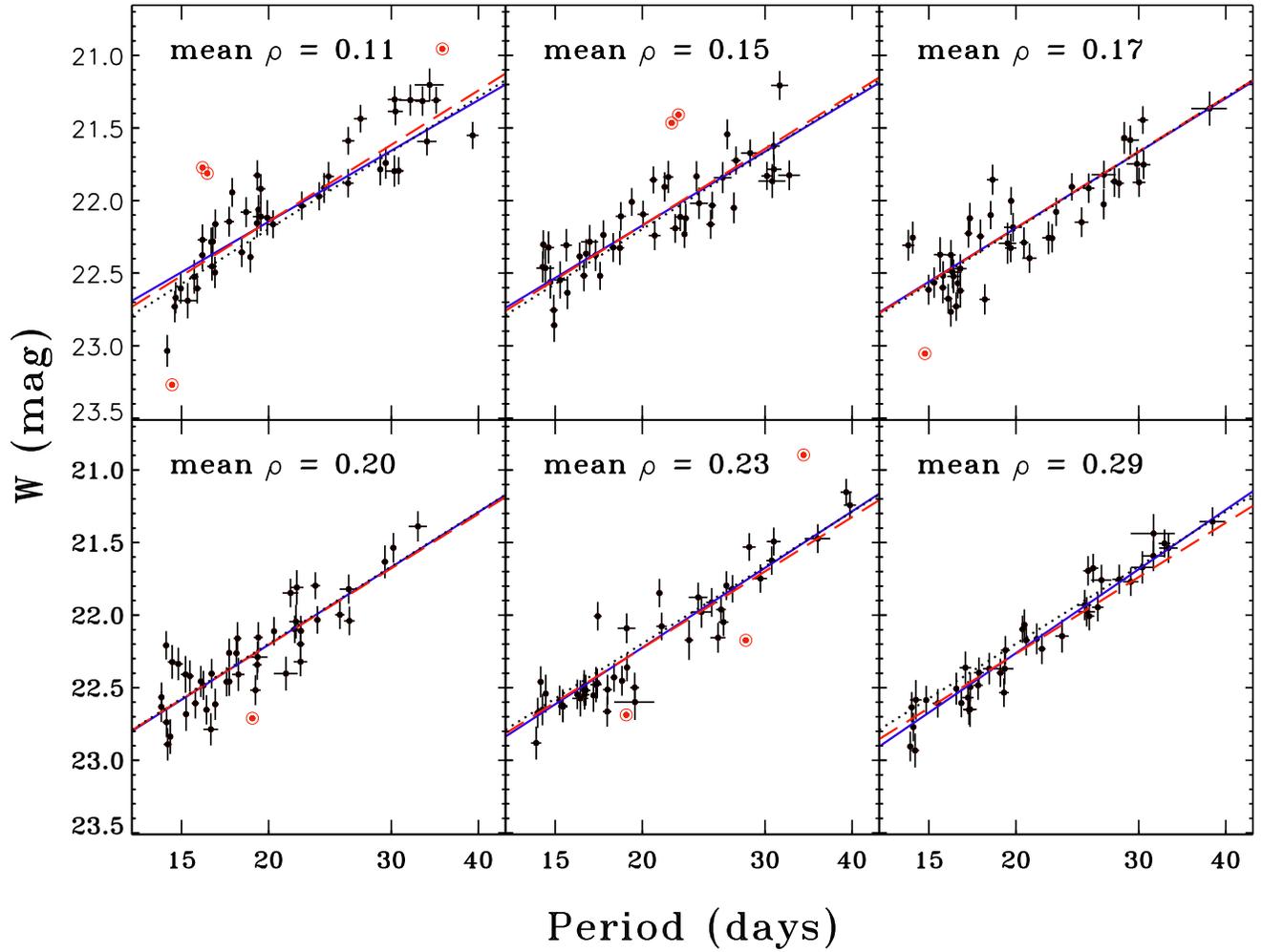}
\caption{
Same as Fig.~\ref{fig:WPLfits} but for the sample Cepheids with a maximum period cut of $P<$ \WPLPerLimitfourty{} days.  Fit parameters shown in Table~\ref{tab:WPLmodels}.   Figure discussed in \S\ref{sec:WPL} and \S\ref{sec:Disgusted}.
}
\label{fig:WPLfitsfourty}
\end{figure}

\begin{figure}[H]
\includegraphics[scale=0.75, angle=90]{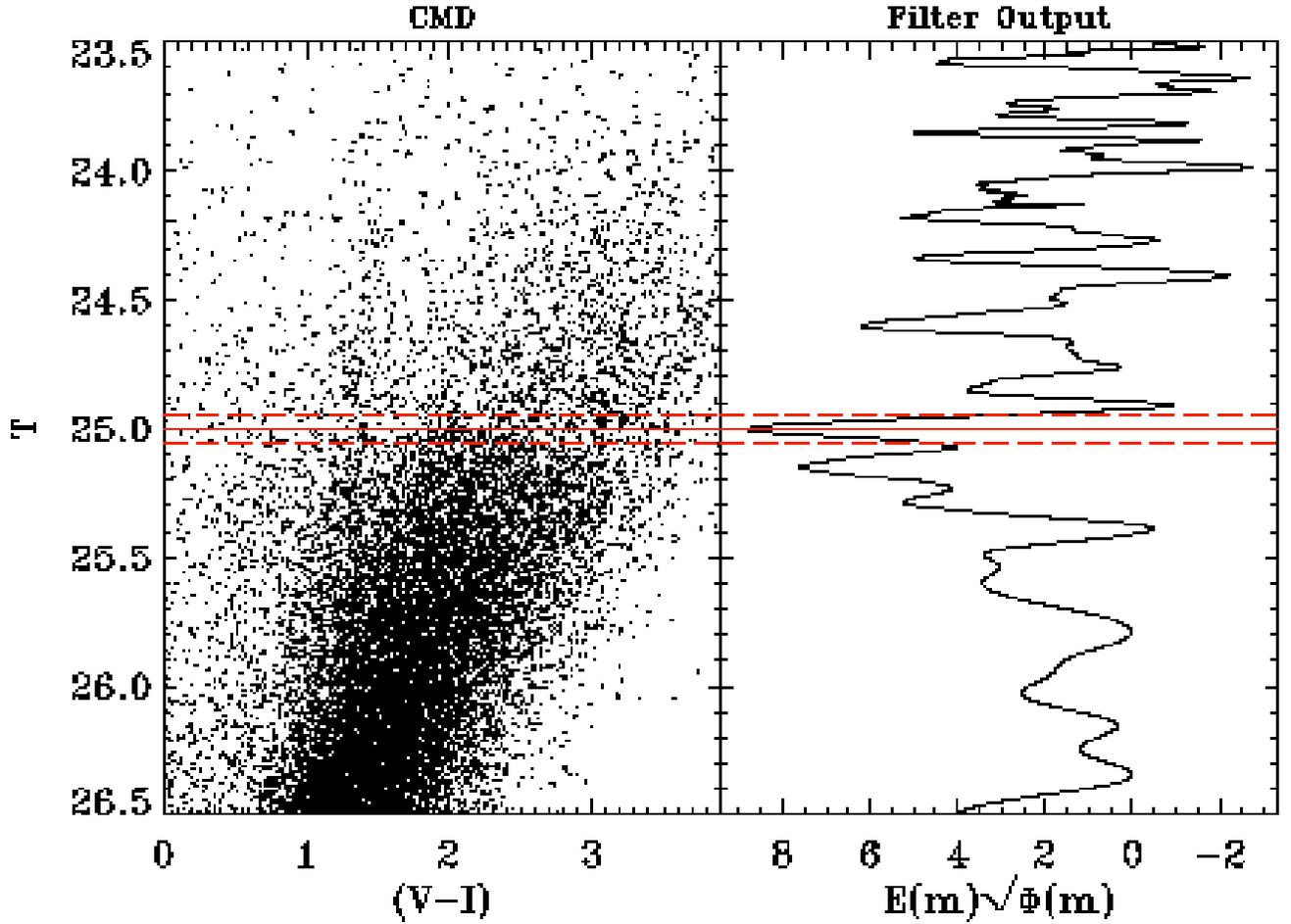}
\caption{
Determination of the \textit{T} TRGB magnitude for M101 where $T=I-0.20[(V-I)-1.5]$ \citep{madore09}.  Left: \textit{T} vs. $(V-I)$ color-magnitude diagram for sources \TRGBRadCut{} from the galactic center of M101.  Right: Edge detection function $E(m)\sqrt{\phi{}(m)}$, where the maximum value at \TRGBedge{}.  Solid red line: Represents the maximum value of the edge detection function.  Dashed red lines: Represent estimated uncertainty of the edge detection based on bootstrap re-sample  described in \S\ref{sec:TRGB}. 
}
\label{fig:TRGBD}
\end{figure}

\begin{figure}[H]
\includegraphics[scale=0.9]{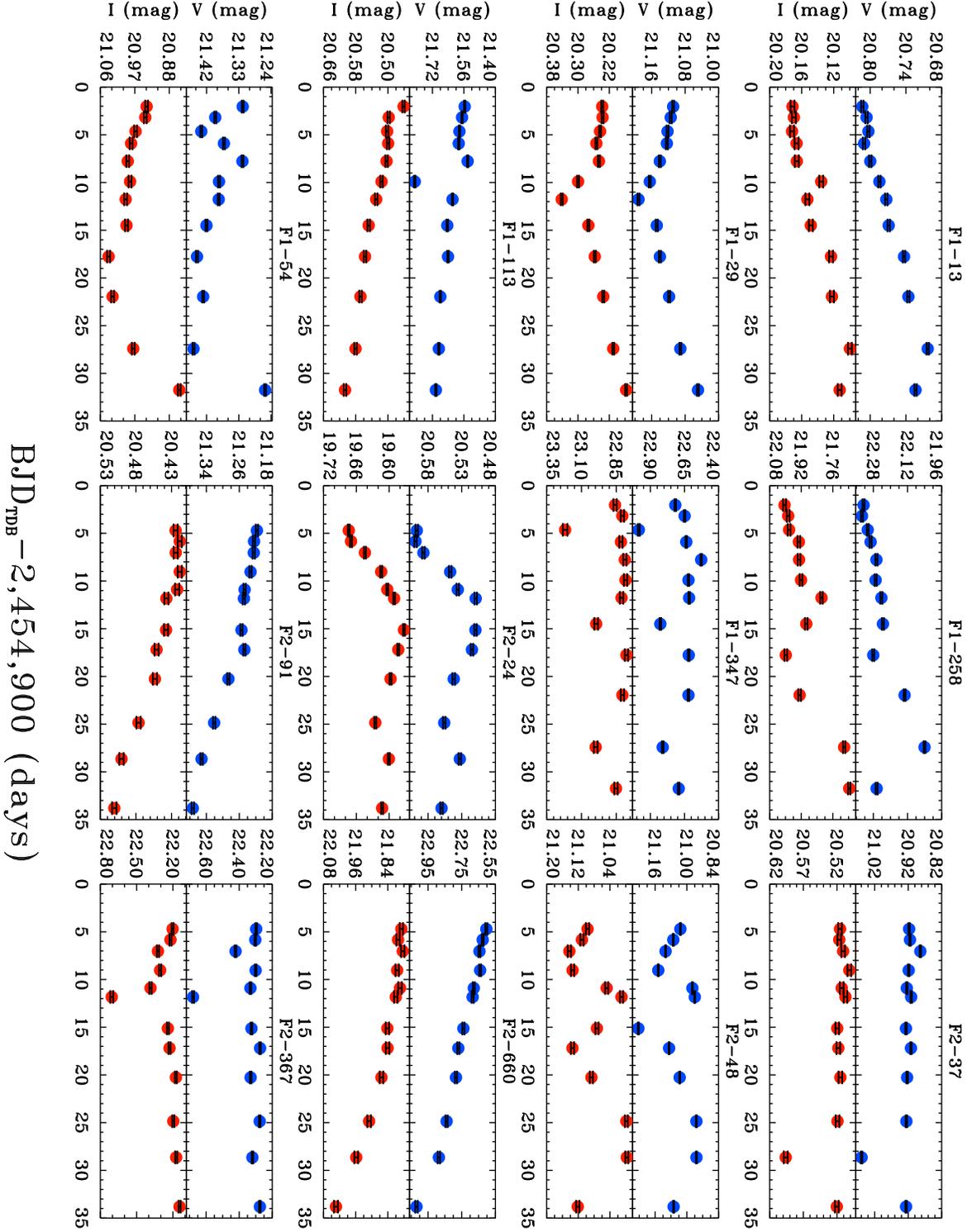}
\caption{
Light curves of interesting luminous variable objects.  Blue (top): \textit{V} filter light curves.  Red (bottom): \textit{I} filter light curves. Photometric errors are shown but usually smaller then the points.  Figure discussed in \S\ref{sec:VarObjs}.
}
\label{fig:VarObjone}
\end{figure}

\begin{figure}[H]
\includegraphics[scale=0.8]{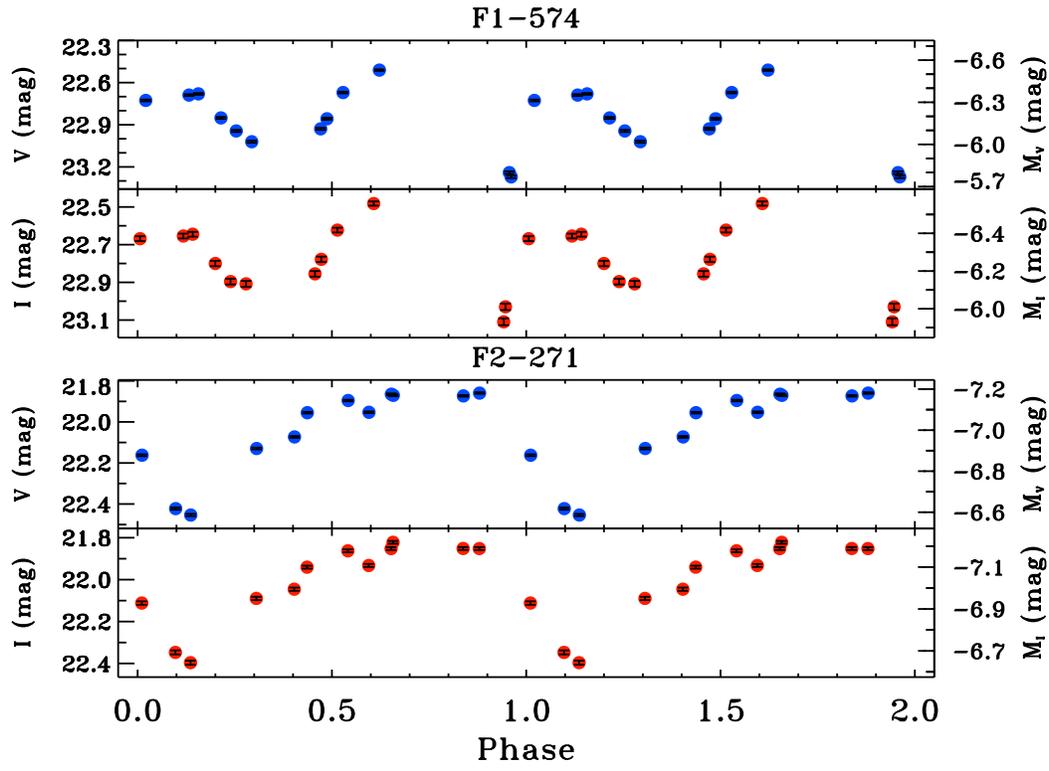}
\caption{
Object F1-574 (top) and F2-271 (bottom) which show possible periodic behavior with P = 0.508 days and P = 10.78 days respectively. Absolute magnitudes on the right side of the figure assume our Cepheid distance modulus and make no reddening correction. Figure discussed in \S\ref{sec:VarObjs}.
}
\label{fig:PerObjone}
\end{figure}

\begin{figure}[H]
\includegraphics[scale=0.75, angle=90]{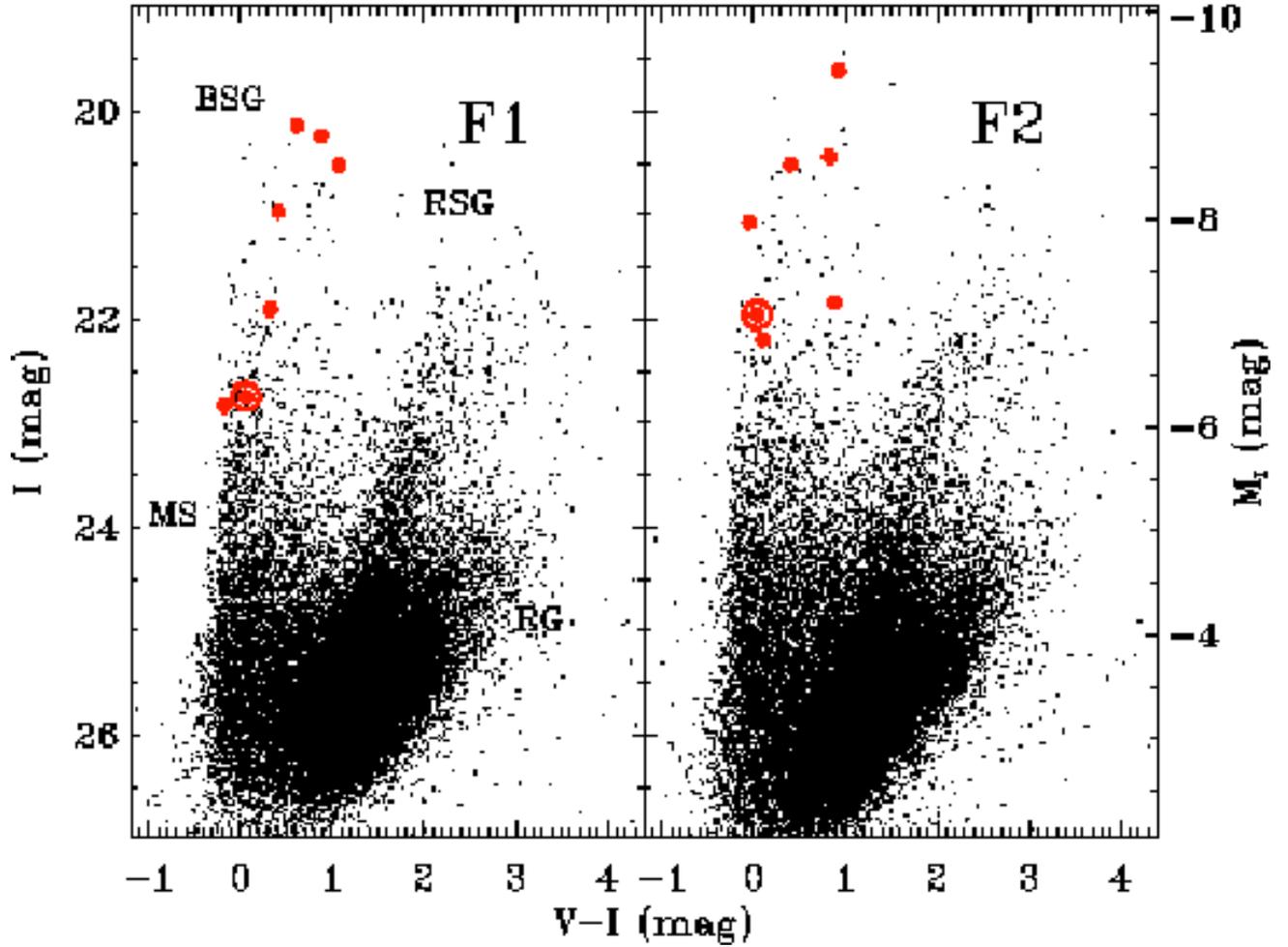}
\caption{
\textit{I} vs. $(V-I)$ color-magnitude diagram for M101 for Field 1 (left) and Field 2 (right) displaying every fifth source. Red circles represent luminous variable object seen in Fig.~\ref{fig:VarObjone} and Fig.~\ref{fig:PerObjone}.  Red annuli mark the possible periodic objects.  MS labels the location of the main sequence.  BSG labels the location of blue super giants. RSG labels the location of red super giants. RG labels the location of red giants.   Absolute magnitudes on the right side of the figure assume our Cepheid distance modulus and make no reddening correction.  Figure discussed in \S\ref{sec:VarObjs}.
}
\label{fig:VarCMD}
\end{figure}

Before creating the ISIS PSF from a DAOPHOT PSF we must first use IRAF to change the units and zero-point of the ACS/WFC drizzled images.  We must have run the modified ISIS interp.csh\footnote{http://www.astronomy.ohio-state.edu/$\sim$siverd/soft/is3/index.html}, the stock script ref.csh and the stock script subtract.csh to create the ISIS reference images and subtracted images.\footnote{The script detect.csh can also be run to create the stacked subtracted images seen in Fig.~\ref{fig:F2Sub}.  However, these images are not used to obtain the differential flux light curves in our analysis.}  Additionally, DAOPHOT must be run on each reference image, carefully selecting a sample of PSF stars.  Besides iteratively rejecting stars which differ from the DAOPHOT model PSF, we must also reject stars which have extended emission or objects with other stars nearby.  Rejecting stars which have other nearby stars can be done using the WATCH PROGRESS option in DAOPHOT. We now have a PSF from DAOPHOT which is free of contamination from non-point sources which we save as PSF.fits.  We also use standard procedures in DAOPHOT and ALLSTAR to create a catalog of source positions and photometry.  We remove non-stellar sources from the catalog using cuts on the catalog. We then use the source list (.als) from ALLSTAR as an object list for ISIS by using the following awk command and removing the first 3 lines:
\begin{verbatim}
awk '{printf"%9.3f %9.3f %5.0f %5.0f %13s %8.3f %6.3f\n",$2-1,$3-1,$2-1,
     $3-1,"lc"$1".data",$4-0.1,$5}' ref.als > ! phot.data
\end{verbatim}
where the $-1$ pixel shift in the awk command accounts for the difference in the coordinate conventions between DAOPHOT and ISIS\footnote{We summarize the different coordinate system conventions between astronomical software in Table~\ref{tab:Pixels}.}.  This object list should be the file associated with the VARIABLES parameter in the ISIS process\_config file.

We then need to manually make two files which ISIS usually creates and put them in the directory with the images to be processed by ISIS.  The first is an $n \times n$ fits image named 'psf\_file0.fits' filled with zeros, where n is specified by psf\_width in the ISIS phot\_config file. The second file must be named 'psf\_table' which will contain one line of the form:
\begin{verbatim}
psf_file0.fits     9     9  4211  4234     0     0,
\end{verbatim}
where the $1^{\textrm{st}}$ element is the PSF file name, the $2^{\textrm{nd}}$ and $3^{\textrm{rd}}$ are the center of the 'psf\_file0.fits' image in ISIS coordinates and the $4^{\textrm{th}}$ and $5^{\textrm{th}}$ are the x,y size of the reference images in pixels. We then use the addstar command in DAOPHOT to put a `fake' star with extremely high signal to noise to the center of psf\_file0.fits.  It is important to account difference in the pixel coordinate system convention of DAOPHOT when creating the PSF.  

Lastly, the stock phot.csh script must be altered so that it does not call the program Bphot, which would usually make the ISIS PSF. The modified phot.csh script can then be run, creating the light curves for every object in the object list created in DAOPHOT.  These light curves will be in differential flux units, but can be transformed to an instrumental magnitude using `fluxtomag' in the VARTOOLS\footnote{VARTOOLS is available from Joel Hartman at https://www.cfa.harvard.edu/$\sim$jhartman/vartools/.} package.


\bibliographystyle{apj}



\begin{deluxetable}{cccccc}
\tablecolumns{6}
\tablewidth{0pc}
\tabletypesize{\scriptsize}
\tablecaption{Log of Observations}
\tablehead{
\colhead{} &
\colhead{} &
\multicolumn{4}{c}{$(\textrm{BJD}_{\textrm{TDB}} - 2,450,000.0)^{a}$ at Mid-exposure} \\
\colhead{Visit Number} &
\colhead{UT Date} &
\colhead{I} &
\colhead{V} &
\colhead{I} &
\colhead{V} 
}
\startdata
          F1-01 &  2006 Dec 23 & 4092.03116 & 4092.03899 & 4092.04805 & 4092.05619  \\
          F1-02 &  2006 Dec 24 & 4093.16580 & 4093.17363 & 4093.18269 & 4093.19083  \\
          F1-03 &  2006 Dec 25 & 4094.62862 & 4094.63644 & 4094.64551 & 4094.65364  \\
          F1-04 &  2006 Dec 26 & 4095.89387 & 4095.90169 & 4095.91076 & 4095.91889  \\
          F1-05 &  2006 Dec 28 & 4097.75827 & 4097.76609 & 4097.77515 & 4097.78329  \\
          F1-06 &  2006 Dec 30 & 4099.88941 & 4099.89724 & 4099.90630 & 4099.91444  \\
          F1-07 &  2007 Jan 01 & 4101.75398 & 4101.76181 & 4101.77087 & 4101.77901  \\
          F1-08 &  2007 Jan 04 & 4104.48410 & 4104.49193 & 4104.50099 & 4104.50913  \\
          F1-09 &  2007 Jan 07 & 4107.74682 & 4107.75464 & 4107.76370 & 4107.77184  \\
          F1-10 &  2007 Jan 11 & 4111.94164 & 4111.94947 & 4111.95853 & 4111.96666  \\
          F1-11 &  2007 Jan 17 & 4117.40136 & 4117.40919 & 4117.41825 & 4117.42639  \\
          F1-12 &  2007 Jan 21 & 4121.72890 & 4121.73673 & 4121.74579 & 4121.75393  \\
          F2-01 &  2006 Dec 25 & 4094.69519 & 4094.70301 & 4094.71207 & 4094.72021  \\
          F2-02 &  2006 Dec 26 & 4095.82724 & 4095.83507 & 4095.84413 & 4095.85227  \\
          F2-03 &  2006 Dec 28 & 4097.02585 & 4097.03367 & 4097.04273 & 4097.05087  \\
          F2-04 &  2006 Dec 30 & 4099.02364 & 4099.03147 & 4099.04053 & 4099.04867  \\
          F2-05 &  2006 Dec 31 & 4100.88828 & 4100.89611 & 4100.90517 & 4100.91331  \\
          F2-06 &  2007 Jan 01 & 4101.82055 & 4101.82837 & 4101.83744 & 4101.84557  \\
          F2-07 &  2007 Jan 05 & 4105.09577 & 4105.10359 & 4105.15438 & 4105.16252  \\
          F2-08 &  2007 Jan 07 & 4107.16210 & 4107.16993 & 4107.21903 & 4107.22716  \\
          F2-09 &  2007 Jan 10 & 4110.22705 & 4110.23487 & 4110.27747 & 4110.28560  \\
          F2-10 &  2007 Jan 14 & 4114.82183 & 4114.82965 & 4114.87171 & 4114.87985  \\
          F2-11 &  2007 Jan 18 & 4118.61321 & 4118.62103 & 4118.66682 & 4118.67496  \\
          F2-12 &  2007 Jan 23 & 4123.79267 & 4123.80049 & 4123.80955 & 4123.81769  \\
\enddata
\tablecomments{ Table discussed in \S\ref{sec:observations}. \\
$^{a}$ See \citet{eastman10} for details.
}

\label{tab:observlog}
\end{deluxetable}

\begin{deluxetable}{lllrrrrrrr}
\tablecolumns{8}
\tablewidth{0pc}
\tabletypesize{\scriptsize}
\tablecaption{Secondary Standards}
\tablehead{
\colhead{} &
\colhead{R.A.} &
\colhead{Dec.} &
\colhead{X} &
\colhead{Y} &
\multicolumn{2}{c}{Magnitudes} &
\colhead{} \\
\colhead{ID} &
\colhead{(J2000)} &
\colhead{(J2000)} &
\colhead{pixels} &
\colhead{pixels} &
\colhead{\textit{V}} &
\colhead{\textit{I}} &
\colhead{$I_V$} 
}
\startdata
      F1-71 &  210.86852031 &   54.35418692 & 2544.1 & 2268.0 &  21.393(005) &  20.532(005) &$ -0.2$\\
      F1-58 &  210.85461416 &   54.39176894 &  281.2 & 3861.6 &  21.333(005) &  20.721(005) &$ -0.4$\\
     F1-162 &  210.89357513 &   54.36263801 & 1573.1 & 1538.3 &  22.011(007) &  20.872(006) &$ -0.3$\\
     F1-208 &  210.84008842 &   54.35383418 & 3033.3 & 3356.5 &  22.016(007) &  20.892(006) &$ -0.5$\\
     F1-137 &  210.82119974 &   54.34762558 & 3754.2 & 3911.8 &  21.821(006) &  21.154(007) &$ -0.8$\\
     F1-151 &  210.85330380 &   54.34252050 & 3566.6 & 2527.9 &  21.856(006) &  21.201(007) &$  0.9$\\
     F1-103 &  210.88537412 &   54.37791726 &  695.1 & 2284.8 &  21.649(005) &  21.735(009) &$  0.4$\\
     F1-185 &  210.85991217 &   54.36690151 & 1842.5 & 2958.0 &  21.931(006) &  21.226(007) &$  0.9$\\
     F1-601 &  210.90181610 &   54.35600312 & 1877.6 & 1033.3 &  22.701(006) &  20.896(006) &$  0.8$\\
     F1-181 &  210.87887687 &   54.37002640 & 1324.4 & 2313.8 &  21.992(006) &  21.477(008) &$ -0.4$\\
\multicolumn{8}{l}{\it Table to be published in its entirety in machine-readable form. }
\enddata

\tablecomments{R.A. and Dec. are in units of degrees.  Errors in mean magnitudes and semiamplitudes are in units of 0.001 mag. Table discussed in \S\ref{sec:dolphot}.}
\label{tab:stdstars}
\end{deluxetable}

\begin{deluxetable}{lrllrrrrrrr}
\tablecolumns{11}
\tablewidth{0pc}
\tabletypesize{\scriptsize}
\tablecaption{Cepheid Variables}
\tablehead{
\colhead{} &
\colhead{Period} &
\colhead{R.A.} &
\colhead{Dec.} &
\colhead{X} &
\colhead{Y} &
\multicolumn{2}{c}{Magnitudes} &
\multicolumn{2}{c}{Amp.} &
\colhead{} \\
\colhead{ID} &
\colhead{(days)} &
\colhead{(J2000)} &
\colhead{(J2000)} &
\colhead{pixels} &
\colhead{pixels} &
\colhead{\textit{V}} &
\colhead{\textit{I}} &
\colhead{\textit{V}} &
\colhead{\textit{I}} &
\colhead{$I_V$} 
}
\startdata
     F1-567 & 40.9345(1.0315) &  210.87019095 &   54.34667242 & 3014.6 & 1992.2 &  22.985(008) &  22.171(006) & 0.158 & 0.091 &  75.3 \\
     F1-773 & 75.3739(2.2386) &  210.89765450 &   54.37240300 &  859.1 & 1655.8 &  22.676(016) &  21.754(017) & 0.315 & 0.206 & 287.1 \\
    F1-1274 & 49.4143(2.3786) &  210.84511130 &   54.35727011 & 2723.2 & 3258.9 &  23.290(019) &  22.413(009) & 0.505 & 0.261 & 342.5 \\
    F1-1552 & 42.6833(3.6532) &  210.85128425 &   54.38254091 &  947.1 & 3730.9 &  23.412(044) &  22.512(026) & 0.519 & 0.310 & 306.6 \\
    F1-1619 & 30.8813(0.7128) &  210.88234973 &   54.36916648 & 1324.6 & 2155.5 &  23.532(014) &  22.700(014) & 0.503 & 0.345 & 668.9 \\
    F1-1721 & 27.2601(0.4105) &  210.86452566 &   54.35514022 & 2546.3 & 2449.2 &  23.626(016) &  22.849(014) & 0.477 & 0.278 & 275.8 \\
    F1-1767 & 21.7926(0.3649) &  210.86424080 &   54.34258320 & 3383.2 & 2107.0 &  23.669(021) &  22.922(024) & 0.335 & 0.202 & 129.5 \\
    F1-1890 & 30.4055(0.7575) &  210.83646784 &   54.36458112 & 2380.2 & 3798.2 &  23.699(015) &  22.755(013) & 0.479 & 0.269 & 501.1 \\
    F1-1975 & 30.4193(0.5237) &  210.86646959 &   54.34309782 & 3312.5 & 2035.4 &  23.728(016) &  22.797(012) & 0.501 & 0.266 & 241.0 \\
    F1-2045 & 22.8373(0.1005) &  210.87188971 &   54.34614160 & 3021.9 & 1911.6 &  23.776(013) &  23.083(015) & 0.446 & 0.302 & 236.3 \\
\multicolumn{11}{l}{\it Table to be published in its entirety in machine-readable form. }
\enddata

\tablecomments{R.A. and Dec. are in units of degrees.  \Astrometry Errors in mean magnitudes and semiamplitudes are in units of 0.001 mag. Table discussed in \S\ref{sec:FTest}.}
\label{tab:cephbasic}
\end{deluxetable}

\begin{deluxetable}{lrllrrrrrrrr}
\tablecolumns{12}
\tablewidth{0pc}
\tabletypesize{\scriptsize}
\tablecaption{Rejected Cepheid Candidates}
\tablehead{
\colhead{} &
\colhead{Period} &
\colhead{R.A.} &
\colhead{Dec.} &
\colhead{X} &
\colhead{Y} &
\multicolumn{2}{c}{Magnitudes} &
\multicolumn{2}{c}{Amp.} &
\colhead{} &
\colhead{} \\
\colhead{ID} &
\colhead{(days)} &
\colhead{(J2000)} &
\colhead{(J2000)} &
\colhead{pixels} &
\colhead{pixels} &
\colhead{\textit{V}} &
\colhead{\textit{I}} &
\colhead{\textit{V}} &
\colhead{\textit{I}} &
\colhead{$I_V$} & 
\colhead{$Reason^{a}$} 
}
\startdata
     F1-381 & 146.449 &  210.86070136 &   54.35415151 & 2674.5 & 2569.1 &  22.525(007) &  22.265(011) & 0.136 & 0.140 &  98.5 & A \\
     F1-757 & 166.814 &  210.84322037 &   54.38390199 &  988.9 & 4080.5 &  22.968(010) &  22.648(014) & 0.281 & 0.226 &  62.6 & A \\
    F1-1005 &  76.313 &  210.87279394 &   54.36153133 & 1987.2 & 2309.6 &  22.883(015) &  21.845(010) & 0.336 & 0.260 & 422.4 & A \\
    F1-1357 &   5.986 &  210.86954527 &   54.33204480 & 3994.6 & 1605.6 &  23.518(009) &  23.128(014) & 0.120 & 0.174 &  40.6 & A \\
    F1-1378 & 168.978 &  210.88086502 &   54.33618592 & 3534.5 & 1284.7 &  22.867(057) &  22.828(070) & 1.279 & 0.295 &  19.2 & A \\
    F1-1576 & 148.915 &  210.88486096 &   54.35099847 & 2487.3 & 1547.2 &  23.457(008) &  23.252(016) & 0.139 & 0.145 &  21.0 & A \\
    F1-2754 &  13.727 &  210.88137081 &   54.37066234 & 1241.5 & 2235.4 &  24.022(013) &  23.661(019) & 0.249 & 0.212 &  62.8 & A \\
    F1-2812 &  34.901 &  210.86097991 &   54.37769971 & 1109.2 & 3220.5 &  23.966(013) &  22.849(009) & 0.314 & 0.278 & 135.8 & A \\
    F1-2883 &  36.574 &  210.87213007 &   54.36013865 & 2090.4 & 2296.1 &  23.930(011) &  23.003(012) & 0.374 & 0.303 & 236.1 & A \\
    F1-4020 &  19.565 &  210.85013743 &   54.37919292 & 1187.8 & 3681.1 &  24.339(016) &  23.475(017) & 0.190 & 0.152 &  25.2 & A \\
\multicolumn{11}{l}{\it Table to be published in its entirety in machine-readable form. }
\enddata

\tablecomments{R.A. and Dec. are in units of degrees.  \Astrometry Errors in mean magnitudes are in units of 0.001 mag. Table discussed in \S\ref{sec:FTest}.\\
$^{a}$ Reasons for the Cepheid canidates rejection; \textbf{A:} Amplitude ratio ratio cut. \textbf{B:} Blue blends.  \textbf{C:} Large Extinction.  \textbf{D:} Population II Cepheids. \textbf{E:} Candidates removed by period cut. \textbf{F:} Candidates removed by iterative sigma clipping.
}
\label{tab:cephreject}
\end{deluxetable}

\begin{deluxetable}{lrr}
\tablecolumns{3}
\tablewidth{0pc}
\tabletypesize{\scriptsize}
\tablecaption{Sample Selection}
\tablehead{
\colhead{ } &
\multicolumn{2}{c}{Field} \\
\colhead{Sample Cuts} &
\colhead{F1} &
\colhead{F2} 
}
\startdata
                          Initial Cepheid Candidates        \dotfill{} &$   632$&$   595$ \\
                          1. Amplitude Ratio                \dotfill{} &$  -139$&$  -121$ \\
                         2. $E(V-I)<(2\sigma{}$ foreground) \dotfill{} &$   -29$&$   -19$ \\
                          3. $E(V-I)>0.5$ mag               \dotfill{} &$   -24$&$   -11$ \\
                        4. $\Delta{}\mu{}_{0} > 11.5$ mag   \dotfill{} &$    -6$&$    -7$ \\
                         5. Period $< 3$ days               \dotfill{} &$    -2$&$    -4$ \\
                          6. $\sigma{}$ clipping            \dotfill{} &$   -21$&$   -17$ \\
                          Final Sample                      \dotfill{} &$   411$&$   416$ \\
\enddata

\tablecomments{Effects of sample selection described in \S\ref{sec:sampleSel}.}
\label{tab:samplecuts}
\end{deluxetable}

\begin{deluxetable}{lrcrrrrc}
\tablecolumns{8}
\tablewidth{0pc}
\tabletypesize{\scriptsize}
\tablecaption{Cepheid Variables: Calculated Quantities}
\tablehead{
\colhead{} &
\colhead{Period} &
\colhead{F-Test} &
\colhead{$\LMCreldist{}$} &
\colhead{$E(V-I)$} &
\colhead{$E(B-V)$} &
\colhead{} &
\colhead{$12+\log{(\textrm{O}/\textrm{H})}$} \\
\colhead{ID} &
\colhead{(days)} &
\colhead{\chisq{} ratio} &
\colhead{(mag)} &
\colhead{(mag)} &
\colhead{(mag)} &
\colhead{$\rho{}/\rho{}_{iso}$} &
\colhead{(dex)}
}
\startdata
     F1-567 & 40.934 &  19.381 &  10.350(074) &  0.020(020) &  0.015(014) & 0.1696 & 8.60  \\
     F1-773 & 75.374 &  28.559 &  10.644(099) &  0.075(074) &  0.054(054) & 0.2551 & 8.52  \\
    F1-1274 & 49.414 &  82.090 &  10.769(101) &  0.066(065) &  0.048(047) & 0.1131 & 8.65  \\
    F1-1552 & 42.683 &  35.455 &  10.626(164) &  0.102(102) &  0.074(074) & 0.1897 & 8.58  \\
    F1-1619 & 30.881 &  59.668 &  10.453(077) &  0.062(062) &  0.045(045) & 0.2160 & 8.56  \\
    F1-1721 & 27.260 &  44.162 &  10.507(072) &  0.017(017) &  0.013(012) & 0.1573 & 8.61  \\
    F1-1767 & 21.793 &  28.944 &  10.303(088) &  0.008(007) &  0.006(005) & 0.1569 & 8.61  \\
    F1-1890 & 30.405 &  44.678 &  10.324(077) &  0.175(175) &  0.128(127) & 0.1097 & 8.65  \\
    F1-1975 & 30.419 &  45.180 &  10.384(072) &  0.163(163) &  0.119(118) & 0.1620 & 8.60  \\
    F1-2045 & 22.837 &  38.497 &  10.610(068) & -0.051(163) & -0.037(118) & 0.1740 & 8.59  \\
\multicolumn{8}{l}{\it Table to be published in its entirety in machine-readable form.}
\enddata

\tablecomments{Errors in the mean colors and the LMC ralative distance moduli are in units of 0.001 mag. Table discussed in \S\ref{sec:FTest}.}
\label{tab:cephprop}
\end{deluxetable}

\tabletypesize{\normalsize}
\tablewidth{0pt}
\begin{deluxetable}{lccc}
\tablecaption{Error Budget of the Cepheid Distance Scale}
\tablehead{\multicolumn{1}{l}{Error source} & \colhead{LMC relative $\mu_{0}$} & \colhead{This Work M101 $\mu_{0}$} & \colhead{Goal$^{\textrm{a}}$ M101 $\mu_{0}$}}
\startdata
{\bf A. Fiducial galaxy}   & \nodata & {\bf NGC 4258} & {\bf NGC 4258}\\
$ \ \ \  $S1a. Distance modulus (sys) \dotfill{} & \nodata & 0.12 & \nodata   \\
$ \ \ \  $S1b. Distance modulus (ran) \dotfill{} & \nodata  & 0.09 & \nodata   \\
$ \ \ \  $S1. Distance modulus (tot) \dotfill{} & \nodata  & 0.15  & 0.07  \\
{\bf B. Photometric calibration}  &              \\
$ \ \ \ \ \  $S2a. $V$ zeropoint     \dotfill{}    & 0.02 & 0.02 & \nodata  \\
$ \ \ \ \ \  $S2b. $I$ zeropoint      \dotfill{}   & 0.02 & 0.02 & \nodata  \\
$ \ \ \  $S2. Photometry (sys)     \dotfill{}  & 0.05 & 0.05 & \nodata  \\
$ \ \ \  $S3. NGC 4258$^{\textrm{b}}$ Photometry (sys)     \dotfill{}  & \nodata & 0.05 & \nodata  \\
$ \ \ \  $R1. Photometry (ran)   \dotfill{}    & 0.03 & 0.03  & 0.03   \\
$ \ \ \  $R2. NGC 4258$^{\textrm{b}}$ Photometry (ran) \dotfill{}      & \nodata & 0.03  & 0.03  \\
{\bf C. Extinction corrections}   &             \\
$ \ \ \  $R3. Uncertainty in $R_V$ \dotfill{}  & 0.02 & 0.02  & 0.02   \\
$ \ \ \  $R4. NGC 4258$^{\textrm{b}}$ Uncertainty in $R_V$ \dotfill{}  & \nodata & 0.02  & 0.02   \\
{\bf D. Corrections to LMC Metallicity}  &              \\
$ \ \ \  $S4. Adopted correction   \dotfill{}  & \MetDistErrSys{} & \MetDistErrSys{} & \MetDistErrSys{}  \\
$ \ \ \  $S5. NGC 4258$^{\textrm{b}}$ Adopted correction \dotfill{}    & \nodata & 0.04 & \MetDistErrSys{}    \\
$ \ \ \  $R5. Zero-point \dotfill{}   & \MetDistErrRan{} & \MetDistErrRan{} & \MetDistErrRan{}    \\
$ \ \ \  $R6. NGC 4258$^{\textrm{b}}$ Zero-point \dotfill{}   & \nodata & \nodata & \MetDistErrRan{}    \\
\tableline
$ \ \ \  $R$_T$. Total random    \dotfill{}    & \LMCranErr{} & \MranErr{}  & \MGOALranErr{}   \\
$ \ \ \  $S$_T$. Total systematic  \dotfill{}  & \LMCsysErr{} & \MsysErr{}  & \MGOALsysErr{}  \\
\tableline
{\bf Combined error (mag)} \dotfill{} & {\bf \LMCtotErr{}} & {\bf \MtotErr{}} & {\bf \MGOALtotErr{}}  \\
{\bf Combined error (\%) } \dotfill{}& \ \ {\bf \LMCpercentErr{}} & \ \ \ \ {\bf \MpercentErr}  & \ \ \ \ {\bf \MGOALpercentErr}   \\
\enddata
\label{tab:err}
\tablecomments{All errors expressed in magnitudes unless otherwise indicated.  Table discussed in \S\ref{sec:FinDistance} and \S\ref{sec:Disgusted}. \\ 
$^{\textrm{a}}$ Goal errors are calculated using the anticipated improvement in the maser distance determination to NGC 4258 \citep{humphreys08}. \\
$^{\textrm{b}}$ This work uses the \citet{macri06} Cepheid NGC 4258 distance.  Goal work errors are based on using the same analysis in this paper on the \citet{macri06} HST/ACS Cepheid observations.
}
\end{deluxetable}
\begin{deluxetable}{lclllc}
\tablecolumns{6}
\tablewidth{0pc}
\tabletypesize{\scriptsize}
\tablecaption{Recent Distance Determinations to M101}
\tablehead{
\colhead{Study} &
\colhead{$\textrm{Method}^{a}$} &
\colhead{LMC $\mu_{0}$} &
\colhead{M101 $\mu_{0}$ } &
\colhead{LMC-relative $\Delta\mu_{0}^{b}$ } &
\colhead{Outer/Inner Field}  \\
\colhead{ } &
\colhead{ } &
\colhead{mag } &
\colhead{mag } &
\colhead{mag } &
\colhead{ } 
}
\startdata

This work                 \dotfill{}  &    Cepheids   &       \MacriLMCdist$  \pm $\MacriLMCdistRan$_{r}  \pm $\MacriLMCdistSys$_{s}$             &   \MetDist$ \pm  $\MranErr$_{r}  \pm $\MsysErr$_{s}$               &  \MetLMCDist$ \ \pm{} \ $\MetDistErrRan$_{r} \ \pm{} $\MetDistErrSys$_{s}$    & inner             \\
This work                 \dotfill{}  &    TRGB       &       \nodata{}             &   \TRGBDistMod$ \pm  $\TRGBBootSig$_{r}  \pm $\TRGBsys$_{s}$          &  \nodata{}           & inner             \\
\citet{saha06}            \dotfill{}  &    Cepheids   &       $18.54^{c}$           &       $29.16\pm{}0.04$    &  $10.62\pm{}0.04$    & inner             \\
\citet{saha06}            \dotfill{}  &    Cepheids   &       $18.54^{c}$           &       $29.18\pm{}0.08$    &  $10.64\pm{}0.08$    & outer             \\
\citet{freedman01}        \dotfill{}  &    Cepheids   &       $18.50$      &       $29.13\pm{}0.11$    &  $10.63\pm{}0.11$    & outer        \\
\citet{macri01}           \dotfill{}  &    Cepheids   &       $18.50$      &       $29.19\pm{}0.08$    &  $10.69\pm{}0.08$    & inner             \\
\citet{macri01}           \dotfill{}  &    Cepheids   &       $18.50$      &       $29.53\pm{}0.03$    &  $11.03\pm{}0.03$    & outer             \\
\citet{stetson98}         \dotfill{}  &    Cepheids   &       $18.50$      &       $29.21\pm{}0.17$    &  $10.71\pm{}0.17$    & inner             \\
\citet{kennicutt98b}      \dotfill{}  &    Cepheids   &       $18.50$      &       $29.21\pm{}0.09$    &  $10.71\pm{}0.09$    & inner             \\
\citet{kennicutt98b}      \dotfill{}  &    Cepheids   &       $18.50$      &       $29.34\pm{}0.08$    &  $10.84\pm{}0.08$    & outer             \\
\citet{kelson96}          \dotfill{}  &    Cepheids   &       $18.50$               &       $29.34\pm{}0.17$    &  $10.84\pm{}0.17$    & outer             \\
\citet{rizzi07}           \dotfill{}  &       TRGB    &       \nodata{}             &       $29.34\pm{}0.09$    &  \nodata{}           & \nodata{}         \\
\citet{sakai04}           \dotfill{}  &       TRGB    &       \nodata{}             &       $29.42\pm{}0.02$    &  \nodata{}           & \nodata{}         \\
\citet{jurcevic00}        \dotfill{}  &     RSV       &       $18.50\pm{}0.10$      &       $29.40\pm{}0.16$    &  $10.90\pm{}0.16$    & \nodata{}         \\
\citet{feldmeier96}       \dotfill{}  &     PNLF      &       \nodata{}             &       $29.42\pm{}0.15$    &  \nodata{}           & \nodata{}         \\

\enddata

\tablecomments{ Table discussed in \S\ref{sec:FinDistance}, \S\ref{sec:TRGB} and \S\ref{sec:Disgusted}.\\
$^{a}$ TRGB: tip of the red giant branch; RSV:red supergiant variable stars; PNLF: planetary nebula luminosity function. \\
$^{b}$ We simply subtract each measurements adopted LMC Distance Modulus from the M101 Distance Modulus to obtain the LMC-relative Distance Modulus for rough comparisons with our measurements. \\
$^{c}$ Zero point also assumes a distance modulus to Pleiads is 5.61 mag and relays on the Baade-Becker-Wesselink distances to 32 Cepheids in \citet{fouque03} and \citet{barnes03}. \\
}

\label{tab:CompDist}
\end{deluxetable}

\begin{deluxetable}{lclllcc}
\tablecolumns{6}
\tablewidth{0pc}
\tabletypesize{\scriptsize}
\tablecaption{Model Wesenheit P-L Fit Parameters}
\tablehead{
\colhead{Model} &
\colhead{$a_{w}^{S}$} &
\colhead{$a_{w}^{I}$} &
\colhead{$b_{w}^{S}$} &
\colhead{$b_{w}^{I}$} &
\colhead{\chisq}  &
\colhead{DoF}  \\
\colhead{ } &
\colhead{$\textrm{mag} \ \log{(\textrm{day})}^{-1}$} &
\colhead{$\textrm{mag} \ \log{(\textrm{day})}^{-1}$} &
\colhead{$\textrm{mag}$} &
\colhead{$\textrm{mag}$} &
\colhead{ } &
\colhead{ }
}
\startdata

Model 1     &    \nodata{}   &  $\WPLawinterAlone ^{+ \WPLawinterUPAlone }_{- \WPLawinterDOWNAlone }$    &  \nodata{}    &   $\WPLbwinterAlone ^{+ \WPLbwinterUPAlone }_{- \WPLbwinterDOWNAlone }$     &  $\WPLAloneChi $ & $\WPLAloneDoF $ \\ 
Model 2     &    \nodata{}   &  $\WPLawinterFixSlope ^{+ \WPLawinterUPFixSlope }_{- \WPLawinterDOWNFixSlope }$     &    $\WPLbwslopeFixSlope ^{+ \WPLbwslopeUPFixSlope }_{- \WPLbwslopeDOWNFixSlope }$  &    $\WPLbwinterFixSlope ^{+ \WPLbwinterUPFixSlope }_{- \WPLbwinterDOWNFixSlope }$    &  $\WPLFixSlopeChi$&  $\WPLFixSlopeDoF$  \\ 
Model 3     &    $\WPLawslopeFixInter ^{+ \WPLawslopeUPFixInter }_{- \WPLawslopeDOWNFixInter }$   &  $\WPLawinterFixInter ^{+ \WPLawinterUPFixInter }_{- \WPLawinterDOWNFixInter }$      &  \nodata{}    &      $\WPLbwinterFixInter ^{+ \WPLbwinterUPFixInter }_{- \WPLbwinterDOWNFixInter }$ &  $\WPLFixInterChi$&  $\WPLFixInterDoF$  \\ 
Model 4     &    $\WPLawslope ^{+ \WPLawslopeUP }_{- \WPLawslopeDOWN }$   &    $\WPLawinter ^{+ \WPLawinterUP }_{- \WPLawinterDOWN }$   &  $\WPLbwslope ^{+ \WPLbwslopeUP }_{- \WPLbwslopeDOWN }$   &    $\WPLbwinter ^{+ \WPLbwinterUP }_{- \WPLbwinterDOWN }$   &  $\WPLFitChi$  &  $\WPLFitDoF$ \\ 

Restricted Model 1     &    \nodata{}   &  $\WPLawinterAlonefourty ^{+ \WPLawinterUPAlonefourty }_{- \WPLawinterDOWNAlonefourty }$    &  \nodata{}    &   $\WPLbwinterAlonefourty ^{+ \WPLbwinterUPAlonefourty }_{- \WPLbwinterDOWNAlonefourty }$     & $ \WPLAloneChifourty$ & $ \WPLAloneDoFfourty$ \\ 
Restricted Model 2     &    \nodata{}   &  $\WPLawinterFixSlopefourty ^{+ \WPLawinterUPFixSlopefourty }_{- \WPLawinterDOWNFixSlopefourty }$     &    $\WPLbwslopeFixSlopefourty ^{+ \WPLbwslopeUPFixSlopefourty }_{- \WPLbwslopeDOWNFixSlopefourty }$  &    $\WPLbwinterFixSlopefourty ^{+ \WPLbwinterUPFixSlopefourty }_{- \WPLbwinterDOWNFixSlopefourty }$    &  $\WPLFixSlopeChifourty $ &  $\WPLFixSlopeDoFfourty $ \\ 
Restricted Model 3     &    $\WPLawslopeFixInterfourty ^{+ \WPLawslopeUPFixInterfourty }_{- \WPLawslopeDOWNFixInterfourty }$   &  $\WPLawinterFixInterfourty ^{+ \WPLawinterUPFixInterfourty }_{- \WPLawinterDOWNFixInterfourty }$      &  \nodata{}    &      $\WPLbwinterFixInterfourty ^{+ \WPLbwinterUPFixInterfourty }_{- \WPLbwinterDOWNFixInterfourty }$ & $ \WPLFixInterChifourty $ & $ \WPLFixInterDoFfourty $\\ 
Restricted Model 4     &    $\WPLawslopefourty ^{+ \WPLawslopeUPfourty }_{- \WPLawslopeDOWNfourty }$   &    $\WPLawinterfourty ^{+ \WPLawinterUPfourty }_{- \WPLawinterDOWNfourty }$   &  $\WPLbwslopefourty ^{+ \WPLbwslopeUPfourty }_{- \WPLbwslopeDOWNfourty }$   &    $\WPLbwinterfourty ^{+ \WPLbwinterUPfourty }_{- \WPLbwinterDOWNfourty }$   &  $\WPLFitChifourty$ &  $\WPLFitDoFfourty$ 

\enddata
\tablecomments{ \\
Restricted Models have a $\WPLPerLimitfourty{}$ day maximum period cut imposed.  \WPLConIntfourty$\%$ uncertainties and confidence intervals are presented. Table discussed in \S\ref{sec:WPL} and \S\ref{sec:Disgusted}. \\
Model 1: constant slope and intercept \\
Model 2: linearly varying intercept with a constant slope \\
Model 3: linearly varying slope with a constant intercept \\
Model 4: linearly varying slope and intercept
}
\label{tab:WPLmodels}
\end{deluxetable}

\begin{deluxetable}{lclll}
\tablecolumns{6}
\tablewidth{0pc}
\tabletypesize{\scriptsize}
\tablecaption{Model Wesenheit P-L Fit Parameters as a function of [O/H]}
\tablehead{
\colhead{Model} &
\colhead{$a_{w}^{S}$} &
\colhead{$a_{w}^{I}$} &
\colhead{$b_{w}^{S}$} &
\colhead{$b_{w}^{I}$} \\
\colhead{ } &
\colhead{$\textrm{mag} \ \log{(\textrm{day})}^{-1} \ \textrm{dex}^{-1} $ }&
\colhead{$\textrm{mag} \ \log{(\textrm{day})}^{-1}$} &
\colhead{$\textrm{mag} \ \textrm{dex}^{-1}$ } &
\colhead{$\textrm{mag}$} 
}
\startdata

Model 1     &    \nodata{}   &  $\WPLawMETinterAlone ^{+ \WPLawMETinterUPAlone }_{- \WPLawMETinterDOWNAlone } $    &  \nodata{}    &   $\WPLbwMETinterAlone ^{+ \WPLbwMETinterUPAlone }_{- \WPLbwMETinterDOWNAlone }$     \\ 
Model 2     &    \nodata{}   &  $\WPLawMETinterFixSlope ^{+ \WPLawMETinterUPFixSlope }_{- \WPLawMETinterDOWNFixSlope }$     &    $\WPLbwMETslopeFixSlope ^{+ \WPLbwMETslopeUPFixSlope }_{- \WPLbwMETslopeDOWNFixSlope }\pm \WPLbwMETslopeerrSysFixSlope_{s}$  &    $\WPLbwMETinterFixSlope ^{+ \WPLbwMETinterUPFixSlope }_{- \WPLbwMETinterDOWNFixSlope }\pm \WPLbwMETintererrSysFixSlope_{s}$    \\ 
Model 3     &    $\WPLawMETslopeFixInter ^{+ \WPLawMETslopeUPFixInter }_{- \WPLawMETslopeDOWNFixInter }\pm \WPLawMETslopeerrSysFixInter _{s}$   &  $\WPLawMETinterFixInter ^{+ \WPLawMETinterUPFixInter }_{- \WPLawMETinterDOWNFixInter }\pm \WPLawMETintererrSysFixInter _{s}$      &  \nodata{}    &      $\WPLbwMETinterFixInter ^{+ \WPLbwMETinterUPFixInter }_{- \WPLbwMETinterDOWNFixInter }$  \\ 
Model 4     &    $\WPLawMETslope ^{+ \WPLawMETslopeUP }_{- \WPLawMETslopeDOWN }\pm \WPLawMETslopeerrSys _{s}$   &    $\WPLawMETinter ^{+ \WPLawMETinterUP }_{- \WPLawMETinterDOWN }\pm \WPLawMETintererrSys _{s}$   &  $\WPLbwMETslope ^{+ \WPLbwMETslopeUP }_{- \WPLbwMETslopeDOWN }\pm \WPLbwMETslopeerrSys _{s}$   &    $\WPLbwMETinter ^{+ \WPLbwMETinterUP }_{- \WPLbwMETinterDOWN }\pm \WPLbwMETintererrSys _{s}$  \\ 

Restricted Model 1     &    \nodata{}   &  $\WPLawMETinterAlonefourty ^{+ \WPLawMETinterUPAlonefourty }_{- \WPLawMETinterDOWNAlonefourty }$    &  \nodata{}    &   $\WPLbwMETinterAlonefourty ^{+ \WPLbwMETinterUPAlonefourty }_{- \WPLbwMETinterDOWNAlonefourty }$      \\ 
Restricted Model 2     &    \nodata{}   &  $\WPLawMETinterFixSlopefourty ^{+ \WPLawMETinterUPFixSlopefourty }_{- \WPLawMETinterDOWNFixSlopefourty }$     &    $\WPLbwMETslopeFixSlopefourty ^{+ \WPLbwMETslopeUPFixSlopefourty }_{- \WPLbwMETslopeDOWNFixSlopefourty } \pm \WPLbwMETslopeerrSysFixSlopefourty_{s}$  &    $\WPLbwMETinterFixSlopefourty ^{+ \WPLbwMETinterUPFixSlopefourty }_{- \WPLbwMETinterDOWNFixSlopefourty }\pm \WPLbwMETintererrSysFixSlopefourty_{s}$     \\ 
Restricted Model 3     &    $\WPLawMETslopeFixInterfourty ^{+ \WPLawMETslopeUPFixInterfourty }_{- \WPLawMETslopeDOWNFixInterfourty }\pm \WPLawMETslopeerrSysFixInterfourty _{s}$   &  $\WPLawMETinterFixInterfourty ^{+ \WPLawMETinterUPFixInterfourty }_{- \WPLawMETinterDOWNFixInterfourty }\pm \WPLawMETintererrSysFixInter _{s}$      &  \nodata{}    &      $\WPLbwMETinterFixInterfourty ^{+ \WPLbwMETinterUPFixInterfourty }_{- \WPLbwMETinterDOWNFixInterfourty }$  \\ 
Restricted Model 4     &    $\WPLawMETslopefourty ^{+ \WPLawMETslopeUPfourty }_{- \WPLawMETslopeDOWNfourty }\pm \WPLawMETslopeerrSysfourty _{s}$   &    $\WPLawMETinterfourty ^{+ \WPLawMETinterUPfourty }_{- \WPLawMETinterDOWNfourty }\pm \WPLawMETintererrSysfourty _{s}$   &  $\WPLbwMETslopefourty ^{+ \WPLbwMETslopeUPfourty }_{- \WPLbwMETslopeDOWNfourty }\pm \WPLbwMETslopeerrSysfourty _{s}$   &    $\WPLbwMETinterfourty ^{+ \WPLbwMETinterUPfourty }_{- \WPLbwMETinterDOWNfourty }\pm \WPLbwMETintererrSysfourty _{s}$ 

\enddata
\tablecomments{ Same as Table~\ref{tab:WPLmodels} but model fits are now a function of metallicity using Eq.\ref{eq:MetEq}. \WPLConIntfourty$\%$ uncertainties and confidence intervals are presented.  Table discussed in \S\ref{sec:WPL} and \S\ref{sec:Disgusted}.
}
\label{tab:WPLmodelsMET}
\end{deluxetable}

\begin{deluxetable}{lcccc}
\tablecolumns{4}
\tablewidth{0pc}
\tabletypesize{\scriptsize}
\tablecaption{Model Wesenheit P-L Comparison}
\tablehead{
\colhead{Models Compared} &
\multicolumn{2}{c}{Full Sample} &
\multicolumn{2}{c}{Restricted Sample} \\
\colhead{} &
\colhead{F-Test Confidence} &
\colhead{$\Delta \chisq$} &
\colhead{F-Test Confidence} &
\colhead{$\Delta \chisq$} 
}
\startdata
Model 2 vs. Model 1   &   \WPLALONEvFSLOPE$\%$ & \WPLALONEvFSLOPEChi & \WPLALONEvFSLOPEfourty$\%$ & \WPLALONEvFSLOPEChifourty \\
Model 4 vs. Model 2   &   \WPLFREEvFSLOPE$\%$ &   \WPLFREEvFSLOPEChi & \WPLFREEvFSLOPEfourty$\%$  & \WPLFREEvFSLOPEChifourty \\
\enddata
\tablecomments{$ $Restricted Sample has a $\WPLPerLimitfourty{}$ day maximum period cut imposed. Table discussed in \S\ref{sec:WPL} and \S\ref{sec:Disgusted}.
}
\label{tab:WPLmodelsComp}
\end{deluxetable}

\begin{deluxetable}{lllrrrrrc}
\tablecolumns{9}
\tablewidth{0pc}
\tabletypesize{\scriptsize}
\tablecaption{Other Objects}
\tablehead{
\colhead{} &
\colhead{R.A.} &
\colhead{Dec.} &
\colhead{X} &
\colhead{Y} &
\multicolumn{2}{c}{Magnitudes} &
\colhead{} &
\colhead{$12+log(O/H)$} \\
\colhead{ID} &
\colhead{(J2000)} &
\colhead{(J2000)} &
\colhead{pixels} &
\colhead{pixels} &
\colhead{V} &
\colhead{I} &
\colhead{$\rho{}/\rho{}_{iso}$} &
\colhead{(dex)}
}
\startdata
      F1-13 &  210.86366905 &   54.35792401 & 2375.8 & 2560.5 &  20.763  &  20.145  & 0.1577 & 8.61  \\
      F1-29 &  210.84194069 &   54.34349859 & 3688.0 & 2994.3 &  21.116  &  20.243  & 0.1012 & 8.66  \\
     F1-113 &  210.88622020 &   54.37159190 & 1100.4 & 2074.4 &  21.601  &  20.522  & 0.2286 & 8.54  \\
      F1-54 &  210.83356976 &   54.33960030 & 4083.5 & 3208.2 &  21.381  &  20.970  & 0.0868 & 8.67  \\
     F1-258 &  210.85598559 &   54.37583152 & 1314.9 & 3360.8 &  22.239  &  21.909  & 0.1771 & 8.59  \\
     F1-347 &  210.81751077 &   54.34294154 & 4125.0 & 3922.7 &  22.660  &  22.840  & 0.0450 & 8.71  \\
      F2-24 &  210.75884312 &   54.34062464 & 1344.9 &  595.3 &  20.549  &  19.616  & 0.1160 & 8.65  \\
      F2-91 &  210.76574754 &   54.35390318 &  350.0 &  685.4 &  21.265  &  20.443  & 0.0919 & 8.67  \\
      F2-37 &  210.70780831 &   54.32347179 & 3292.9 & 2118.5 &  20.917  &  20.519  & 0.2704 & 8.51  \\
      F2-48 &  210.69712403 &   54.35234385 & 1533.4 & 3312.8 &  21.029  &  21.079  & 0.2601 & 8.52  \\
     F2-660 &  210.70992763 &   54.34032812 & 2134.3 & 2490.6 &  22.718  &  21.845  & 0.2367 & 8.54  \\
     F2-367 &  210.70670420 &   54.34630827 & 1785.8 & 2777.3 &  22.297  &  22.200  & 0.2392 & 8.53  \\
\enddata

\tablecomments{R.A. and Dec. are in units of degrees.  \Astrometry Table discussed in \S\ref{sec:VarObjs}.}
\label{tab:VarObj}
\end{deluxetable}

\begin{deluxetable}{lrllrrrrrc}
\tablecolumns{10}
\tablewidth{0pc}
\tabletypesize{\scriptsize}
\tablecaption{Other Possible Periodic Objects}
\tablehead{
\colhead{} &
\colhead{Period} &
\colhead{R.A.} &
\colhead{Dec} &
\colhead{X} &
\colhead{Y} &
\multicolumn{2}{c}{Magnitudes} &
\colhead{} &
\colhead{$12+\log{(\textrm{O}/\textrm{H})}$} \\
\colhead{ID} &
\colhead{(days)} &
\colhead{(J2000)} &
\colhead{(J2000)} &
\colhead{pixels} &
\colhead{pixels} &
\colhead{V} &
\colhead{I} &
\colhead{$\rho{}/\rho{}_{iso}$} &
\colhead{(dex)}
}
\startdata
     F1-574 &  0.508 &  210.82228318 &   54.34545471 & 3880.3 & 3808.9 &  22.806  &  22.747  & 0.0516 & 8.70  \\
     F2-271 & 10.779 &  210.75024732 &   54.35297493 &  656.0 & 1263.2 &  21.998  &  21.962  & 0.1291 & 8.63  \\
\enddata

\tablecomments{R.A. and Dec. are in units of degrees. \Astrometry Table discussed in \S\ref{sec:VarObjs}.}
\label{tab:PerObj}
\end{deluxetable}

\begin{deluxetable}{lcc}
\tablecolumns{3}
\tablewidth{0pc}
\tabletypesize{\normalsize}
\tablecaption{Pixel Coordinate Conventions}
\tablehead{
\colhead{Software} &
\multicolumn{2}{l}{Coordinate of Center of Lower Left Pixel} \\
\colhead{} &
\colhead{x pixel} &
\colhead{y pixel} 
}
\startdata

DOLPHOT  \dotfill & 0.5 & 0.5 \\ 
IRAF     \dotfill & 1.0 & 1.0 \\
DAOPHOT  \dotfill & 1.0 & 1.0 \\
ISIS     \dotfill & 0.0 & 0.0 \\
DS9$^{a}$      \dotfill & 1.0 & 1.0 \\  

\enddata
\tablecomments{Table discussed in \S\ref{sec:isis_psf}.\\
$^{a}$ \ \citet{ds9}
}

\label{tab:Pixels}
\end{deluxetable}

\end{document}